\documentclass[a4paper,fleqn]{cas-sc}
\usepackage[authoryear]{natbib}
\bibpunct{(}{)}{;}{a}{,}{,}
\usepackage{multirow}
\usepackage{amsmath}
\usepackage{rotating,float,caption}
\usepackage{subcaption}

\usepackage{chemformula}
\usepackage[dvipsnames]{xcolor}
\usepackage{comment}
\usepackage{longtable}
\def\tsc#1{\csdef{#1}{\textsc{\lowercase{#1}}\xspace}}
\tsc{WGM}
\tsc{QE}

\definecolor{KA}{rgb}{0.76, 0.13, 0.28}
\definecolor{KA}{rgb}{0.11, 0.56, 0.33}
\definecolor{SA}{rgb}{0.15, 0.66, 0.13}

\begin{document}
\let\WriteBookmarks\relax
\def\floatpagepagefraction{1}
\def\textpagefraction{.001}

\newcommand{\ex}[0]{\varepsilon_z}
\newcommand{\vs}[0]{\upsilon}
\newcommand{\fk}[0]{f_\kappa}
\newcommand{\xil}[0]{\xi_\lambda}

\newcommand{\aj} {"Astron. J."} 
\newcommand{\actaa} {"Acta Astron."}
\newcommand{\araa} {"Annu. Rev. Astron. Astrophys."}
\newcommand{\apj}  {"Astrophys. J."}
\newcommand{\apjl} {"Astrophys. J. Lett."}
\newcommand{\apjs} {"Astrophysical Journal Supplement Series"}
\newcommand{\ao} {"Appl. Optics"}
\newcommand{\apss} {"Astrophys. Space Sci."}
\newcommand{\aap} {"Astronom. Astrophys."}
\newcommand{\aapr} {"Astron. Astrophys Rev"}
\newcommand{\aaps} {"Astron. Astrophys. Sup."}
\newcommand{\azh} {"Astron. Zh+"}
\newcommand{\caa} {"Chinese Astron. Astr."}
\newcommand{\icarus} {"Icarus"}
\newcommand{\jcap} {"J. Cosmol. Astropart. Phys."}
\newcommand{\jrasc} {"J. Roy. Astron. Soc. Can."}
\newcommand{\memras} {"Memoirs of the RAS"}
\newcommand{\mnras} {"Mon. Not. R. Astron. Soc."}
\newcommand{\na} {"New Astron."}
\newcommand{\nar} {"New Astron. Rev."}
\newcommand{\pra} {"Phys. Rev. A"}
\newcommand{\prb} {"Phys. Rev. B"}
\newcommand{\prc} {"Phys. Rev. C"}
\newcommand{\prd} {"Phys. Rev. D"}
\newcommand{\pre} {"Phys. Rev. E"}
\newcommand{\prl} {"Phys. Rev. Lett"}
\newcommand{\pasa} {"Publ. Astron. Soc. Aust."}
\newcommand{\pasp} {"Publ. Astron. Soc. Pac."}
\newcommand{\pasj} {"Publ. Astron. Soc. Jpn."}
\newcommand{\rmxaa} {"Rev. Mex. Astron. Astr."}
\newcommand{\rjras} {"Q. J. Roy. Astron. Soc."}
\newcommand{\skytel} {"Sky Telescope"}
\newcommand{\solphys} {"Sol. Phys."}
\newcommand{\sovast} {"Sov. Astron."}
\newcommand{\ssr} {"Space Sci. Rev."}
\newcommand{\zap} {"Zeitschrift fuer Astrophysik"}
\newcommand{\nat} {"Nature"}
\newcommand{\iaucirc} {"IAU Cirulars"}
\newcommand{\gca} {"Geochim. Cosmochim. Ac."}
\newcommand{\grl} {"Geophys. Res. Lett."}
\newcommand{\jcp} {"J. Chem. Phys."}
\newcommand{\jgr} {"J. Geophys. Res."}
\newcommand{\jqsrt} {"J. Quant. Spectrosc. RA"}
\newcommand{\nphysa} {"Nucl. Phys. A"}
\newcommand{\physrep} {"Phys. Rep."}
\newcommand{\physscr} {"Phys. Scrip."}
\newcommand{\planss} {"Planet. Space Sci."}
\newcommand{\baas} {"Bull. Aust. Acoust. Soc"}
\newcommand{\aplett} {"Astrophys. Lett."}
\newcommand{\procspie} {"Proc. SPIE"}
\newcommand{\cjaa} {"Chinese J. Astron. Ast."}
\newcommand{\fcp} {"Fundam. Cosm. Phys."}
\newcommand{\memsai} {"Mem. Soc. Astron. Ital."}
\newcommand{\bain} {"Bull. Astron. Inst. Neth., Suppl. Ser."}
\newcommand{\psj} {"PSJ"}
\shorttitle{Extent of formation of molecules in cometary comae}
\shortauthors{Ahmed and Acharyya}

\title[mode = title]{The extent of formation of organic molecules in the comae of comets showing relatively high activity}

\author[1,2]{Sana Ahmed}[orcid=0000-0002-1258-6032]






\affiliation[1]{organization={Planetary Sciences Division, Physical Research Laboratory},
	addressline={University Area}, 
	city={Ahmedabad},
	  citysep={}, 
	postcode={380009}, 
	state={Gujarat},
	country={India}}

\affiliation[2]{organization={School of Physics, Trinity College Dublin},
	addressline={College Green}, 
	city={Dublin},
	  citysep={}, 
	postcode={2}, 
	country={Ireland}}
 
\author[1]{Kinsuk Acharyya}[orcid=0000-0002-0603-8777]







\begin{abstract}
Comets are a rich reservoir of complex organic molecules. Ground and space-based observatories have 
recently greatly enhanced the cometary molecular inventory. Although these molecules' origin is believed 
to be the cometary nucleus, they can be partially synthesised in the coma. We studied organic molecules' 
nucleus versus coma origins for various initial conditions, using a multifluid chemical-hydrodynamical 
model and an updated chemical network. For the study, we considered four comets [C/1996 B2 (Hyakutake), 
C/2012 F6 (Lemmon), C/2013 R1 (Lovejoy), and C/2014 Q2 (Lovejoy)] due to their relatively high activity 
and observation of large number organics species. We emphasised on the C-H-O and N-bearing species, including 
the simplest amino acid, glycine. We discuss the formation pathways of the organics and the conditions for 
their formation in the coma and find that the abundance varies from one comet to another due to differences 
in the initial conditions, relative abundances of the reactants and temperature. We compare the organic 
abundances when they are present as parent volatiles to their formation solely due to gas-phase chemistry. 
Their abundance purely due to the coma chemistry is moderately to significantly lower compared to that when 
they are parent volatiles. However, we find that the production rates of some of the coma-synthesised organic 
molecules can reach peak values of $\sim 10^{22} - 10^{26}$ molecules s$^{-1}$, which is in the realm of 
detection by in situ/space-based observations, and can therefore be important considering future missions to 
comets. We also compare our modeled abundances with those observed in 67P/C-G by {\it Rosetta}, which 
detected several organics at a large heliocentric distance and low production rate.
\end{abstract}


\maketitle

\section{Introduction} \label{section:intro}
Planetary systems like ours are formed due to the gravitational collapse of interstellar matter. 
In various phases of this grand journey, simple through complex molecules are abundantly found, 
implying that the increase in molecular complexity is a part of this journey. Comets are made up of the 
leftover volatile material that are the relics of the protoplanetary disk that formed the Solar 
System. The frozen volatile ices in the cometary interiors can represent the oldest and largely 
unprocessed material in the Solar System. Therefore, comets are the best candidates to study the 
early formation history of the Solar System. Although measurements of the D/H ratio indicate that comets 
contributed to a relatively small fraction of the terrestrial water, yet they probably  delivered 
organics and prebiotic material to the planet by impact processes \citep{Chyba1990,Ehrenfreund2002}. 
Organics are the seeds for creating molecules of biological interest, and investigations of their formation 
in comets provide clues in understanding the prebiotic chemistry.

	
The \textit{Rosetta} mission (with on-board instruments of unprecedented sensitivity) to the comet 67P/Churyumov-Gerasimenko 
(henceforth 67P/C-G) resulted in the identification of many new molecules, particularly organic molecules, and the number of 
known molecular species in comets more than doubled as a result of this mission \citep{Altwegg2019, Morse2019}. The current number 
of volatile parent species (namely those present in the ices in the interior of the cometary nucleus), including tentative 
detections, stands at 72, out of which 37 are complex organic molecules. ROSINA/\textit{Rosetta} also detected organic 
species with abundance $<$ 0.01 \% with respect to water \citep{LeRoy2015}, and considering the lower activity of 67P/C-G, 
it proves that the detection of bigger molecules with very low abundance ($\sim$ 10$^{20}$ molecules s$^{-1}$) is possible, be 
it a parent volatile or produced via coma chemistry. In this context, it is also relevant to inquire about how we can know 
if a species is sublimating from a cometary nucleus or produced via coma chemistry.
Coma mapping has revealed that molecules including \ch{C2}, \ch{CN}, \ch{HNC} and \ch{H2CO} have distributed sources in
the coma. These species show extended and relatively flat spatial profiles \citep{AHearn1995, Cottin2008, Cordiner2014}, as opposed to species releasing directly from the nucleus showing
a strong centrally peaked spatial profiles.

The most abundant volatile in a majority of comets is water, followed by carbon monoxide and carbon dioxide, that are present 
at the $1-20\%$ level compared to \ch{H2O} \citep{Cochran2015,Biver2019}. Organic species such as methane (\ch{CH4}), 
ethane (\ch{C2H6}), methanol 
(\ch{CH3OH}) and hydrogen cyanide (\ch{HCN}) have been observed in comets since the 1990s at infrared and 
submillimeter wavelengths \citep{Mumma2011, DelloRusso2016}. Simpler radicals like diatomic carbon (\ch{C2}) 
and cyanide (\ch{CN}), detected at optical wavelengths, are most likely to be released by the photodissociation of 
acetylene (\ch{C2H2}) and \ch{HCN}, though their origin is still not understood completely. Complex 
molecules such as ethanol (\ch{C2H5OH}), formamide (\ch{NH2CHO}), acetaldehyde (\ch{CH3CHO}), glycolaldehyde 
(\ch{CH2OHCHO}), ethylene glycol [\ch{(CH2OH)2}] and methyl formate (\ch{HCOOCH3}) have been detected in comets 
at millimeter and submillimeter wavelengths \citep{BockeleeMorvan2000,BockeleeMorvan2004, Crovisier2004a,
Mumma2011,Biver2014,Biver2015}. Isotopic measurements of the dust samples of comet 81P/Wild 2 collected by \textit{Stardust} 
revealed the presence of extraterrestrial glycine (\ch{NH2CH2COOH}), which is the simplest amino acid \citep{Elsila2009}. 
Later, glycine was also detected by \textit{Rosetta} \citep{Altwegg2016}, confirming its presence in comets. The formation of 
these molecules in comets is yet to be understood, and hence their study is of paramount importance.
	
We can take clues from the proposed pathways for the formation of COMs in the interstellar medium and apply them for 
the explanation of these molecules in comets. One of the most widely used mechanisms proposed to explain the formation of organics 
in the interstellar medium is their synthesis on the surface of dust grains.
However, in a recent study, \cite{Cordiner2021} found that \ch{HC3N} (cyanoacetylene) and 
\ch{NH2CHO} (formamide) can be efficiently produced via two-body neutral-neutral reactions in cometary comae. Thus, all of 
the molecules observed in the coma of comets may not originate from the ices inside the cometary nucleus. A similar scenario 
exists for organics in the interstellar medium as well. For example, the detection of acetaldehyde (\ch{CH3CHO}), 
dimethyl ether (\ch{CH3OCH3}), methyl formate (\ch{HCOOCH3}), and ketene (\ch{CH2CO}) in the cold (T $\le$ 10 K) 
prestellar core L1689B \citep{Bacmann2012}, and COMs in the cold core B1-b \citep{Cernicharo2012}, the solar-type protostar 
IRAS 16293-2422 \citep{Jaber2014} and the prestellar core L1544 \citep{Jimenez2016} are difficult to explain if the 
hypothesis that the COMs are exclusively formed on the dust-grain surface is correct. For all these sources, the involvement 
of gas-phase pathways is required, which resulted in the revisitation of gas-phase formation pathways of COMs \citep{Balucani2015, 
Skouteris2018}.
	
The work by \cite{Giguere1978} was one of the first attempts to comprehensively model 
the coma chemistry. \cite{Giguere1978} showed that several species such as \ch{H3O+}, \ch{NH4+} and \ch{O2}, which are not simple 
photo-products of parent species, can reach large abundances; a subsequent work by \cite{Huebner1980} demonstrated the importance 
of photodissociative ionization. In this context, it is pertinent to note that \textit{Rosetta} measurements show that \ch{O2} 
can be a parent species \citep{Bieler2015}. 
These models also demonstrated that the solar UV radiation field cannot penetrate into the
inner regions of the coma, and optical depth calculations need to be taken into account for greater accuracy. In addition to the 
common parent volatiles namely \ch{H2O}, \ch{CO}, \ch{CO2}, \ch{N2}, \ch{NH3} and \ch{CH4}, \cite{Mitchell1981} present chemical 
model results of cometary comae that also include other volatile species having composition similar to that of interstellar molecular 
clouds. \cite{Mitchell1981} find that chemical coma models that include a more complex mixture of parent volatiles can better predict 
the observed abundances of neutral radicals such as \ch{C2}, \ch{C3}, \ch{CH} and \ch{NH2}. MHD studies by \cite{Wegmann1987} and 
\cite{Schmidt1988}, that also included chemistry, were made for the gas and plasma flow around 1P/Halley during the \textit{Giotto} encounter. 
Subsequently, gas-phase formation of organic molecules observed in comet Hale-Bopp was modeled by \cite{Rodgers2001}. 
\cite{Glinski2004} explored the extent to which the photodissociation products of water affect the overall oxygen/hydrogen chemistry 
in the inner coma, including their effect on small hydrocarbons. \cite{Pierce2010} examined the extent to which the presence of organic 
molecules such as \ch{H2CO} and \ch{CH3OH} affect the \ch{CO} chemistry in the inner coma. 
A global model by \cite{Boice2017}, that 
included chemical kinetics, gas-dust interaction, and subsurface sublimation from the comet nucleus, can give an improved knowledge 
of the relationship of the coma with the ices in the nucleus. 

Since the last decade has seen the detection of COMs 
in more comets, including the \textit{Rosetta} measurements,
the gas-phase formation of organic molecules in 
the coma cannot be ruled out completely and it is pertinent to revisit it comprehensively and delineate the contribution from 
gas-phase coma chemistry. This requires an extensive chemical network which should include the proposed formation pathways 
for COMs and the chemical reactions traditionally used to study the cometary coma. In this work, we have used a combined 
chemical-hydrodynamical multifluid coma model to study four comets, namely, C/1996 B2 (Hyakutake), C/2012 F6 (Lemmon), C/2013 
R1 (Lovejoy), and C/2014 Q2 (Lovejoy), which show moderate to high activity. Besides, many organic molecules have been detected 
in these comets, making them excellent candidates for the task. The close approach to Earth of C/1996 B2 (Hyakutake) led 
to a significant improvement in our knowledge of cometary compositions, with the first cometary identifications of \ch{HNC}, 
\ch{HNCO} and \ch{CH3CN} at millimeter wavelengths \citep{Dutrey1996,Irvine1996,Lis1997}, and the first IR detection of 
the hydrocarbons \ch{CH4}, \ch{C2H6} and \ch{C2H2} \citep{Brooke1996,Mumma1996}. \ch{NH2CHO}, \ch{(CH2OH)2} and \ch{CH3CHO} 
were re-detected in C/2012 F6 (Lemmon) and C/2013 R1 (Lovejoy) after their initial discovery in comet Hale-Bopp \citep{Biver2014}. 
The spectral survey of C/2014 Q2 (Lovejoy) resulted in the first confirmed identification of \ch{C2H5OH} and \ch{CH2OHCHO} 
in a comet \citep{Biver2015} and presented evidence of the first cometary detection of \ch{C2H4} from a ground-based 
facility \citep{DelloRusso2022}.
	
Our main goal for this work is to determine the extent of molecule complexity that can occur in the cometary 
coma and its dependence on the initial production rates of various parent species, especially smaller parent 
species producing COMs. Section \ref{section:model} briefly describes the multifluid model used and the comets 
that have been studied in this work. The results and discussions are presented in Section \ref{section:reslt} 
and Section \ref{section:discu} respectively, and the concluding remarks are made in Section \ref{section:concl}. 
An overview of organic molecules' gas phase formation mechanisms in the coma, including new reactions, is given 
in the Appendix.

	
\section{Coma Model} \label{section:model}
\subsection{Hydrodynamic Model}
We use a combined hydrodynamic and chemical model suitable for a spherically symmetric outflow of the coma gas  
in the steady state, which is described in detail in \cite{Ahmed2021} and the main aspects are briefly described here.
The model assumes a hydrodynamic flow in a coma that is collisionally dominated. 
This assumption is valid in the inner region of the coma; for a Halley-type comet, it is $\sim$ 10$^4$ km at 1 au \citep{Rodgers2004}.
The solar wind does not penetrate this region and the ions are predominantly of cometary origin.
As the comet approaches the sun, the size of the collisional region will increase since it is proportional to the total gas 
production rate. It can be approximately calculated by considering the cometocentric distance `d' at which the mean free path of the particle 
is equal to `d'.


While early cometary models (for example, \citealp{Oppenheimer1975, Giguere1978, Huebner1980, Mitchell1981, Biermann1982}) 
established the importance of ion-neutral and recombination reactions towards coma chemistry, 
these models assumed constant temperature and velocity to reduce complexity. Susbsequent models illustrated 
the non-viable nature of the assumptions of constant temperature and velocity (for example, 
\citealp{Marconi1982, Marconi1983, Marconi1986, Ip1983, Gombosi1986, Combi1988}). The initial adiabatic 
expansion of the gas outwards from the nucleus leads to a drop in temperature in the inner coma regions 
while photochemical heating increases the temperature (particularly that of the electrons) further outwards. 
The rates of the recombination reactions of electrons with ions is thus reduced in the outer coma, due to 
the high temperature of the electrons. Hence, the chemistry and physics of the coma are intimately
linked, and a complete description of the coma requires a combined chemical-hydrodynamical multifluid 
model (for example, \citealp{Korosmezey1987, Wegmann1987, Schmidt1988, Rodgers2002, Weiler2012, Ahmed2024}).
Though Monte Carlo techniques give a more accurate description of the coma dynamics (for example, 
\citealp{Combi1988, Hodges1990}), they are computationally very expensive, and extensive chemistry 
cannot be included in such coma models. Since our aim is to study the coma chemistry of organic 
molecules in detail, the fluid approximation works well for us. In the inner collisional coma, 
where most of the chemistry takes place, the coma properties derived from Monte Carlo models and 
fluid models are sufficiently well-matched. \citep{Combi2004, Rodgers2004}. Multifluid models can 
be used to model the coma with production rates as high as $10^{29} - 10^{31}$ molecules s$^{-1}$ \citep{Rodgers2001, Rodgers2005}.

The chemical species in the coma interact with one another, resulting in an active gas-phase coma chemistry, 
and some of these chemical reactions include collisional reactions between ions and neutrals, recombination reactions, and 
electron impact reactions.  The energy released due to exothermic chemical reactions is distributed non-uniformly among the 
product species, leading to unequal heating of the species involved in the reaction. Thus, we use a multifluid approach, 
whereby the neutral species, the ionic species, and the electrons are treated as separate fluids, each of them having different 
temperatures. Determining the temperature is essential since most of the chemical reaction rates are temperature-dependent. All 
three fluids are assumed to flow outwards with a single bulk velocity.
There is exchange of energy between the three fluids due to elastic (ion-neutral, 
electron-neutral and electron-ion) collisions, and inelastic (electron-neutral and neutral-neutral) scattering processes. The 
ion-neutral elastic collisions are treated as a special case of the ion-neutral reactions and the subsequent energy exchange between 
the two fluids is calculated by the scheme suggested by \cite{Draine1986}. The electron-ion collisions are due to Coulomb interactions 
and the energy thus exchanged was given by \cite{Draine1980}. For electron-neutral elastic scattering, the collisions of electrons with 
\ch{H2O}, \ch{CO} and \ch{CO2} are considered, as they are the most abundant neutral species in the coma. The collision cross sections of 
electrons with these neutral gases was compiled by \cite{Itikawa2002, Itikawa2015} and \cite{Itikawa2005}, and the energy exchange is 
calculated in a manner similar to the ion-neutral collisions, by considering electrons instead of the ions. The inelastic electron-neutral 
scattering results in the rotational or vibrational excitation of the neutral species, leading to cooling of the electron fluid, and we 
have considered the excitation of \ch{H2O} and \ch{CO}. The analytical expressions given by \cite{Cravens1986} are used to calculate the 
cooling of the electron fluid due to rotational and vibrational excitation of water molecules. In \cite{Ahmed2021}, we calculated the 
exchange of energy due to vibrational excitation of \ch{CO} by electrons, in the manner prescribed by \cite{Waite1981}. The excitation of 
\ch{CO} to higher electronic states by electron impacts is also included, following \cite{Schmidt1988}. Water-water inelastic collisions 
are included as the neutral-neutral inelastic scattering process, and the resultant energy lost by the neutral fluid is given by \cite{Shimizu1976}.  
	
The governing equations for our model are obtained by using the principles of conservation of number density, mass, momentum and energy, 
for each fluid \citep{Rodgers2002, Weiler2006Thesis, Holscher2015Thesis}. These are a set of coupled first-order differential equations 
that can be numerically integrated for some given boundary conditions, to obtain the species number densities, fluid temperatures, and 
outflow velocity. We use Rosenbrock methods for numerical integration and adopt the Fortran \texttt{stiff} integrator based on the Kaps-Rentrop 
algorithm as the numerical scheme \citep{Press1992}.
	
\subsection{Chemical Network}
The chemistry occurring in the gas-phase cometary coma is driven by ion-neutral and neutral-neutral reactions and dissociative 
recombination of ions with electrons. Many of these ions originate from the photoionization and photodissociative ionization reactions undergone 
by the parent species. Protonated species are also found in abundance in the coma, and these are formed when a hydrogen cation attaches to a 
neutral molecule. When collision occurs between an \ch{H+}-bearing ion (protonated species) and a neutral molecule having 
higher proton affinity, transfer of \ch{H+} cation takes place to the neutral molecule. A proton transfer "chain" is set up by successive 
transfer of \ch{H+} from one molecule to another, moving in the direction of increasing proton affinity, and this is a significant 
feature of the coma chemistry. Dissociative recombination reactions, wherein ions combine with electrons to form neutral molecules, are also 
important in forming new species. In this section we will not repeat the detailed discussions on ion chemistry 
and formation of simple molecules which can be found in existing literature \citep{Marconi1983, Gombosi1986, Schmidt1988, Rodgers2002, 
Weiler2006Thesis, Holscher2015Thesis}. 

The rate coefficients (cm$^{3}$ s$^{-1}$) of bimolecular 
processes are expressed in the form of a "so-called" modified Arrhenius formula:

\begin{equation}
	k(T) = \alpha \left(\frac{T}{300\text{ K}}\right)^{\beta}e^{-\gamma/T}
	\label{eq:arrhenius}
\end{equation}
where $\alpha$ , $\beta$ and $\gamma$ are parameters obtained from literature and databases. 

\cite{Woon2009} used the 
Su-Chesnavich approach to calculate the rate coefficients for those ion-neutral reactions whose rates 
are unmeasured in the laboratory. The rate coefficient for ion-neutral reactions involving a non-polar 
neutral species is given by the Langevin expression $k_L = 2\pi e \sqrt{\alpha_P / \mu}$, where $e$ is 
the electronic charge, $\alpha_P$ is the average dipole polarizability of the neutral species, and $\mu$ 
is the reduced mass of the ion and neutral species. For neutral species possessing a dipole moment $\mu_D$, 
a unitless parameter $x$ is defined to delineate the temperature ranges, such that 
$x = \mu_D / \sqrt{2\alpha_P k_B T}$, $k_B$ and $T$ being the the Boltzmann constant and temperature, 
repsectively. If all the quantities are in cgs-esu units, $x\ge 2$ and $x<2$ correspond to the low and 
high temperature ranges, respectively. In this case, the rate coefficients for ion-dipole collisional 
reactions in terms of the parameters $x$ and $k_L$ are:

\begin{equation}
\frac {k(T)} {k_L} = 0.4767x + 0.62 \quad \text{if $x\ge 2$}
\label{eq:ionpol1_kL}
\end{equation}

\begin{equation}
\frac {k(T)} {k_L} = \frac {(x+0.5090)^2} {10.526} + 0.9754 \quad \text{if $x< 2$}.
\label{eq:ionpol2_kL}
\end{equation}

Equations \ref{eq:ionpol1_kL} and \ref{eq:ionpol2_kL} can be re-written using the parameters $\alpha$, 
$\beta$ and $\gamma$, as Equations \ref{eq:ionpol1} and \ref{eq:ionpol2}, respectively.

\begin{equation}
k(T)=\alpha \beta \left[0.62+0.4767\gamma  \left(\frac {300 \text{ K}} T\right)^{0.5}\right]
\label{eq:ionpol1}
\end{equation}
\begin{equation}
k(T)=\alpha \beta \left[1+0.0967\gamma \left(\frac {300 \text{ K}} T\right)^{0.5}+\frac {\gamma^2} {10.526} 
\frac {300 \text{ K}} T\right].
\label{eq:ionpol2}
\end{equation}
Here, $\alpha$ is the branching ratio of the reaction, $\beta$ is the Langevin rate and $\gamma$ is the value 
of $x$ at 300 K. For $x=0$, the rate coefficient reduces to the Langevin expression. Since the ions and 
neutrals have different temperatures ($T_i$ and $T_n$, respectively), we calculate the rate coefficients 
at an effective temperature $T_{\text {eff}}$, as suggested by \cite{Flower1985}. 

\begin{equation}
T_{\text {eff}} = \frac {m_i T_n + m_n T_i} {m_i + m_n},
\end{equation}
where $m_i$ and $m_n$ are the masses of the reacting ion and neutral species, respectively. 
We also use the effective temperature in the Arrhenius formula (Equation \ref{eq:arrhenius}) if the reacting
species belong to different fluids. The reactions having electrons as one of the reactants have
$T_{\text {eff}} \approx T_e$, since the mass of the electron is much less than the mass of the other 
reacting species.

Some abundant cometary neutrals and ions do not react chemically with each other. These include the parent
molecules \ch{H2O}, \ch{CO2} and \ch{CO} that do not undergo chemical reactions with the cometary ions \ch{H3O+}, 
\ch{NH4+}, \ch{CH3OH+}, \ch{CH3OH2+}, and \ch{HCNH+}. These species undergo elastic scattering with each other, 
where the ion and neutral species do not get altered chemically, but exchange energy during collisions. We model 
these as ion-neutral `pseudo reactions' for which \cite{Weiler2006Thesis} calculated the rate coefficients using hard-sphere collision theory, providing an expression for the rate as follows \citep{Connors1990}:

\begin{equation}
	k(T) = (r_i + r_n)^2 \left(\frac {8\pi k_B T} {\mu}\right)^{1/2} e^{-E/k_B T},
\end{equation}
where $r_i$ and $r_n$ are the radii of the ion and neutral species, respectively, $\mu$ is the reduced mass and the temperature to be used is the effective temperature. The energy threshold $E$
is assumed to be zero. Using typical molecular and ion radii, one obtains the Arrhenius
coefficients of the elastic collision rates:
$\alpha = 10^{-10}$ cm$^3$ s$^{-1}$, $\beta = 0.5$ and $\gamma = 0$. 
This treatment is a simplification because it does not take into account the intermolecular forces between the ion and the neutral species, for example Coulomb interactions between the ion and the induced dipole on the neutral molecule. This is especially true for the case of elastic collisions of ions with water, as \ch{H2O} has a large dipole moment.  
Alternatively, \cite{Rodgers2004} suggest the use of the Langevin rate ($\sim 10^{-9}$ cm$^3$ s$^{-1}$) for the ion-neutral elastic scattering.

Most of the chemical reactions and the parameters required to calculate the reaction rates using Equations \ref{eq:arrhenius}, 
\ref{eq:ionpol1} or \ref{eq:ionpol2} are taken from the KIDA database (\citealp{Wakelam2015}, \url{http://kida.astrophy.u-bordeaux.fr/}). 
Additional reactions involving organics are taken from \cite{Skouteris2018}, the OSU chemical network \citep{Garrod2007}, and the 
UMIST (\citealt{McElroy2013}, \url{http://udfa.ajmarkwick.net/index.php}) and NIST (\citealp{Manion2008}, \url{https://kinetics.nist.gov/}) 
databases. The reaction rates for the electron impact excitation and radiative de-excitation of the electronic states of \ch{CO} are taken 
from \cite{Schmidt1988}. The photochemical reactions are taken from \cite{Weiler2006Thesis}, \cite{Huebner2015} and \cite{Heays2017}. The dissociation 
of the major parent species \ch{H2O}, \ch{CO}, \ch{CO2} and \ch{N2} by solar UV photons and by photoelectrons creates metastable excited 
states of atomic  oxygen, carbon, and nitrogen. We have included these reactions in our network, as well as the de-excitation of the 
metastable states due to radiative or collisional quenching; the rates for these processes are taken from \cite{Raghuram2013} and 
\cite{Raghuram2020}. Most other reactions in the network that involve atomic \ch{O}, \ch{C} and \ch{N} are for the ground state of 
the atomic species; in case the atom is in a metastable excited state, this is specified in the network.  In total we have about 
480 species which are connected by nearly 5000 reactions. An overview of organic molecules' gas phase formation mechanisms in the coma, 
including new reactions is given in Appendix \ref{App_A}.
	
\subsubsection{Calculation of photolytic rates}
The UV radiation field does not remain constant throughout the coma and optical depth effects are present in the regions closer to the 
nucleus. This alters the rates of the photochemical reactions, and they need to be recalculated by accounting for the optical depth. 
If $\phi_{i,\infty}$ is the UV flux at wavelength $\lambda_i$ reaching the outer coma, then the UV flux at a cometocentric distance 
$r$ is given by

\begin{equation}
	\phi_i(r)=\phi_{i,\infty}e^{-\tau_i(r)},
\end{equation}
where $\tau_i$ is the optical depth at wavelength $\lambda_i$, and can be calculated from the wavelength-dependent photochemical 
reaction cross sections $\sigma_i$ (taken from the PHIDRATES database available at \url{https://phidrates.space.swri.edu/}) and 
model-calculated species number densities. Once the wavelength-dependent and distance-dependent UV fluxes are known, the photolytic 
rate coefficient in the wavelength interval $\lambda_i$ and $\lambda_i+\Delta \lambda_i$, and at a distance $r$ can be written as

\begin{equation}
	k_i(r) = \int_{\lambda_i}^{\lambda_i +\Delta \lambda_i}\sigma(\lambda)\phi(\lambda,r) d\lambda.
	\label{eq:photorate}
\end{equation}
This integration can be approximated as  $k_i(r)=\sigma_i \Phi_i(r)$, where  $\sigma_i$ is the wavelength-averaged cross section 
in the $i$-th bin that has a width of $\Delta \lambda_i$, and $\Phi_i(r)$ is the attenuated spectral photon flux integrated over 
the same wavelength bin. The total rate coeffiecient for any photolytic process can then be found out by summing over all the 
wavelength bins, such that
	
\begin{equation}
	k(r)=\sum_i k_i(r).
	\label{eq:photo}
\end{equation}
The intensity of the solar UV flux varies over the 11 year solar cycle. In order to calculate the photolytic rates in the 
coma of a particular comet, we require the spectral UV flux at the time of observation of the comet. This wavelength-dependent solar 
flux data set is available with the LASP Interactive Solar Irradiance Data Center (\url{https://lasp.colorado.edu/lisird/}). The 
FISM2 and the NRLSSI2 flux models are used to derive the daily solar spectral irradiances in the wavelength range $0.1-190$ nm and 
$> 190$ nm, respectively. These fluxes are available at 1 au, and we have scaled them to the heliocentric distance $r_h$ at which 
the cometary observations are carried out, by using the multiplicative factor $r_h^{-2}$. We collect the daily spectral flux data 
for the time over which a comet is observed, and then average out this flux over the cometary observational epoch, which we then 
use to calculate the photochemical rates using Equation \ref{eq:photorate}. Since the UV flux variation with the solar cycle can 
be known from FISM2 and NRLSSI2, we can calculate the corresponding variation in the photochemical reaction rates at different 
points in the solar cycle when the cometary observation is carried out. For those organic molecules whose photolytic 
rates are not listed in the PHIDRATES database, we use the rates given by \cite{Heays2017}. These rates are calculated for a 
UV radiation field defined by \cite{Draine1978}, that is an intensity of $2.6 \times 10^{-6}$ W m$^{-2}$ for 
the radiation integrated between $91.2 - 200$ nm, 
and need to be used with an appropriate scale factor for the solar radiation field. \cite{Heays2017} 
calculate the scale factor to be 37700 for the approximate solar intensity at 1 au during the quiet period of
the solar cycle.

\subsubsection{Calculation of species formation rates} \label{subsubsection:reaction_rates}
The reaction rate per unit volume $\mathcal R _{ij}$ (cm$^{-3}$ s$^{-1}$) of the $j$-th chemical reaction forming the $i$-th 
species is calculated by multiplying the rate coefficient of the process obtained from the relevant rate equation (Equations 
\ref{eq:arrhenius}, \ref{eq:ionpol1}, \ref{eq:ionpol2} or \ref{eq:photo}), by the number densities of the reacting species. 
The reaction rate thus depends upon the abundance of the reacting species, the fluid temperature, and the cometocentric distance 
(for photolytic processes). The net coma formation rate per unit volume $P_i$ for the $i$-th species is obtained by summing the 
reaction rates per unit volume of all the chemical processes contributing to the formation of that species, such that

\begin{equation}
	P_i = \sum_j \mathcal{R}_{ij}. 
	\label{eq:formationrate}
\end{equation}
The relative reaction rate for each chemical process can be calculated as $\mathcal{R}_{ij}/P_i$. The relative reaction rate is a measure 
of the fractional contribution of each chemical process towards the net creation rate per unit volume of a species in the coma.

\subsection{Comet Compositions and Model Runs}
We selected a sample of four Oort cloud comets to study organics' formation in the gas-phase comae. 
These comets are, in chronological order of their discovery, C/1996 B2 (Hyakutake), C/2012 F6 (Lemmon), C/2013 R1 (Lovejoy) 
and C/2014 Q2 (Lovejoy). There are two main reasons for choosing these comets. First, all these four comets 
show high production rates near perihelion \citep{Combi2014, Combi2018, Combi2019} which will make coma chemistry 
more competitive and second, there was detection of a lot of organic 
molecules in these comets, while sensitive upper limits were obtained for some other organics. 
Also, most comets' relative abundances of complex organic molecules vary within one order 
of magnitude \citep{Biver2019}. The organic abundances in short-period Jupiter family comets (JFCs) do not differ significantly 
from Oort cloud comets; however, there are lesser measurements of organics available for JFCs, because they generally exhibit 
comparatively lower activity. Therefore these four comets can be used as templates for other comets.
	
	\begin{table}
		\caption{Input parameters used}	
		\label{table:input}
		\setlength{\tabcolsep}{12pt}
		\renewcommand{\arraystretch}{1.2}	
		\begin{tabular}{ lcccc} 
			\hline
			Comet	              & $Q_{\ch{H2O}}$ (mol s$^{-1}$)         & $r_h$ (au)             & Radius (km) &$T_0$ (K)\\
			\hline 
			C/1996 B2 (Hyakutake) & $1.7\times 10^{29}$		      & 1.06		       & 2.1	     & 233.2	\\ 
			C/2012 F6 (Lemmon)    & $6.7\times 10^{29}$		      & 0.78	               & 5	     & 311.9	\\ 
			C/2013 R1 (Lovejoy)   & $9.13\times 10^{28}$		      & 0.92	               & 1.3	    & 256.3	\\
			C/2014 Q2 (Lovejoy)   & $4.07\times 10^{29}$		      & 1.3		       & 4.3     	& 226.1	\\  
			\hline
			\multicolumn{5}{l}{Visual albedo $A_v=0.04$, infrared emissivity $\epsilon_{\text {IR}}=0.9$} \\
			\hline
		\end{tabular}
		
		\textbf{Notes.} \\
		Reference for $Q_{\ch{H2O}}$: \citealt{Mumma1996} (C/1996 B2), \citealt{Combi2014} (C/2012 F6), \citealt{Combi2018} 
		(C/2013 R1 and C/2014 Q2). \\
		Reference for radius: \url{https://ssd.jpl.nasa.gov/} (C/1996 B2), \citealt{Paradowski2020} (C/2013 R1, C/2014 Q2). 
	\end{table}
	
\begin{table}
\begin{center}
\caption{Species abundance as a percentage with respect to \ch{H2O} ($n_i$) and the corresponding production rate ($Q_i$) in molecules s$^{-1}$ of 
parent volatiles outgassing from the nucleus for the different cometary compositions. 
}	
\label{table:ratio}
\renewcommand{\arraystretch}{1.2}	
\footnotesize
\begin{tabular}{ |l|c|c|c|c|c|c|c|c| } 
\hline
\multirow{2}{*}{Molecule}                  &
\multicolumn{2}{c|}{C/1996 B2 (Hyakutake)} &
\multicolumn{2}{c|}{C/2012 F6 (Lemmon)}    &
\multicolumn{2}{c|}{C/2013 R1 (Lovejoy)}   &
\multicolumn{2}{c|}{C/2014 Q2 (Lovejoy)} \\		\cline{2-9}
              & $n_i$     & $Q_i$                  & $n_i$    & $Q_i$                  & $n_i$    & $Q_i$                  & $n_i$ & $Q_i$                  \\
\hline
\ch{H2O}      & 100       & 1.7 $\times$ 10$^{29}$ &  100     & 6.7 $\times$ 10$^{29}$ &  100     & 9.13$\times$ 10$^{28}$ &  100  & 4.07 $\times$ 10$^{29}$ \\
\ch{CO}       & 19        & 3.2 $\times$ 10$^{28}$ &  4       & 2.7 $\times$ 10$^{28}$ &  7.2     & 6.6 $\times$ 10$^{27}$ &  1.8  & 7.3 $\times$ 10$^{27}$ \\
\ch{CO2}      &  4        & 6.8 $\times$ 10$^{27}$ & ---      & ---                    &  ---     & ---                    &  ---  &    ---                 \\
\ch{CH4}      & 0.8       & 1.4 $\times$ 10$^{27}$ & 0.67     & 4.5 $\times$ 10$^{27}$ & 0.91     & 8.3 $\times$ 10$^{26}$ & 0.75  & 3.1 $\times$ 10$^{27}$ \\
\ch{C2H6}     & 0.6       & 1.0 $\times$ 10$^{27}$ & 0.31     & 2.1 $\times$ 10$^{27}$ & 0.69     & 6.3 $\times$ 10$^{26}$ & 0.68  & 2.8 $\times$ 10$^{27}$ \\
\ch{C2H2}     & 0.5       & 8.5 $\times$ 10$^{26}$ & 0.08$^a$ & 5.4 $\times$ 10$^{26}$ & 0.07$^a$ & 6.4 $\times$ 10$^{25}$ & 0.11  & 4.5 $\times$ 10$^{26}$ \\
\ch{C2H4}     & ---       &  ---                   &    ---   &    ---                 &    ---   &      ---               & 0.22  & 9.0 $\times$ 10$^{26}$ \\
\ch{H2CO}     &  1        & 1.7 $\times$ 10$^{27}$ & 0.7      & 4.7 $\times$ 10$^{27}$ & 0.7      & 6.4 $\times$ 10$^{26}$ & 0.3   & 1.2 $\times$ 10$^{27}$ \\
\ch{CH3OH}    &  2        & 3.4 $\times$ 10$^{27}$ &  1.6     & 1.1 $\times$ 10$^{28}$ & 2.6      & 2.4 $\times$ 10$^{27}$ &  2.4  & 9.8 $\times$ 10$^{27}$ \\
{\bf \ch{HCOOH}}    & ---       & ---                    & 0.07$^a$ & 4.7 $\times$ 10$^{26}$ & 0.12     & 1.1 $\times$ 10$^{26}$ & 0.028 & 1.1 $\times$ 10$^{26}$ \\
{\bf \ch{(CH2OH)2}} & ---       & ---                    & 0.24     & 1.6 $\times$ 10$^{27}$ & 0.35     & 3.2 $\times$ 10$^{26}$ & 0.07  & 2.8 $\times$ 10$^{26}$ \\
{\bf \ch{HCOOCH3}}  & ---       & ---                    & 0.16$^a$ & 1.1 $\times$ 10$^{27}$ & 0.2$^a$  & 1.8 $\times$ 10$^{26}$ & 0.08  & 3.3 $\times$ 10$^{26}$ \\
{\bf \ch{CH3CHO}}   & 0.12$^a$  & 2.0 $\times$ 10$^{26}$ & 0.07$^a$ & 4.7 $\times$ 10$^{26}$ & 0.1      & 9.1 $\times$ 10$^{25}$ & 0.047 & 1.9 $\times$ 10$^{26}$ \\
{\bf \ch{CH2OHCHO}} & ---       & ---                    & 0.08$^a$ & 5.4 $\times$ 10$^{26}$ & 0.07$^a$ & 6.4 $\times$ 10$^{25}$ & 0.016 & 6.5 $\times$ 10$^{25}$ \\
{\bf \ch{C2H5OH}}   & 0.21$^a$  & 3.6 $\times$ 10$^{26}$ & ---      & ---                    & ---      & ---                    & 0.12  & 4.9 $\times$ 10$^{26}$ \\
\ch{H2O2}     & 0.055$^a$ & 9.4 $\times$ 10$^{25}$ & ---      & ---                    & ---      & ---                    & ---   & ---                    \\
\ch{HCN}      & 0.2       & 3.4 $\times$ 10$^{26}$ & 0.14     & 9.4 $\times$ 10$^{26}$ & 0.26     & 2.4 $\times$ 10$^{26}$ & 0.09  & 3.7 $\times$ 10$^{26}$ \\
\ch{HNC}      & 0.01      & 1.7 $\times$ 10$^{25}$ & ---      & ---                    & ---      & ---                    & 0.004 & 1.6 $\times$ 10$^{25}$ \\
\ch{NH3}      & 0.5       & 8.5 $\times$ 10$^{26}$ & 0.58     & 3.9 $\times$ 10$^{27}$ & 0.1      & 9.1 $\times$ 10$^{25}$ & 0.64  & 2.6 $\times$ 10$^{27}$ \\
\ch{HNCO}     & 0.07      & 1.2 $\times$ 10$^{26}$ & 0.025    & 1.7 $\times$ 10$^{26}$ & 0.021    & 1.9 $\times$ 10$^{25}$ & 0.009 & 3.7 $\times$ 10$^{25}$ \\
{\bf \ch{CH3CN}}    & 0.01      & 1.7 $\times$ 10$^{25}$ & ---      & ---                    & ---      & ---                    & 0.015 & 6.1 $\times$ 10$^{25}$ \\
{\bf \ch{NH2CHO}}   & 0.1$^a$   & 1.7 $\times$ 10$^{26}$ & 0.016    & 1.1 $\times$ 10$^{26}$ & 0.021    & 1.9 $\times$ 10$^{25}$ & 0.008 & 3.3 $\times$ 10$^{25}$ \\
{\bf \ch{HC3N}}     & 0.08$^a$  & 1.4 $\times$ 10$^{26}$ & ---      & ---                    & ---      & ---                    & 0.002 & 8.1 $\times$ 10$^{24}$ \\			
\hline 
\end{tabular}
\end{center}
	
\textbf{Notes.} \\
$^a$Indicates the upper limit. \\
Reference. C/1996 B2: \cite{McPhate1996,Lis1997,BockeleeMorvan2004}, C/2012 F6: 
\cite{Biver2014,Biver2016,Paganini2014a, Lippi2020}, C/2013 R1: \cite{Biver2014,Paganini2014b}, C/2014 Q2: \cite{Biver2015,DelloRusso2022}. 
\end{table}
	
We obtained coma model runs for these four comets, namely C/1996 B2, C/2012 F6, C/2013 R1 and C/2014 Q2. The water production rate, size of 
the cometary nucleus and other relevant parameters that we used as inputs for each of these comets are shown in Table \ref{table:input}. We did not 
find the radius of the nucleus of C/2012 F6 in literature, and assumed a value of 5 km for this comet, which is about the average 
cometary size for long period comets. Our assumed value of the visual albedo is widely used to calculate the size of the cometary nucleus 
\citep{Paradowski2020}, while the infrared emissivity value is taken from \cite{Marshall2019}. The heliocentric distances at which the photolytic 
reaction rates are calculated for the different comets are also given in Table \ref{table:input}. $T_0$ is the initial temperature of the coma gas, 
which is calculated as described in Section \ref{subsection:reslt_temp}. The relative abundances $n_i$ (as a percentage) with respect to \ch{H2O} 
for the parent volatiles outgassing from the nucleus, as reported by different authors, are given in Table \ref{table:ratio}. 
The production rate 
$Q_i$ for each species for a particular comet can be found by multiplying these relative abundances with the respective water production rate. 
These species production rates are given in Table \ref{table:ratio}, and they are used as the initial input 
composition for the model runs, 
i.e., they are coming from the nucleus. Notably, a fraction of this production rate can also be due to their formation 
in the coma.
	
We ran two sets of models; in one, we used the water and other parent volatile production rates shown in Tables \ref{table:input} and \ref{table:ratio} 
as inputs, and the model outputs provide the evolution of all the species as a function of cometocentric distance. The resulting species abundances in 
these models are shown in the species flux plots (Section \ref{subsection:reslt_spec}) using dashed lines. In the second set of models, we excluded a 
particular organic molecule as a parent species, i.e., its sublimation from the nucleus, while the other species abundances were used as given in 
Table \ref{table:ratio}. The resulting species abundance as a function of cometocentric distance only via coma chemistry is shown by the solid 
lines in the flux plots. For example, when we study \ch{HC3N}, in the first set of models, we study its evolution as a parent species, whereas, 
in the second set of models, we do not consider its production from the nucleus and only study its formation via coma chemistry. 
The same procedure is adopted for the other organic molecules that are marked in bold in Table \ref{table:ratio}. It is to 
be noted that some species' abundances are reported as upper limits in literature. However, this will not affect the second set of models since 
their initial abundances are set to zero.
	
	
\section{Results} \label{section:reslt}
We first discuss the coma temperature profiles in Section \ref{subsection:reslt_temp}.
Then in Sections \ref{subsection:reslt_spec} and \ref{subsection:reslt_spec2}, we 
evaluate the formation and evolution of a number of organic molecules in diverse cometary conditions using
a comprehensive chemical network. We see how a parent organic molecule evolves as a function of
the cometocentric distance (first set of model runs; see previous section). We also evaluate the cometocentric 
distance variation of this molecule when it is not present as a parent but solely synthesised
due to the gas-phase chemistry in the coma (second set of model runs). Since the volatile abundance
changes from one comet to another, as can be seen in the four test comets we have used, the second
set of models assess how the production of COMs are affected due to the variation in reactant species
abundances, which produce them. Here, the abundance ratios of parent volatiles with respect to water
are treated as representative values; a similar set of models may be run for other cometary compositions
by altering the abundance ratios.
It is also noted that the non-detection of a volatile in a particular comet can be either due to its absence
or its abundance being below the 
instrument's detection limit. We considered the non-detection of a species to imply its absence in the comet for simplicity in the discussion. 
This also implies that if a species is not detected, then our modeled abundance due to gas-phase chemistry (second set of model runs) can be treated as a lower limit.

In Table \ref{table:ratio}, the abundances of some of the molecules are given as upper limits. The species whose gas phase 
formation is under study has its initial abundance set to zero, and upper limits are not a cause for concern in that situation. 
However, the formation rate of a species may be affected by other molecules whose abundances are given as upper limits. Thus, 
if these molecules contribute to the formation of another organic species, then the corresponding formation rates are to be 
taken as upper limits, while the actual formation rate is likely to be lower.
We have indicated these effects wherever applicable.

\subsection{Temperature Profiles} \label{subsection:reslt_temp}
\begin{figure}
	\begin{center}
		\includegraphics[width=0.8\textwidth]{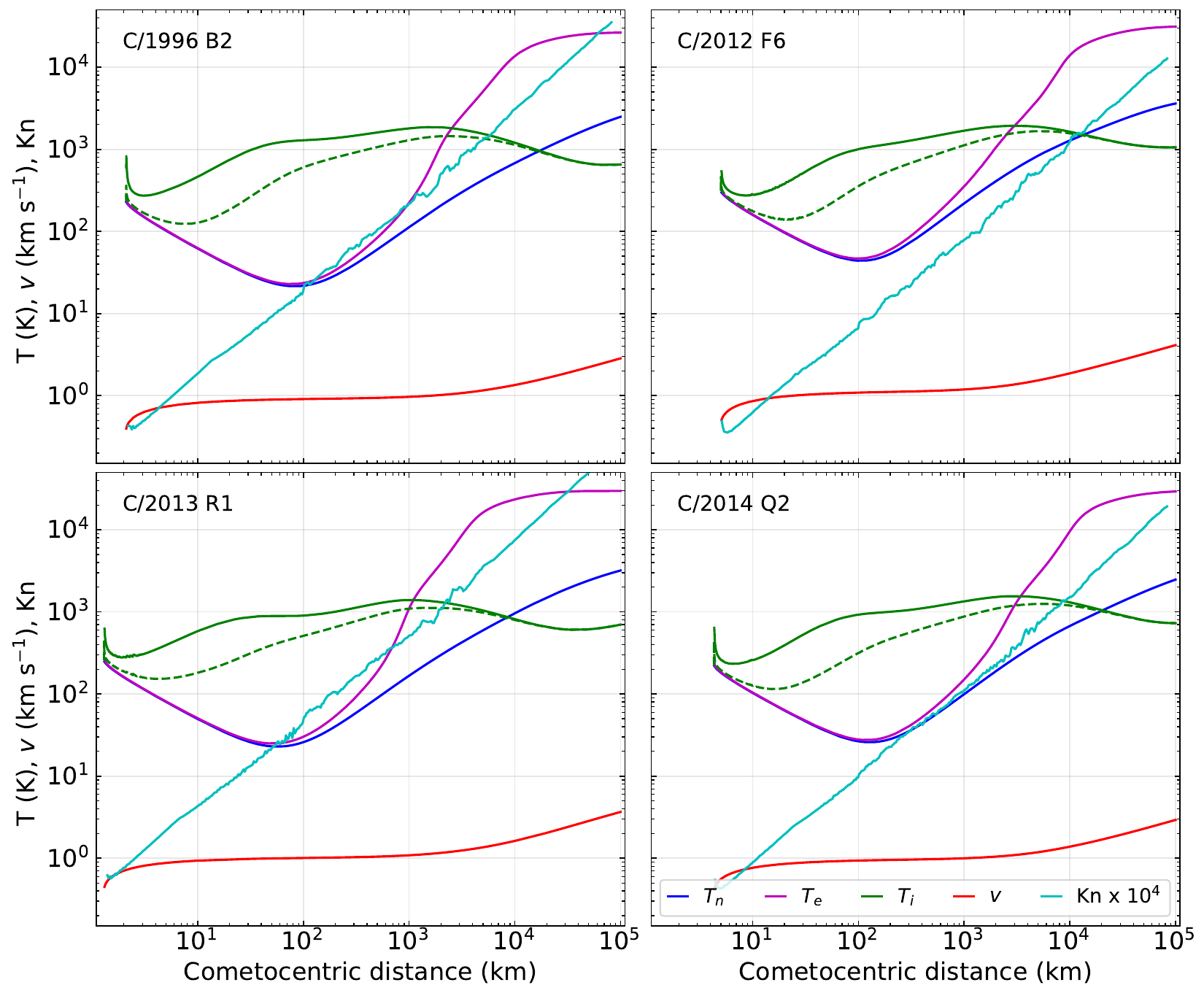}
		\caption{Variation of the fluid temperatures, common fluid velocity and Knudsen number with cometocentric 
			distance. $T_n$, $T_e$, and $T_i$ denote the temperatures of the neutral, electron, and ion fluids, respectively. The dashed lines denote the ion temperature when we use the Langevin rate for the ion-neutral elastic scattering. $v$ denotes the common velocity and Kn denotes the Knudsen number.}
		\label{fig:temperature}
	\end{center}
\end{figure}

\begin{figure}
	\begin{center}
		\includegraphics[width=0.5\textwidth]{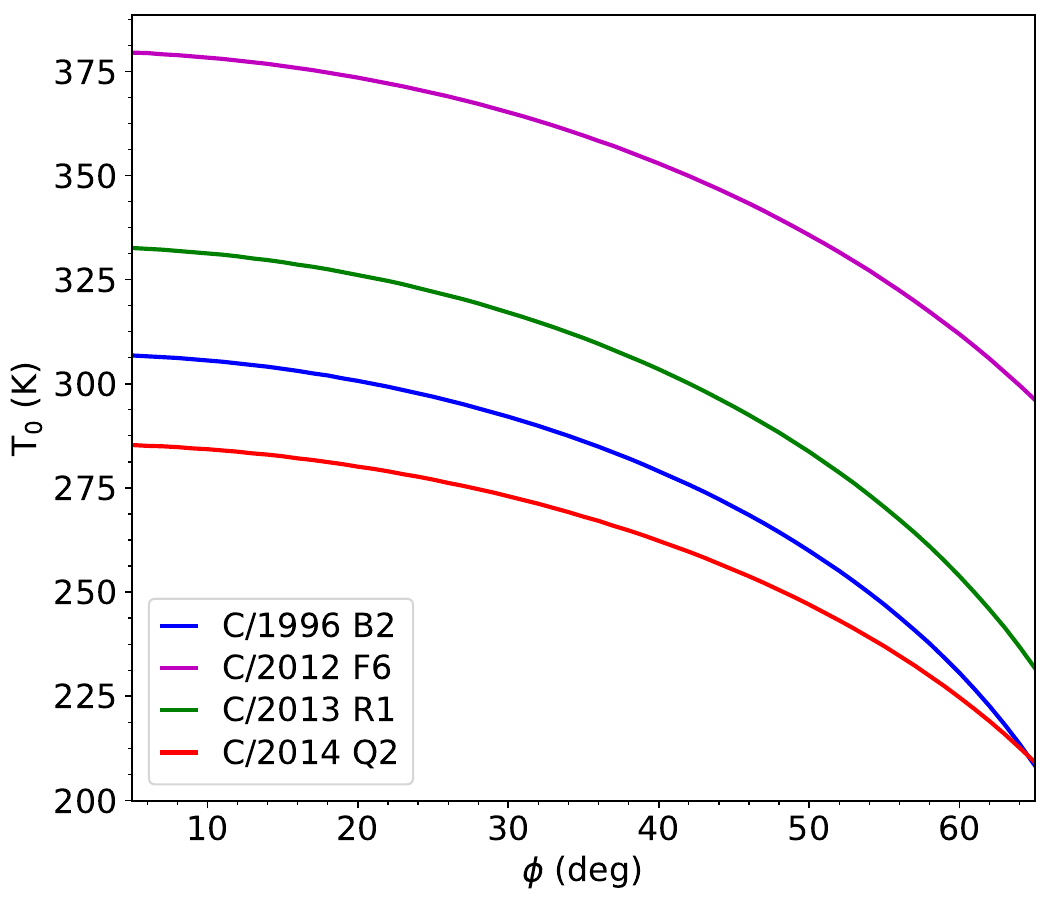}
		\caption{Variation of the initial temperature as a function of the solar zenith angle ($\phi$).}
		\label{fig:zenith}
	\end{center}
\end{figure}

The temperature profiles are shown 
in Figure \ref{fig:temperature}. The Knudsen number Kn indicates the region in which the fluid approximation holds (fluid: Kn$<0.01$; transitional: $0.01 \le $ Kn $\le 100$; \citealp{Marschall2020}).
The initial temperature $T_0$ of the coma gas is related to the surface temperature $T_s$ as

\begin{equation}
T_0 = \frac {T_s} {1+\frac 1 2 (\gamma -1)},
\end{equation}
$\gamma$ being the adiabatic exponent. $T_0$ can be calculated using a simple sublimation model from \cite{Knollenberg1993Thesis} in which a pure ice surface is considered. The ice temperature $T_s$ is calculated
using the Clausius-Clapeyron equation, to relate the vapor pressure and the temperature of the
sublimating ice \citep{Fanale1984}, and then the reservoir outflow analogy of \cite{Knollenberg1993Thesis} is used. For the cometary case, we can also find the surface temperature by solving the energy balance equation:
	
\begin{equation}
\frac {F_{\odot}(1-A_v)} {r_h^2} \cos (\phi)  =\epsilon_{\text {IR}}\sigma_B T_s^4 +HZ(T).
\label{eq:energy_balance}
\end{equation}
Here, $F_{\odot}$ is the incident solar flux, which is scaled by the heliocentric distance $r_h$, $A_v$ is the visual 
albedo of the comet nucleus, $\phi$ is the solar zenith angle, $\epsilon_{\text {IR}}$ is the infrared emissivity, 
$\sigma_B$ is the Stefan-Boltzmann constant, $H$ is the latent heat of sublimation of the ice, and $Z(T)$ is the surface 
sublimation rate. We assume a mean value of $\phi = 60^{\circ}$, $A_v=0.04$ and $\epsilon_{\text {IR}}=0.9$, while the 
latent heat of sublimation is assumed to be that of \ch{H2O} ice ($\sim 5\times10^{11}$ ergs mole$^{-1}$). The calculated 
initial temperatures for the different comets are shown in 
Table \ref{table:input}. The initial temperature of the gas can vary depending on the method that is employed to calculate 
it. Here we have solved the energy balance equation (see, for example, \citealt{Steckloff2015,Marshall2019}), though some 
authors may calculate it using a scaling relation with the heliocentric distance (see, for example, \citealt{Rodgers2002}).
It is also to be noted that the model does
not include the effect of an insulating dust layer that quenches the outgassing activity and the values of $T_0$ (shown in Table \ref{table:input} and Figure \ref{fig:zenith}) are not the temperature of the outgassing layer; the effect of quenching by dust is often overcome by defining an active area fraction such that only a fraction of the surface is active.

The initial temperature for C/2012 F6 is greater than 300 K, since it is modeled at the least heliocentric distance, 
while the temperatures for the other three comets lie in the range $\sim 225 - 255$ K. 
When the modeled 
initial temperatures are compared with in situ measurements for other comets, we find that they are in a similar range. For example,
the surface temperature measurements at comet 1P/Halley were made by the IKS instrument on \textit{Vega 1} when the comet was 
at 0.8 au from the sun \citep{Emerich1987}. A mean temperature of 320 K and a maximum temperature lying between $360-400$ K 
was deduced. \cite{Tosi2019} report widely on the surface temperature maps of 67P/C-G obtained by \textit{Rosetta}'s VIRTIS 
instrument, at heliocentric distances between 3.62 and 3.31 au. For solar incidence angles $<20^{\circ}$ (that is, around 
local noon) and emission angles $<80^{\circ}$, an average temperature of $213\pm 3$ K was recorded on the dayside of 67P/C-G, 
while values reaching upto 230 K were recorded in two VIRTIS observations.
	
The temperature profiles shown in Figure \ref{fig:temperature} are qualitatively similar to such profiles obtained previously 
by gas-phase coma modeling \citep{Weiler2006Thesis, Ahmed2021}. As shown by earlier models, there is a coupling between the 
electron and neutral temperatures upto some distance from the nucleus (for example, \citealt{Weiler2006Thesis, Ahmed2021}). The energetic 
electrons that are created due to 
photochemical reactions lose their energy to the neutral fluid (specifically \ch{H2O} and \ch{CO}; see Section \ref{section:model}) 
due to inelastic collisions, resulting in the rotational and vibrational excitation of the neutral molecules. As we move outwards 
in the coma, the coupling is lost and the electron temperature rises rapidly. On increasing the production rate, 
the electron temperature decouples from the neutral temperature at a larger distance because of higher coma density, and the rise in the
electron temperature is also comparatively less steep. 
	
Many ion-neutral reactions are exothermic; this energy is distributed amongst the product species, 
depending on their masses. The KIDA database contains enthalpies (at 298 K) of most of the ion-neutral reactions in our 
network. We assume that the enthalpies are constant over temperature and use these values in our model. 
{We also assume that the reaction energy is converted completely to kinetic energy of the products. This is a simplification since part of the energy would also be distributed as internal energies of the products which would be released as radiation. For example, energy distributed as internal vibrational energies would be released as infrared radiation, and this can make the gas cooler. Estimating how the internal energy is distributed among the different rotational and vibrational modes would require extensive quantum mechanical calculations. Thus, we make the assumption that the energy is completely kinetic (further comments in Appendix \ref{App_D}). 

The ion temperature 
we obtain is higher than that computed by \cite{Weiler2006Thesis}, by factors of $2-4$ at distances $<1000$ km, and less than a 
factor of 2 beyond this distance. In the network used by \cite{Weiler2006Thesis}, the energy released due to ion-neutral reactions 
is set to zero. 
Several authors have presented model results for the ion temperature. For example, \cite{Rubin2014} predict a 
temperature of $500 - 1000$ K for light cometary ions in 1P/Halley for distances of $10^3 - 10^4$ km, which are in agreement 
with \textit{Gitto's} IMS measurements \citep{Altwegg1993}. \cite{Haerendel1987},  \cite{Korosmezey1987}, and \cite{Schmidt1988} 
also predict ion temperatures approaching 1000 K  at a distance of about $10^4$ km.
Additionally, $T_i$ also depends on the ion-neutral elastic collision rate that we use in our model. 
Instead of the hard-sphere scattering rate suggested by \cite{Weiler2006Thesis}, if we use the Langevin rate \citep{Rodgers2004} for the ion-neutral elastic scattering, then $T_i$ will reduce by factors $\sim 2-4$ for distances $<1000$ km, as shown by the dashed lines in Figure \ref{fig:temperature}. Beyond this distance, the frequency of elastic scattering is reduced due to a lower coma density; in this case, $T_i$ resulting from the use of the Langevin rate tends towards that resulting from the hard-sphere scattering rate.
	
We also looked into the effect of varying the solar zenith angle ($\phi$) on the initial temperature $T_0$, and this is shown in Figure \ref{fig:zenith}. 
The decrease in the solar zenith angle leads to an increase in the surface temperature. 
The gas temperature is higher for lower values of $\phi$, and the particles move away from the surface with a higher initial velocity. 
This will lead to a reduction in the initial gas density (more discussions in Appendix \ref{App_C}).

\subsection{Formation of Organic Molecules} \label{subsection:reslt_spec}

	\subsubsection{\ch{HC3N} and \ch{NH2CHO}}
	Figures \ref{fig:HC3N-NH2CHO}(a) and \ref{fig:HC3N-NH2CHO}(b) respectively show the flux of \ch{HC3N} and \ch{NH2CHO} as a 
	function of the cometocentric distance. The flux is calculated by multiplying the species number density with the factor 
	$4\pi r^2v$, where $r$ is the cometocentric distance and $v$ is the gas velocity. This removes the $r^{-2}$ dependence of 
	the number density, and the stretching or compression effects due to acceleration or deceleration of the coma gas.
	It should of course be kept in mind that the molecular flux is the net flux, that is, production minus the loss.
	The solid lines in Figure \ref{fig:HC3N-NH2CHO} indicate the abundance of \ch{HC3N} or \ch{NH2CHO} when the outgassing 
	of these species from the nucleus is not considered and they form due to gas-phase coma chemistry alone. The dashed lines
	indicate the net species abundance in the coma resulting from both nucleus outgassing and coma chemistry.  Since \ch{HC3N} 
	is not considered as a parent volatile due to its non-detection in C/2012 F6 and C/2013 R1 (which can happen due to its abundance below 
	detection limit or its absence), its coma abundance in these two comets is solely due to its formation by gas-phase chemical reactions.
	
\begin{figure}
\begin{center}
	\includegraphics[height=7cm, width=0.49\textwidth]{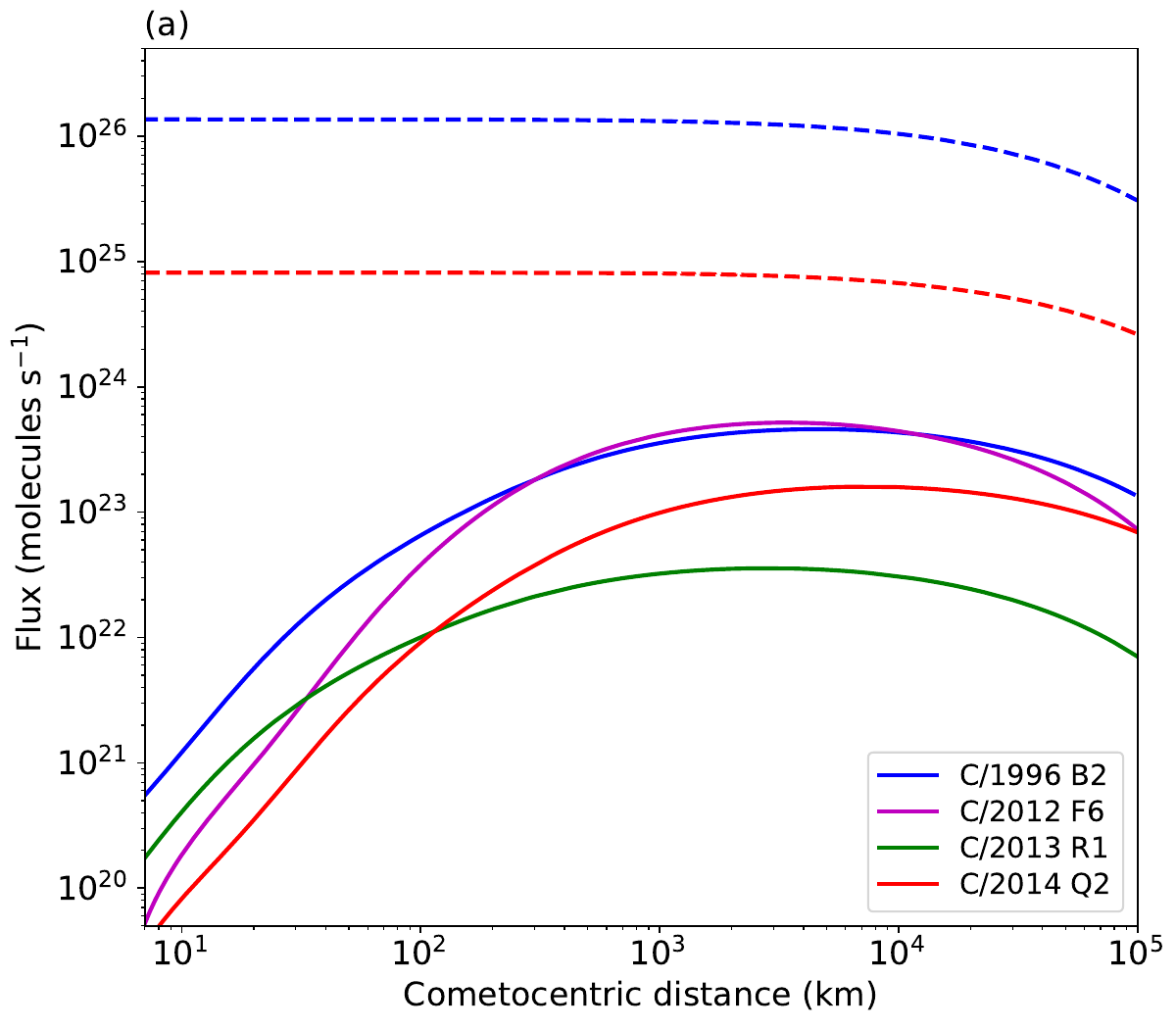}
	\includegraphics[height=7cm, width=0.49\textwidth]{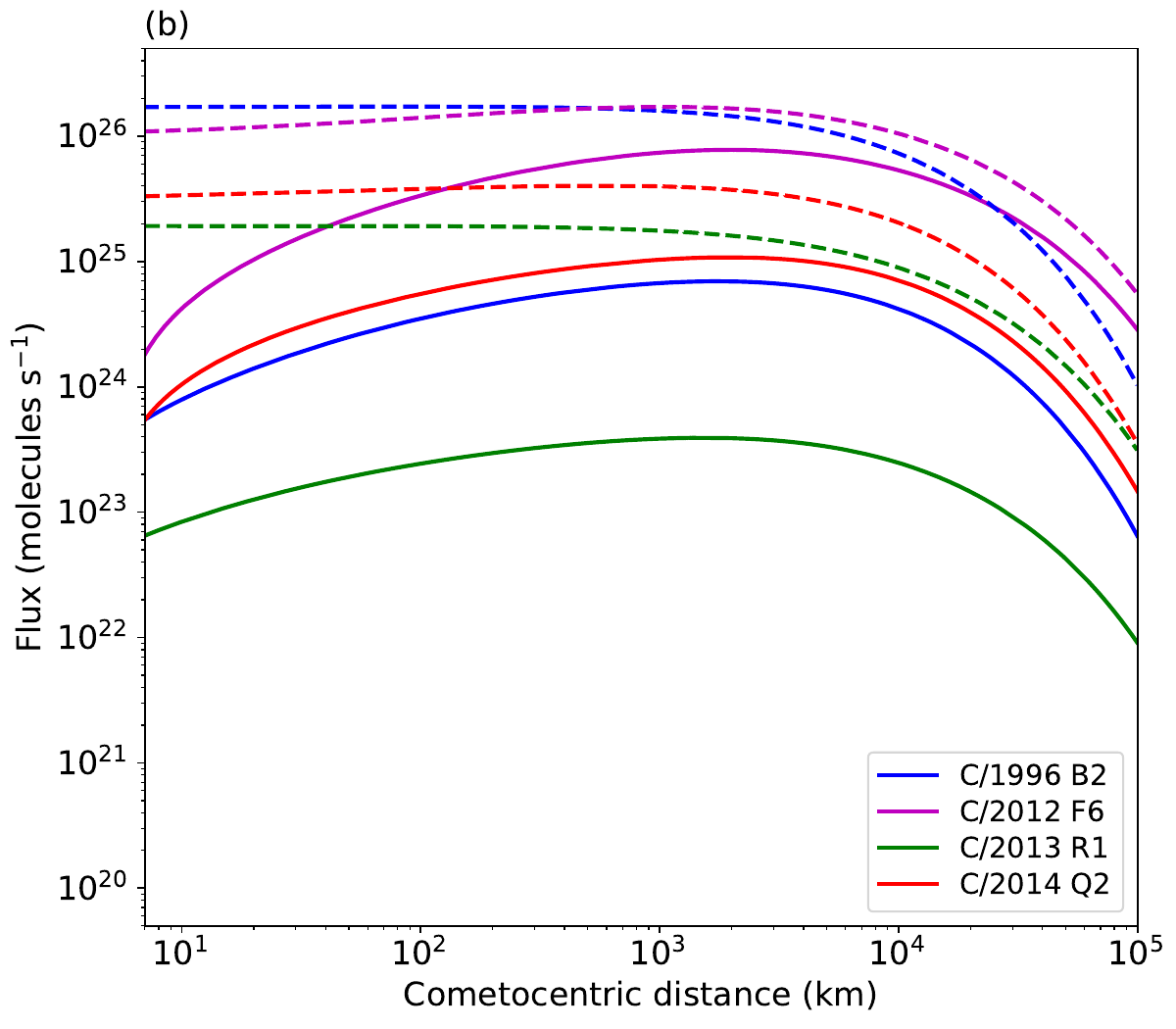}
	\caption{Coma flux profiles of (a) \ch{HC3N} and (b) \ch{NH2CHO}. Solid lines indicate abundance due to coma chemistry while dashed lines indicate net abundance 
due to nucleus outgassing and coma chemistry.}
	\label{fig:HC3N-NH2CHO}
\end{center}
\end{figure}
	
\begin{figure}
\centering
\includegraphics[height=7cm, width=0.32\textwidth]{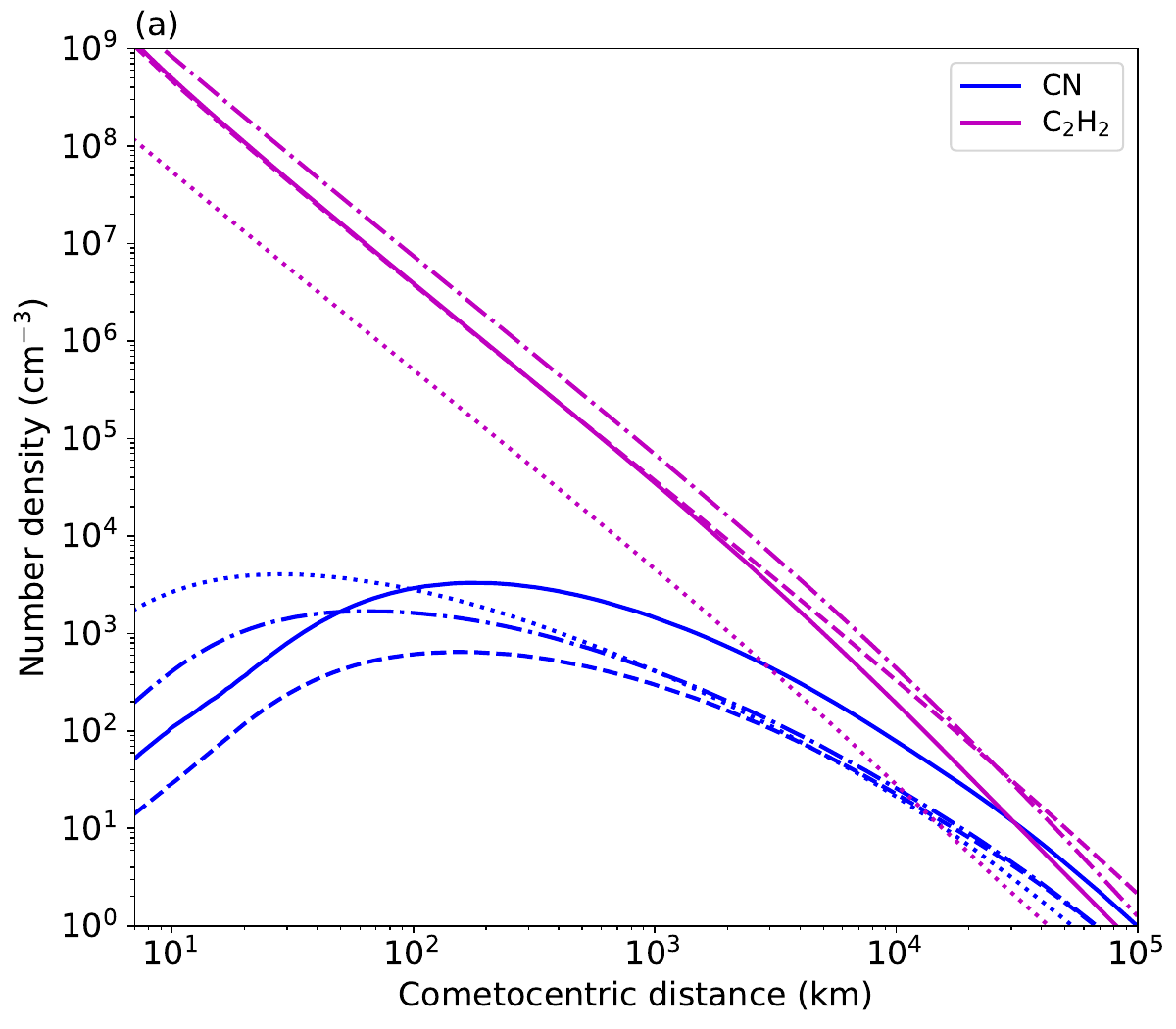}	
\includegraphics[height=7cm, width=0.32\textwidth]{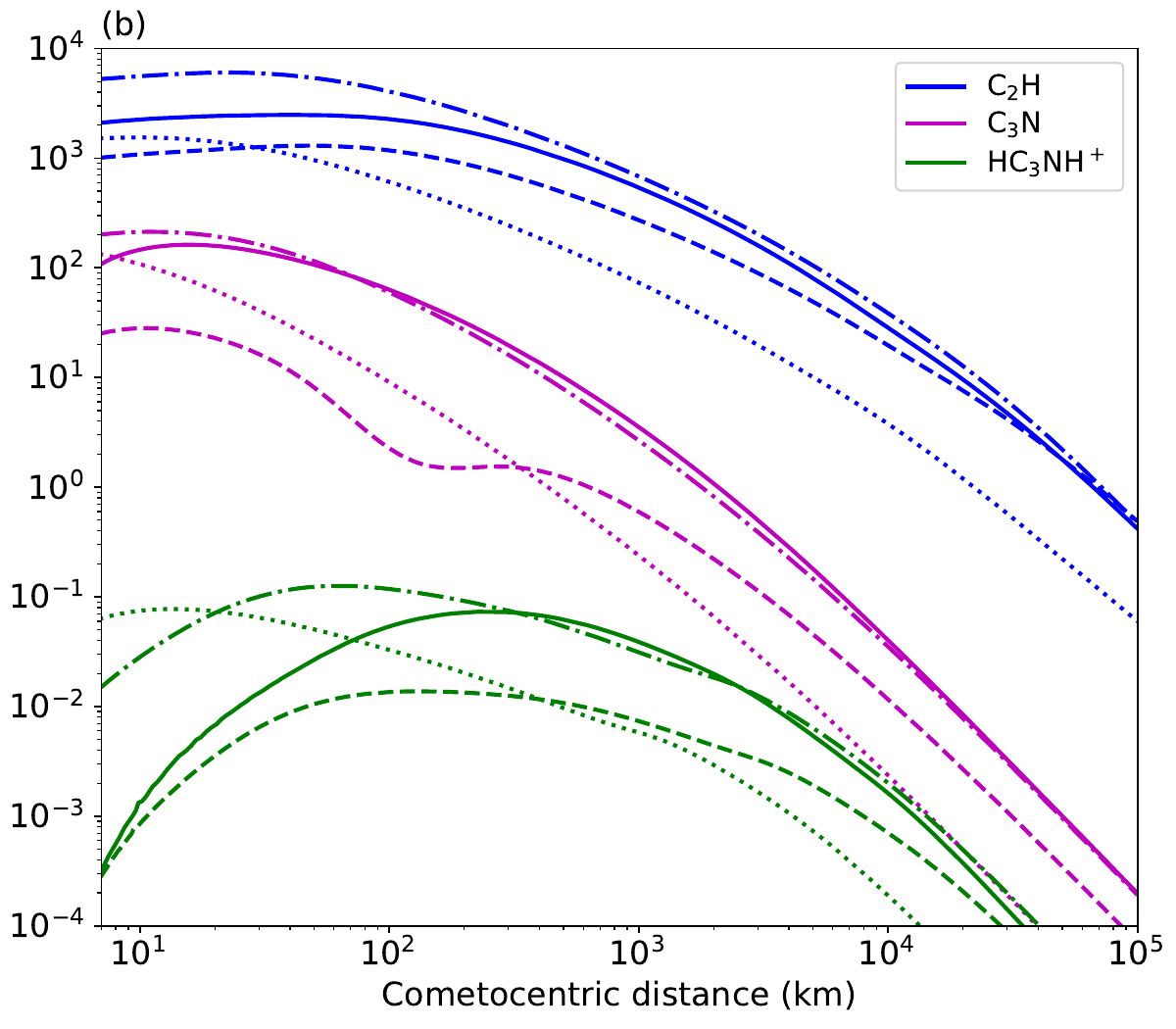}	
\includegraphics[height=7cm, width=0.32\textwidth]{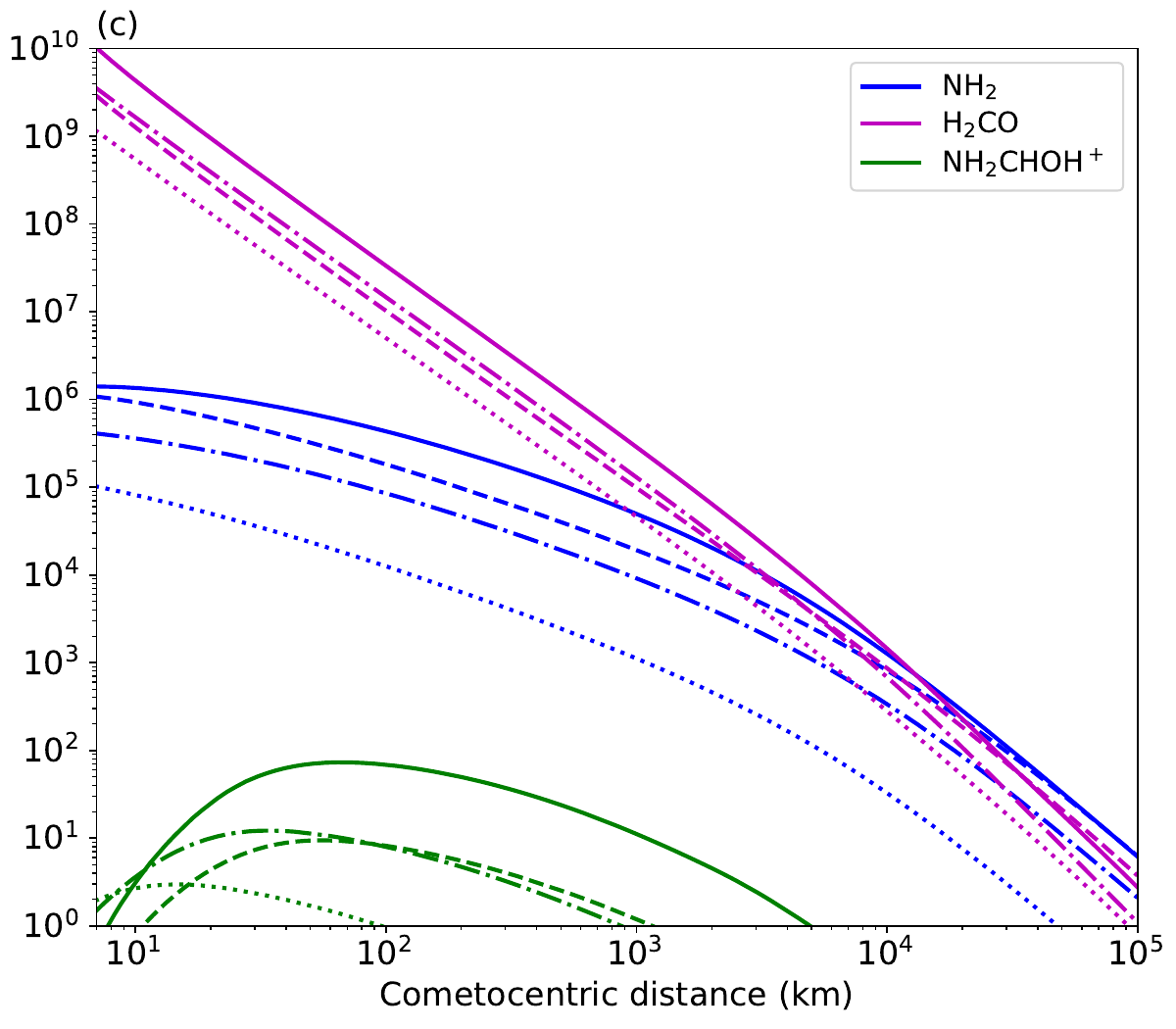}	
		
\caption{Panels (a) and (b) show the cometocentric distance variation of the number densities of species that lead to the formation of \ch{HC3N}, 
while panel (c) shows the same for \ch{NH2CHO}. The line styles indicate different cometary compositions (dotted-dashed: C/1996 B2, solid: C/2012 F6, 
dotted: C/2013 R1 and dashed: C/2014 Q2).}
\label{fig:HC3N}
\end{figure}
	
The flux of \ch{HC3N} outgassing from the nucleus of C/1996 B2 is $1.4 \times 10^{26}$ molecules s$^{-1}$, while it is $8.1 
\times 10^{24}$ molecules s$^{-1}$ in the case of C/2014 Q2. Coma formation of \ch{HC3N} results in peak abundances lying in 
the range $3.5 \times 10^{22}-5.2\times 10^{23}$ molecules s$^{-1}$ for different cometary compositions, achieved at cometocentric 
distances between $\sim 3000-5000$ km. The parent molecular flux of \ch{NH2CHO} lies in the range $1.9\times10^{25} - 
1.7\times 10^{26}$ molecules s$^{-1}$ for the different comets. The flux of \ch{NH2CHO} forming in the coma acquires a value lying 
in the range $3.9\times10^{23} - 7.8\times 10^{25}$ molecules s$^{-1}$ beyond $\sim 1000$ km. The gas-phase formation of \ch{HC3N} 
and \ch{NH2CHO} by neutral-neutral reactions was studied by \cite{Cordiner2021}, for a water production rate of $5\times 10^{29}$ 
molecules s$^{-1}$. They also consider distributed coma sources of \ch{CN} and \ch{H2CO}, which is not included in our models.
The water production rates in our comet models lie in the range $9.13 \times 10^{28}-6.7\times 10^{29}$ molecules s$^{-1}$. 
	
The main chemical process that leads to the formation of \ch{HC3N} in most regions of the coma is the following neutral-neutral reaction:
	
	\begin{equation}
		\ch{CN} + \ch{C2H2} \rightarrow \ch{HC3N} + \ch{H}.
		\label{eq:HC3N}
	\end{equation}
\cite{Cordiner2021} also found the above pathway to be the most dominant.
Figure \ref{fig:HC3N}(a) shows the number densities of \ch{CN} and \ch{C2H2} for the different cometary compositions.  The main source of \ch{C2H2} is 
sublimation from the nucleus, while \ch{CN} forms in the coma primarily due to the photodissociation of \ch{HCN} (see Figure \ref{fig:HC3Na}(b) in Appendix \ref{App_B}).
However, it is to be noted that the \ch{C2H2} abundance is reported as an upper limit in C/2012 F6 and
C/2013 R1, so the modeled \ch{HC3N} abundance due to coma chemistry is also to be treated as an upper limit in these two
cometary compositions.  
The photodissociation of \ch{HNC} also creates \ch{CN} radicals, though the contribution of this reaction is only a few percent. This is due to the fact that \ch{HNC} 
is not a parent volatile in C/2012 F6 and C/2013 R1, and its relative abundance is less than that of \ch{HCN} by more than an order of magnitude in C/1996 B2 
and C/2014 Q2. In some regions of the coma, chemical reactions with energetic electrons (dissociative recombination of \ch{HCNH+} and electron impact ionization of \ch{HNC}) 
contributes upto $\sim 10\%$ to the \ch{CN} formation rate.
	
The flux of \ch{HC3N} is the highest in C/1996 B2 since $P_{\ch{HC3N}}$ (see Figure \ref{fig:HC3Na}(a) in Appendix \ref{App_B}) is the highest in this comet. 
Since \ch{CN} and \ch{C2H2} are the main reactants in the formation of \ch{HC3N}, their abundances in different regions of the coma drive the resultant 
flux of \ch{HC3N} molecules. The number density of \ch{C2H2} in C/1996 B2 is higher than that in C/2012 F6 and C/2014 Q2 by nearly a 
factor of 2. \ch{C2H2} in C/1996 B2 is also higher than that in C/2013 R1 by more than an order of magnitude. Thus, the flux of \ch{HC3N} 
is the highest in C/1996 B2. The abundance of \ch{CN} radicals depends on the production rate of \ch{HCN}, and higher \ch{HCN} production 
rate in C/2012 F6 results in higher density of \ch{CN} at distances $>100$ km, and the \ch{HC3N} abundance in C/2012 F6 becomes similar 
to that in C/1996 B2 beyond this distance. Below this distance, optical depth effects due to the attenuation of the incoming UV flux leads 
to a reduction in the photodissociation rate, and this effect is weaker in comets with lower density. Since C/2013 R1 has the lowest 
volatile production rate and lowest density, the UV flux suffers the least attenuation and \ch{CN} has the highest density in this comet 
at distances $<100$ km. Thus, at distances close to the nucleus, the flux of \ch{HC3N} molecules is slightly higher in C/2013 R1 as compared 
to C/2012 F6 and C/2014 Q2. In the absence of distributed sources, \cite{Cordiner2021} find that that the effective production rate for 
\ch{HC3N} is $1.5\times10^{24}$ molecules s$^{-1}$. The peak \ch{HC3N} abundance that we obtain for C/2012 F6 (comet with the highest production 
rate) is lower, since we have 0.08\% \ch{C2H2} as compared to their value of 0.3\%. Although C/1996 B2 has 0.5\% \ch{C2H2}, its production rate 
is lower than that of \cite{Cordiner2021}, and hence the \ch{HC3N} abundance in C/1996 B2 is lower than their values.
	
The photodissociation of \ch{C2H2} creates the carbon chain radicals \ch{C2} and \ch{C2H}, and \ch{C2} reacts with \ch{HCN} and \ch{HNC} 
to produce \ch{C3N}. \ch{HC3N} can be produced by the reaction of \ch{C2H} with \ch{HNC}, and by \ch{C3N} reacting with the hydrocarbons 
\ch{CH4} and \ch{C2H6}. The contribution of these reactions to $P_{\ch{HC3N}}$ can reach upto 50\% close to the nucleus. This is the region 
where the number density of \ch{CN} is comparatively lower. At higher distances, the number density of \ch{C2H} and \ch{C3N} falls faster 
as compared to \ch{CN}, and their contribution to the formation of \ch{HC3N} reduces. The formation of \ch{HC3N} by the dissociative recombination 
of \ch{HC3NH+} is temperature dependent, and it can contribute $10-20\%$ towards $P_{\ch{HC3N}}$ in the outer region of the coma, where temperatures 
are high. \ch{HC3NH+} forms at varying relative rates in different regions of the coma by processes that include protonation of \ch{HC3N} and 
ion-neutral reactions of \ch{HCN+} with \ch{C2H2}, \ch{C2H2+} with \ch{HCN}, and \ch{HC3N+} with the parent species \ch{H2O}, \ch{CH4} and \ch{C2H4}.

The neutral-neutral reaction betwen \ch{NH2} and \ch{H2CO} is the most dominant pathway to form \ch{NH2CHO} (see Figure \ref{fig:NH2CHOa}(a) 
in Appendix \ref{App_B}), as also found by \cite{Cordiner2021}:
	
	\begin{equation}
		\ch{NH2} + \ch{H2CO} \rightarrow \ch{NH2CHO} + \ch{H}.
		\label{eq:NH2CHO}
	\end{equation}
	Figure \ref{fig:HC3N}(c) shows the number densities of \ch{NH2} and \ch{H2CO} for the different cometary compositions. While \ch{H2CO} 
	is a parent molecule, the main source of \ch{NH2} in the coma is the photodissociation of \ch{NH3}. The dissociative recombination of 
	\ch{NH4+} and \ch{NH3+} ions (created respectively by the 
	protonation and photoionization of parent \ch{NH3}) also create \ch{NH2}, though these reactions only contribute a few percent to $P_{\ch{NH2}}$.
	$P_{\ch{NH2CHO}}$ is the highest in C/2012 F6, resulting in the highest flux of \ch{NH2CHO} amongst all four comets, as seen in 
	Figure \ref{fig:HC3N-NH2CHO}(b). This is because the number densities of \ch{H2CO} and \ch{NH2} (created from \ch{NH3}) is the highest in 
	this comet, resulting in its highest value of $P_{\ch{NH2CHO}}$. The number densities of \ch{H2CO} in C/1996 B2 and C/2014 Q2 are nearly equal. 
	However, the flux of \ch{NH2CHO} in C/2014 Q2 is nearly double that of C/1996 B2, since the number density of \ch{NH2} is higher in C/2014 Q2. 
	The number densities of \ch{H2CO} and \ch{NH2} in C/2013 R1 is lower than that in C/2012 F6 by nearly an order of magnitude, resulting in 
	$P_{\ch{NH2CHO}}$ being lower by nearly two orders of magnitude. The effective production rate for \ch{NH2CHO} estimated by \cite{Cordiner2021} 
	in the absence of distributed sources is $4.7\times10^{24}$ molecules s$^{-1}$, while this value is $5.81\times10^{25}$ molecules s$^{-1}$ in 
	the presence of a large unknown precursor molecule that breaks down to produce \ch{H2CO}. Our peak abundance for C/2012 F6 is higher than the 
	latter value by nearly a factor of 1.5. The \ch{H2CO} and \ch{NH3} outgassing rates that we use are higher, and since C/2012 F6 is modeled at a 
	heliocentric distance of 0.78 au, the rate at which \ch{NH2} is created by the dissociation of \ch{NH3} is higher. In addition, the rate 
	coefficient (cm$^{-3}$ s$^{-1}$) for the formation of \ch{NH2CHO} in our chemical network (from KIDA) is higher than the value used by 
	\cite{Cordiner2021}.
	
	We found that the protonation of \ch{NH2CHO} by \ch{H2O+} and \ch{HCO+} ions leads to the formation of \ch{NH2CHOH+}, which cycles back to create \ch{NH2CHO} 
	by dissociative recombination, as described by the following reactions:
	
	\begin{equation}
		\begin{split}
			&\ch{NH2CHO} + \ch{H3O+} \rightarrow \ch{NH2CHOH+} + \ch{H2O} \\
			&\ch{NH2CHO} + \ch{HCO+} \rightarrow \ch{NH2CHOH+} + \ch{CO} \\
			&\ch{NH2CHOH+} + \ch{e-} \rightarrow \ch{NH2CHO} + \ch{H}. \\
		\end{split}
	\end{equation}
	The high flux of \ch{NH2CHO} in C/2012 F6 results in its faster conversion into \ch{NH2CHOH+}, and reconversion back into 
	\ch{NH2CHO} (see Figures \ref{fig:NH2CHOa}(a) and \ref{fig:NH2CHOa}(c) in Appendix \ref{App_B}). Since the rates of these reactions increase 
	with temperature, the relative reaction rate of the formation \ch{NH2CHO} by dissociative recombination of \ch{NH2CHOH+} 
	increases on moving outwards in the coma, and contributes substantially to $P_{\ch{NH2CHO}}$ in C/2012 F6. In the other 
	three comets, the interconversion rate is lower due to lower flux of \ch{NH2CHO}.
	
	From the results shown in Figures \ref{fig:HC3N-NH2CHO} and \ref{fig:HC3N}, it can be seen that the abundances and formation 
	rates of \ch{HC3N} and \ch{NH2CHO} (and other molecules to be discussed subsequently) are controlled both by the net
	production rate and the relative abundance of the reacting species with respect to water. $Q_{\ch{H2O}}$ of C/1996 B2 
	is lower than that of C/2014 Q2, though the former has 0.5\% \ch{C2H2} compared to 0.1\% \ch{C2H2} in the latter. Thus, 
	the density of \ch{C2H2} is higher in C/1996 B2, resulting in a higher flux of coma-produced \ch{HC3N}. 
	The \ch{HCN} abundance in C/2012 F6 is 0.14\%, which is less than 0.2\% \ch{HCN} in C/1996 B2. However, C/2012 F6 has nearly 
	four times higher water production rate, resulting in its higher production rate of \ch{HCN}, which leads to higher \ch{CN} 
	density beyond 100 km. Although the density of \ch{C2H2} is lower in C/2012 F6 than in C/1996 B2, the higher \ch{CN} density 
	in C/2012 F6 drives up the rate of \ch{HC3N} formation, and the flux of coma-produced \ch{HC3N} at distances $\gtrsim 100$ km 
	becomes nearly equal in both these comets. Thus, variation in the relative abundances of reactants, 
	can give significant variations in the coma formation rates of molecules for both the molecules. 
	
	\subsubsection{\ch{HCOOCH3}} \label{subsubsection:HCOOCH3}

	\begin{figure}
		\centering
		\includegraphics[height=7cm, width=0.49\textwidth]{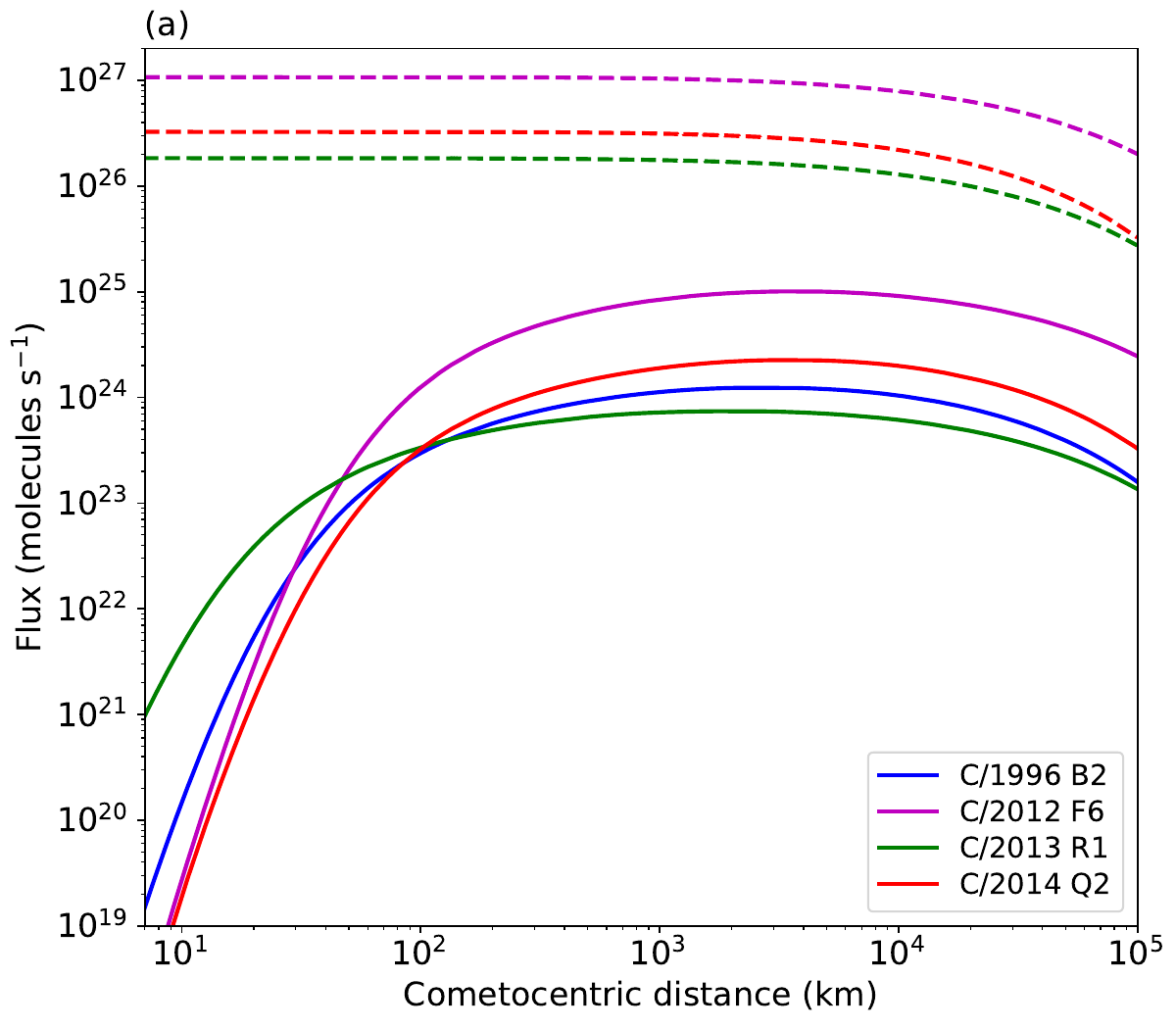}
		\includegraphics[height=7cm, width=0.49\textwidth]{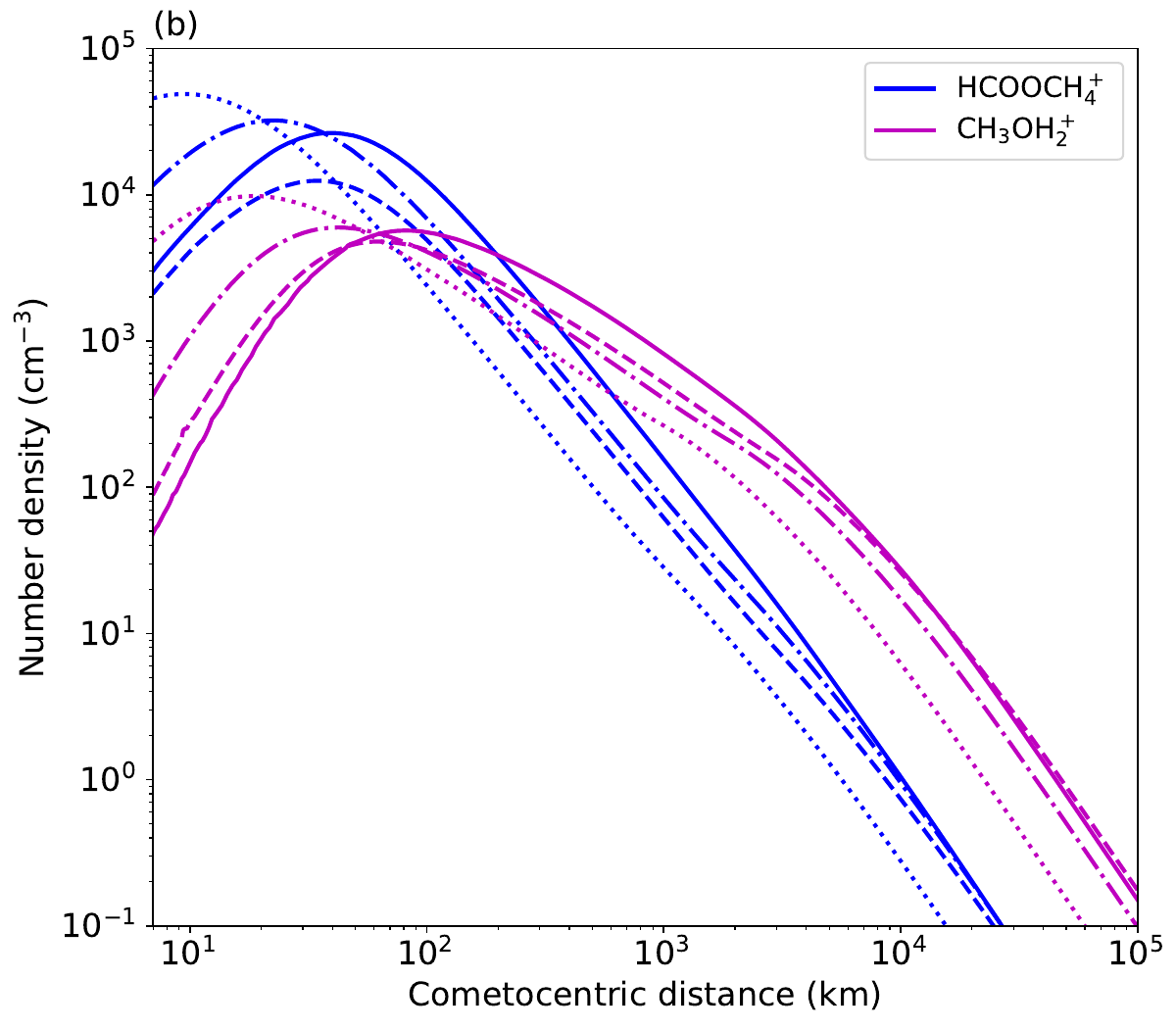}	
		\caption{(a) The coma flux profile of \ch{HCOOCH3}. Solid lines indicate abundance due to coma chemistry while 
			dashed lines indicate net abundance due to nucleus outgassing and coma chemistry.
			(b) The cometocentric distance variation of the number densities of species that lead to the formation of \ch{HCOOCH3} 
			(line styles: same as Figure \ref{fig:HC3N}).}		
		\label{fig:HCOOCH3}
	\end{figure}
	
	Figure \ref{fig:HCOOCH3}(a) shows the flux of \ch{HCOOCH3} as a function of the cometocentric distance. The parent molecular 
	flux of \ch{HCOOCH3} lies in the range $1.8\times 10^{26} - 1.1\times10^{27}$ molecules s$^{-1}$ for the different comets. 
	The coma produced flux of \ch{HCOOCH3} acquires a nearly constant value beyond $\sim 1000$ km, ranging from $7.4\times10^{23} - 1.0\times 10^{25}$ 
	molecules s$^{-1}$. The ion-neutral reaction of protonated methanol with formaldehyde creates protonated methyl formate or \ch{HCOOCH4+}, 
	which then undergoes dissociative recombination to create \ch{HCOOCH3} in the coma (for variation of formation rates, see Figure
	\ref{fig:HCOOCH3a} in Appendix \ref{App_B}).
	
	\begin{equation}
		\begin{split}
			&\ch{H2CO} + \ch{CH3OH2+} \rightarrow \ch{HCOOCH4+} + \ch{H2} \\
			&\ch{HCOOCH4+} + \ch{e-} \rightarrow \ch{HCOOCH3} + \ch{H}. \\
		\end{split}
	\end{equation}
	
	At cometocentric distances $>100$ km, $P_{\ch{HCOOCH3}}$ is the highest in C/2012 F6 , since the number density of \ch{HCOOCH4+} is the 
	highest. This is due to the highest production rate of \ch{CH3OH} (which undergoes protonation to form \ch{CH3OH2+}), and of \ch{H2CO} 
	in this comet, as seen from Table \ref{table:ratio}. Next in terms of abundance of \ch{HCOOCH3} are C/2014 Q2, C/1996 B2 and C/2013 R1, 
	respectively, consistent with the production rates of \ch{H2CO} and \ch{CH3OH}. In C/2012 F6, \ch{CH3OH2+} acquires a peak density at 
	$\sim 100$ km, which decreases on moving closer to the nucleus. This is because \ch{CH3OH2+} forms by successive proton transfer from 
	\ch{H2O+} ions, and the attenuation of the UV flux decreases the rate at which the photoionization of water into \ch{H2O+} occurs. The 
	higher coma density of C/2012 F6 results in stronger UV attenuation, which is also seen previously in the case of formation of \ch{CN} 
	by the photodissociation of \ch{HCN}. A reduction in the density of \ch{CH3OH2+} leads to a reduction in the density of \ch{HCOOCH4+}, 
	resulting in a sharp fall in the flux profile of \ch{HCOOCH3} at low cometocentric distances. This effect is also present in the other 
	comets, though it is weakest in C/2013 R1 since it has the lowest coma density, indicating the least attenuation of the UV flux. At 
	large cometocentric distances, the protonation of \ch{HCOOCH3} into \ch{HCOOCH4+}, and the reconversion of \ch{HCOOCH4+} into \ch{HCOOCH3} 
	takes place. This is similar to what is seen in the case of \ch{NH2CHO}, and the interconversion rate is the highest in C/2012 F6.
	
	\subsubsection{\ch{HCOOH}} \label{subsubsection:HCOOH}
	
	\begin{figure}
		\centering
		\includegraphics[height=7cm, width=0.49\textwidth]{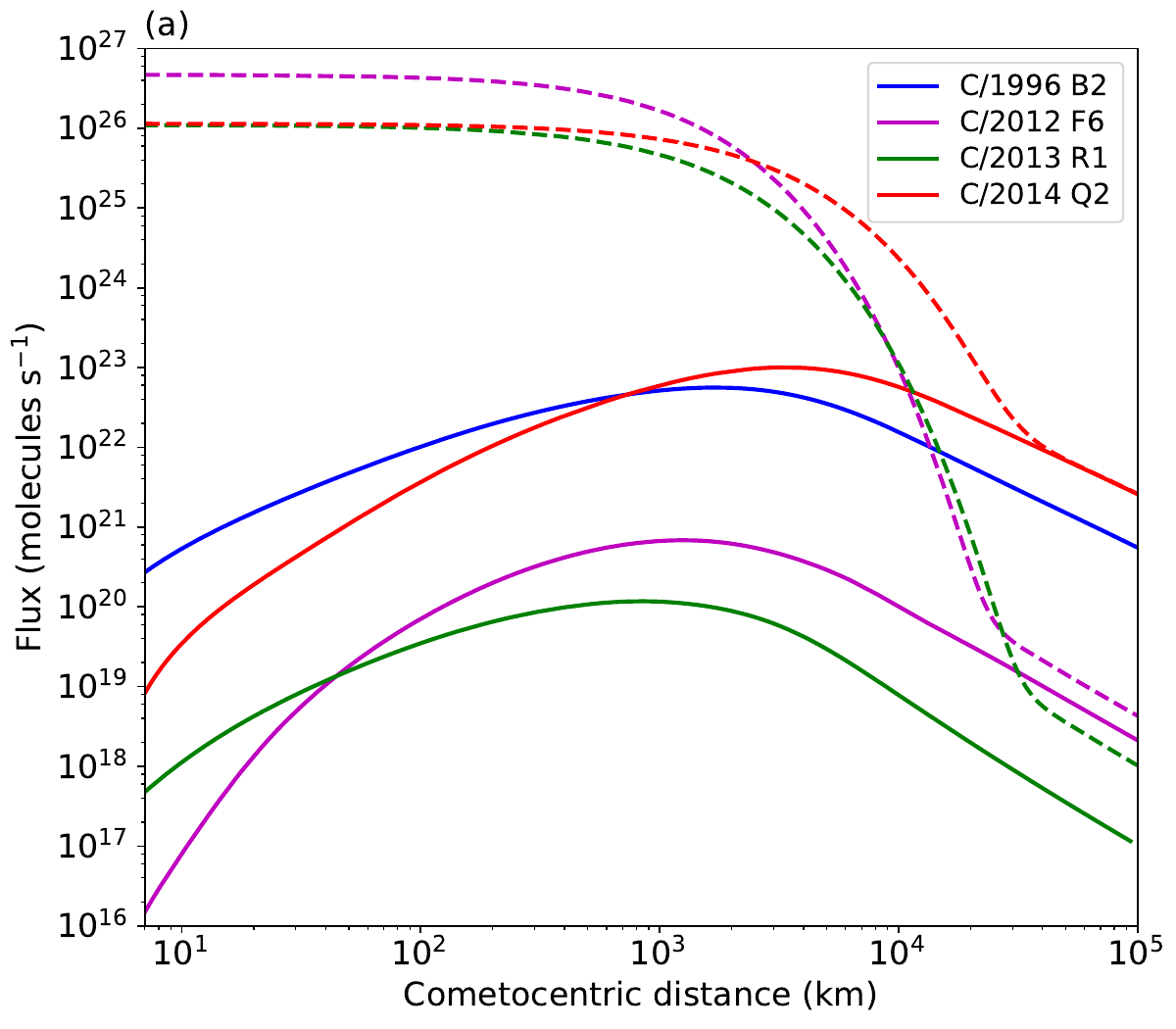}
		\includegraphics[height=7cm, width=0.49\textwidth]{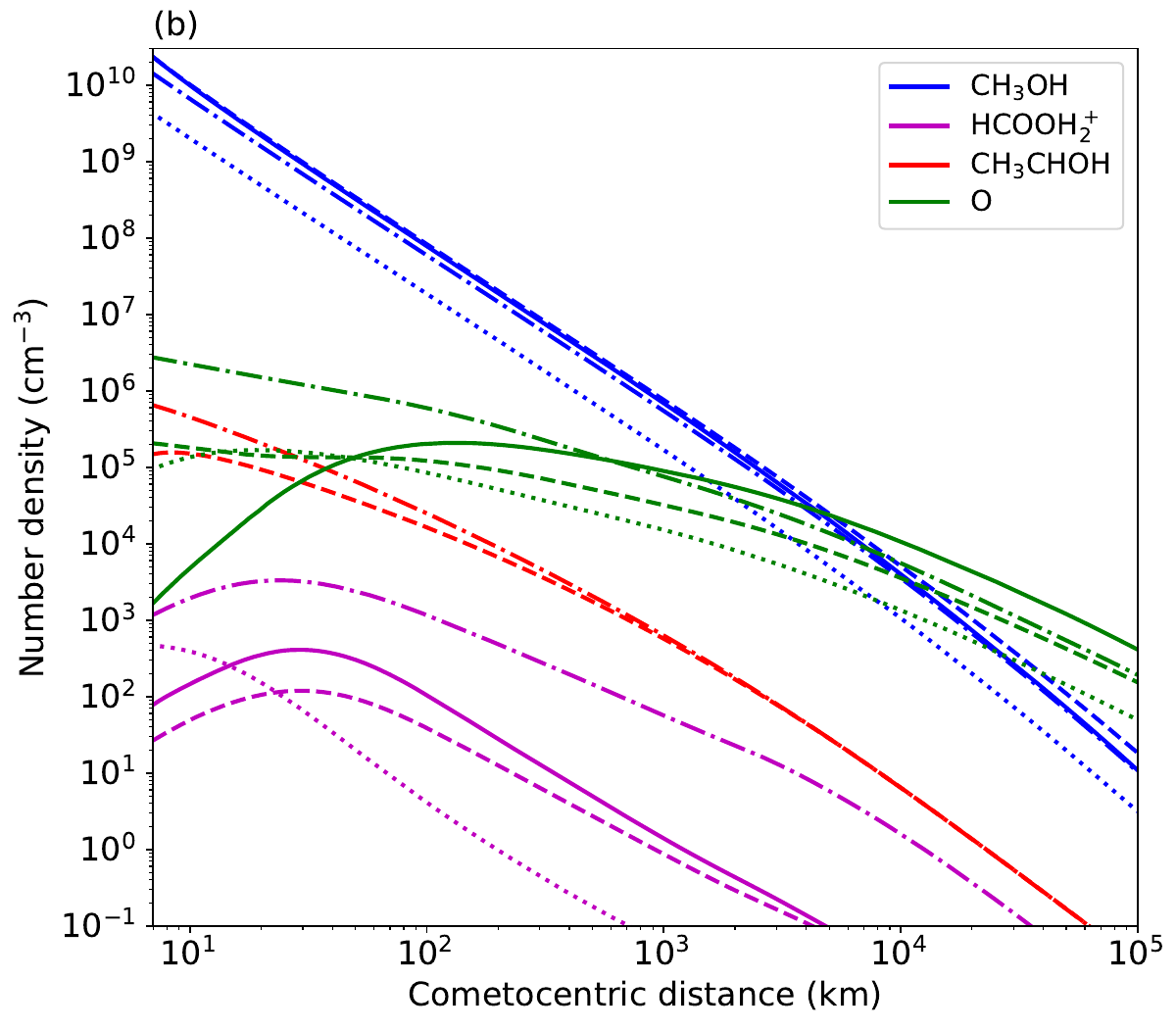}
		
		\caption{(a) The coma flux profile of \ch{HCOOH}. Solid lines indicate abundance due to coma chemistry while 
			dashed lines indicate net abundance due to nucleus outgassing and coma chemistry.
			(b) The cometocentric distance variation of the number densities of species that lead to the formation of \ch{HCOOH} 
			(line styles: same as Figure \ref{fig:HC3N}).}		
		\label{fig:HCOOH}
	\end{figure}
	
Figure \ref{fig:HCOOH}(a) shows the flux of \ch{HCOOH} as a function of the cometocentric distance. The production rate of \ch{HCOOH} 
from the nucleus varies between $1.1\times 10^{26} - 4.7\times 10^{26}$ molecules s$^{-1}$ for the different cometary compositions. 
Beyond $\sim3000$ km, this flux drops sharply due to rapid photodissociation. The flux of \ch{HCOOH} molecules created in the coma 
peaks around $1000-3000$ km, acquiring a value ranging from $10^{20} - 10^{23}$ molecules s$^{-1}$. Most of the \ch{HCOOH} molecules 
originating from the nucleus are destroyed in the outer region of the coma, yet they continue to form due to gas phase chemistry. 
Thus, in a spatially resolved coma, the \ch{HCOOH} molecules in the outer coma are primarily those originating from gas-phase chemistry.
	
The radiative association of \ch{HCO+} ions with water molecules forms \ch{HCOOH2+}, which then undergoes proton transfer reactions 
with molecules of higher proton affinity (such as \ch{CH3OH} and \ch{NH3}) to form \ch{HCOOH} (see Figure \ref{fig:HCOOHa}(a) in Appendix \ref{App_B}). 
The dissociative recombination of \ch{HCOOH2+} also creates \ch{HCOOH}, and this 
reaction becomes dominant in the outer regions when the electron temperature is high. Another process that creates \ch{HCOOH} in 
the coma is the reaction of atomic oxygen with the \ch{CH3CHOH} radical. \ch{CH3CHOH} is created when ethanol undergoes H-abstraction 
reactions with the atomic and molecular radicals O, H and OH (see Figure 
\ref{fig:HCOOHa}(b) in Appendix \ref{App_B}). The formation rate of \ch{CH3CHOH} is high in C/1996 B2 and C/2014 Q2, in which ethanol is a parent
molecule (actually for C/1996 B2, the upper limit for \ch{C2H5OH} is reported; see the following Section \ref{subsubsection:C2H5OH}). Methanol (which undergoes a proton transfer reaction with \ch{HCOOH2+} ions to form HCOOH)
is also a parent molecule in these comets and its number density in the coma is higher than \ch{CH3CHOH}.
However, the number density of \ch {HCOOH2+} ions is low, which reduces the rate of the proton transfer
reaction.
Thus, when ethanol is outgassing from the nucleus, then the dominant process for the creation of 
\ch{HCOOH} is by the reaction of \ch{CH3CHOH} with oxygen (which is present in abundance due to photodissociation of \ch{H2O}). 
$P_{\ch{HCOOH}}$ is higher in C/1996 B2 and C/2014 Q2 by more than two orders of magnitude, since ethanol is present as a parent in these two comets. 
	
\begin{figure}
\centering
\includegraphics[height=7cm, width=0.49\textwidth]{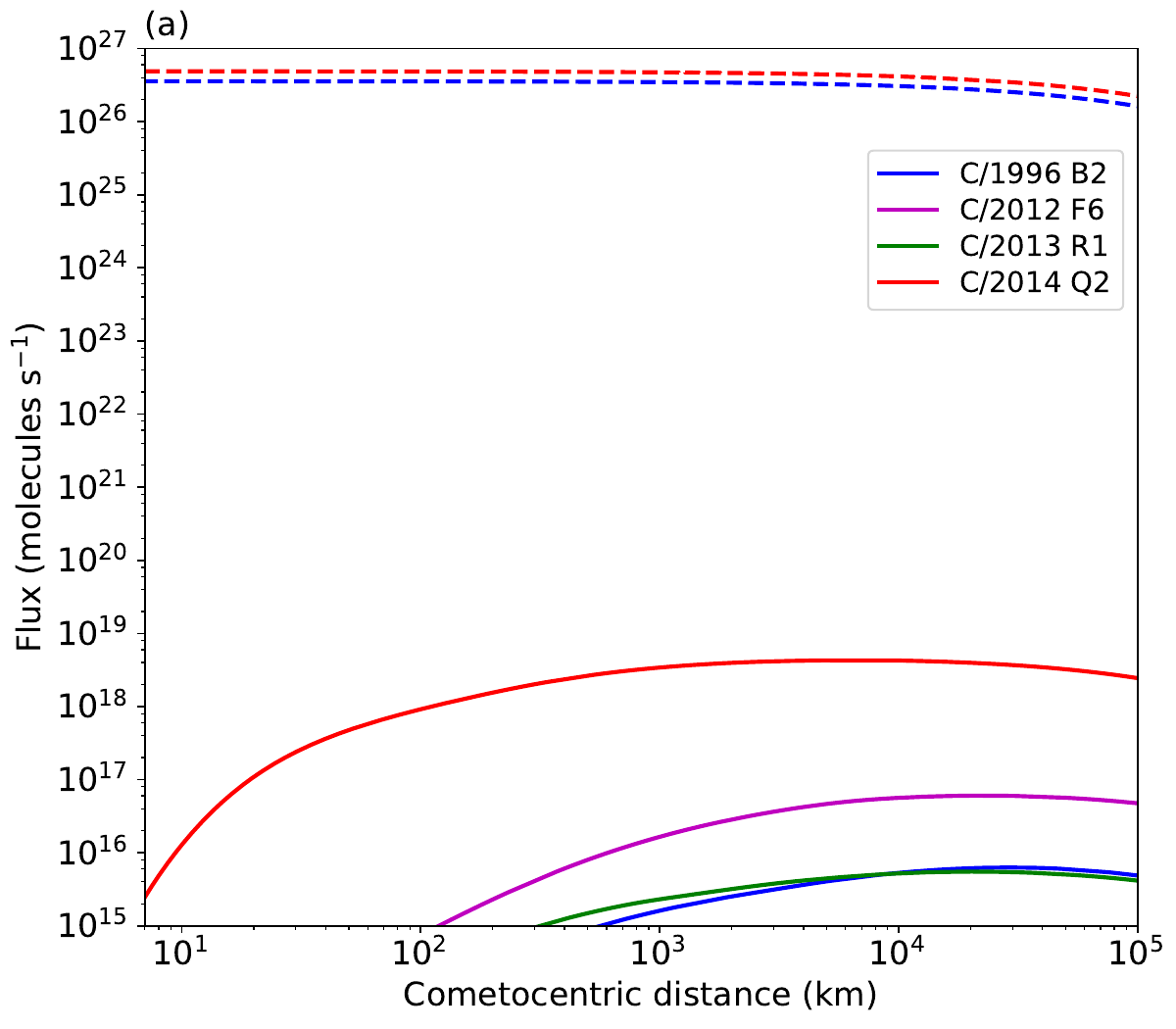}
\includegraphics[height=7cm, width=0.49\textwidth]{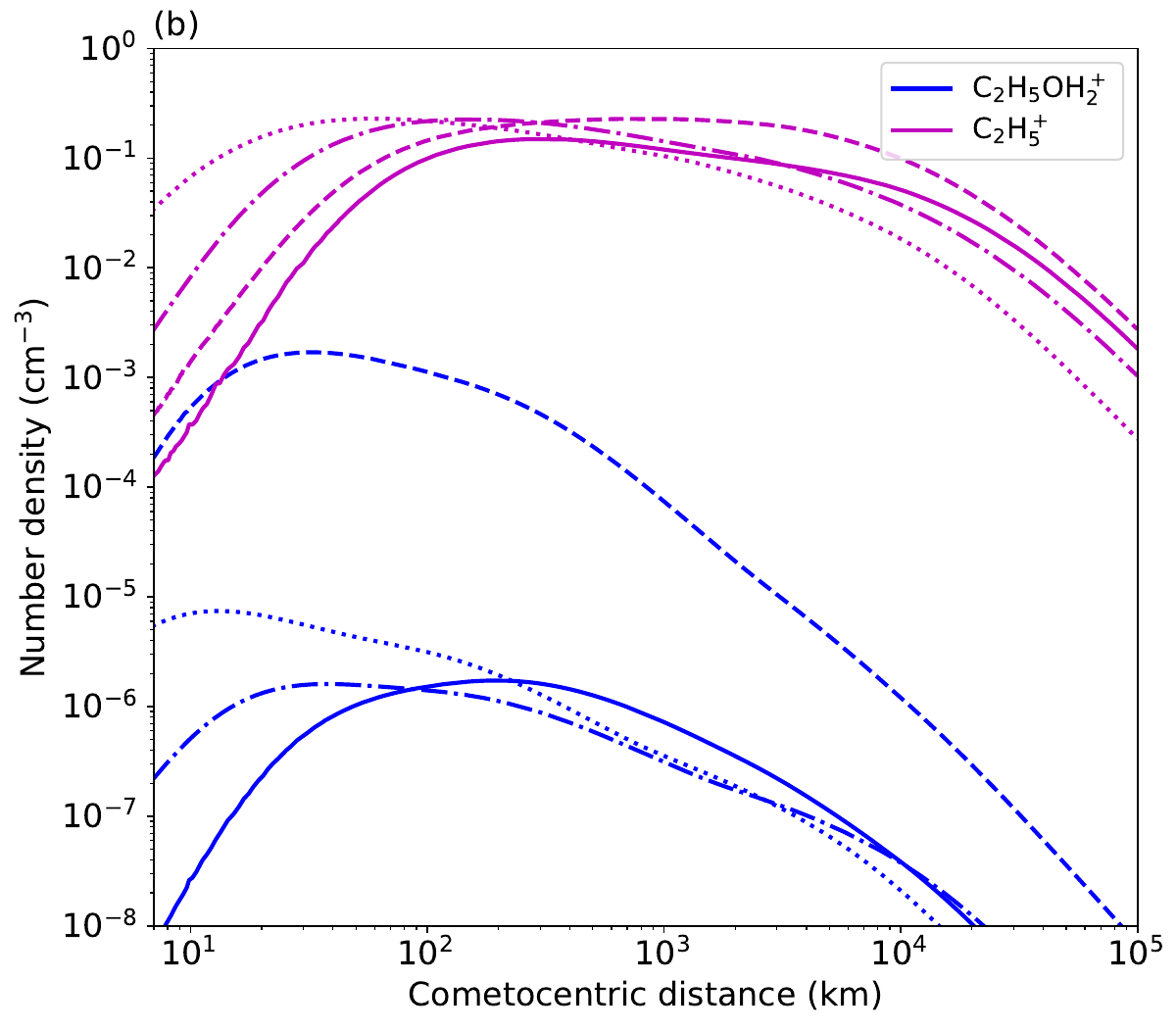}
\caption{(a) The coma flux profile of \ch{C2H5OH}. Solid lines indicate abundance due to coma chemistry while 
dashed lines indicate net abundance due to nucleus outgassing and coma chemistry.
(b) The cometocentric distance variation of the number densities of species that lead to the formation of \ch{C2H5OH} 
(line styles: same as Figure \ref{fig:HC3N}).}		
\label{fig:C2H5OH}
\end{figure}
	
\begin{figure}
\centering
\includegraphics[height=7cm, width=0.49\textwidth]{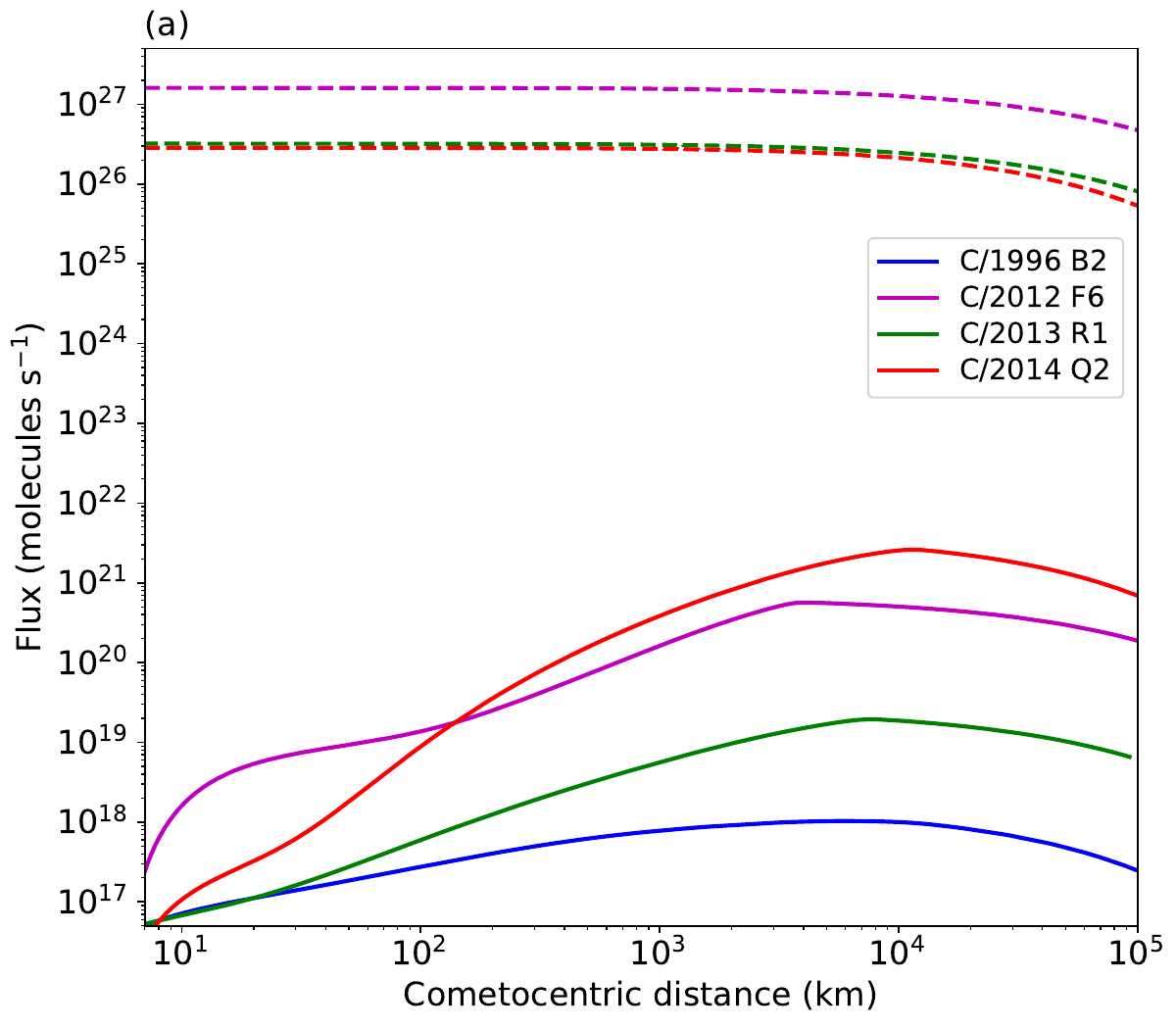}
\includegraphics[height=7cm, width=0.49\textwidth]{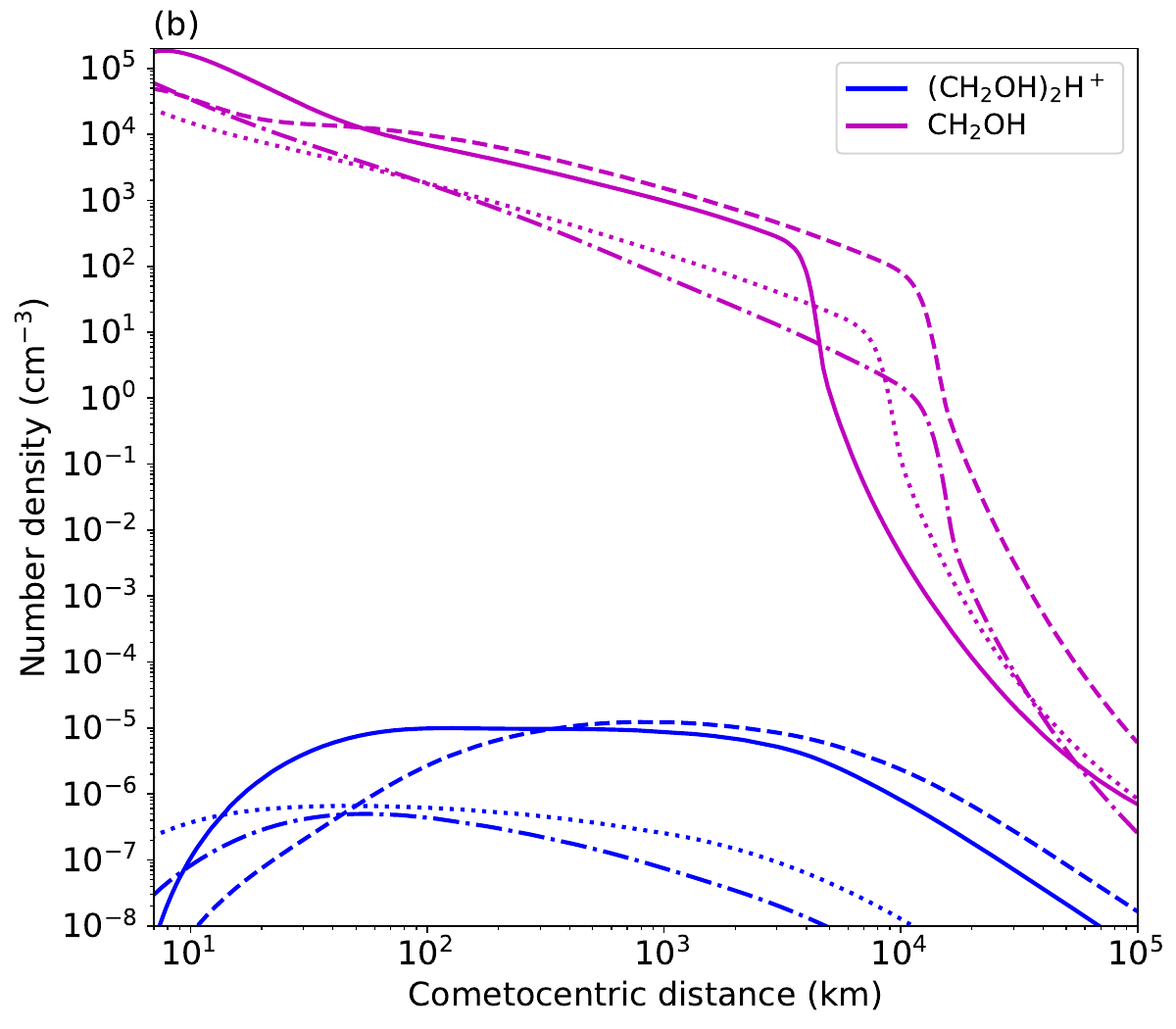}
\caption{(a) The coma flux profile of \ch{(CH2OH)2}. Solid lines indicate abundance due to coma chemistry while 
dashed lines indicate net abundance due to nucleus outgassing and coma chemistry.
(b) The cometocentric distance variation of the number densities of species that lead to the formation of \ch{(CH2OH)2}
(line styles: same as Figure \ref{fig:HC3N}).}		
\label{fig:CH2OH_2}
\end{figure}
	
\subsubsection{\ch{C2H5OH}} \label{subsubsection:C2H5OH}
Figure \ref{fig:C2H5OH}(a) shows the flux of \ch{C2H5OH} as a function of the cometocentric distance. \ch{C2H5OH} is 
parent molecule only in two comets, namely C/1996 B2 and C/2014 Q2 (upper limit is reported in C/1996 B2), with production rates of $3.6\times10^{26}$ and $4.9\times10^{26}$ 
molecules s$^{-1}$ respectively. The production rate of \ch{C2H5OH} due to coma chemistry, acquires a constant value in the outer 
region of the coma, lying in the range $5.5\times 10^{15} - 4.3\times10^{18}$ molecules s$^{-1}$. This is lower than the nucleus 
outgassing rate, by many orders of magnitude. Thus, gas-phase chemistry is not efficient in producing \ch{C2H5OH} in the coma. 
	
Proton transfer and dissociative recombination reactions 
undergone by \ch{C2H5OH2+} lead to the formation of \ch{C2H5OH} in the coma (see Figure \ref{fig:C2H5OHa}(a) in Appendix \ref{App_B}). The number 
density of \ch{C2H5OH2+} in the coma is extremely low, which is the reason for the low values of $P_{\ch{C2H5OH}}$. \ch{C2H5OH2+} 
forms by the radiative association of \ch{H2O} with \ch{C2H5+} (see Figure \ref{fig:C2H5OHa}(b) in Appendix \ref{App_B}). 
Even though water is the most abundant parent volatile, the low formation rate of \ch{C2H5OH2+} is due to the low density of \ch{C2H5+} 
ions in the coma and the slow rate at which \ch{H2O} and \ch{C2H5+} react (Arrhenius coefficient $\alpha \sim 10^{-16}$). In C/2014 Q2, 
the density of \ch{C2H5OH2+} is higher than the other comets by $3-4$ orders of magnitude. This is due to the outgassing of \ch{C2H4} 
from the nucleus of C/2014 Q2, which undergoes radiative association with \ch{H3O+} ions to produce \ch{C2H5OH2+} ions. Though this 
reaction has a higher rate than the previous one, it is still not fast enough to produce sufficient quantities of \ch{C2H5OH2+} (and 
in turn \ch{C2H5OH}) by gas-phase chemistry. 

In C/1996 B2, the ethanol production is reported as an upper limit; thus 
the blue dashed line in Figure \ref{fig:C2H5OH}(a), which shows the net abundance due to nucleus outgassing and
coma chemistry in C/1996 B2, is an upper limit.
This implies that the abundance of \ch{CH3CHOH} (formed by the H-abstraction of ethanol) in 
C/1996 B2 is also an upper limit. 
Hence, the abundance of \ch{HCOOH} (which is created from \ch{CH3CHOH}) in Figure \ref{fig:HCOOH}(a) is also to be treated as an upper limit. The same can be said for the formation of \ch{CH2CH2OH},
\ch{CH2OHCHO} and \ch{CH3COOH} in C/1996 B2, as described later in Sections \ref{subsubsection:CH2OHCHO} and 
\ref{subsubsection:isomer}. 
	
\subsubsection{\ch{(CH2OH)2}} \label{subsubsection:CH2OH_2}
	
Figure \ref{fig:CH2OH_2}(a) shows the flux of \ch{(CH2OH)2} as a function of the cometocentric distance. The parent molecular flux 
of \ch{(CH2OH)2} lies in the range of $2.8\times10^{26} - 1.6\times10^{27}$ molecules s$^{-1}$ for the different comets. The production 
rate of \ch{(CH2OH)2} due to chemical reaction in the coma acquires a peak value lying in the range of $10^{18} - 2.6\times10^{21}$ molecules s$^{-1}$. 
Similar to ethanol, methanol also undergoes H-abstraction reactions with neutral atoms and radicals to form \ch{CH2OH} radicals. The combination reaction of two \ch{CH2OH} radicals leads to the formation of \ch{(CH2OH)2}  (see Figure 
\ref{fig:CH2OH_2a}(a) in Appendix \ref{App_B}). 
At distances $> 5000$ km, there is a sharp decrease in the number density of \ch{{CH2OH}}, which leads to a sudden fall in $P_{\ch{(CH2OH)2}}$. In 
the outer region of the coma, where the temperature of the neutral fluid is high, \ch{CH2OH} is rapidly destroyed by water molecules. This is the 
reason for the exponential decrease of \ch{CH2OH} abundance. The density of \ch{CH2OH} is higher in C/2012 F6 and C/2014 Q2, since they have the highest 
production rates of \ch{CH3OH}. The \ch{CH2OH} abundance begins to fall off earlier in C/2012 F6, as compared to the other comets, 
since it has the highest production rate of \ch{H2O}, which destroys \ch{CH2OH}. In the outer region of the coma, \ch{(CH2OH)2} forms by the 
dissociative recombination of \ch{(CH2OH)2H+}. This is another instance of the interconversion of a species and its protonated form, as discussed 
previously for \ch{NH2CHO}.

\subsubsection{\ch{CH2OHCHO}} \label{subsubsection:CH2OHCHO}
	
\begin{figure}
\centering
\includegraphics[height=7cm, width=0.49\textwidth]{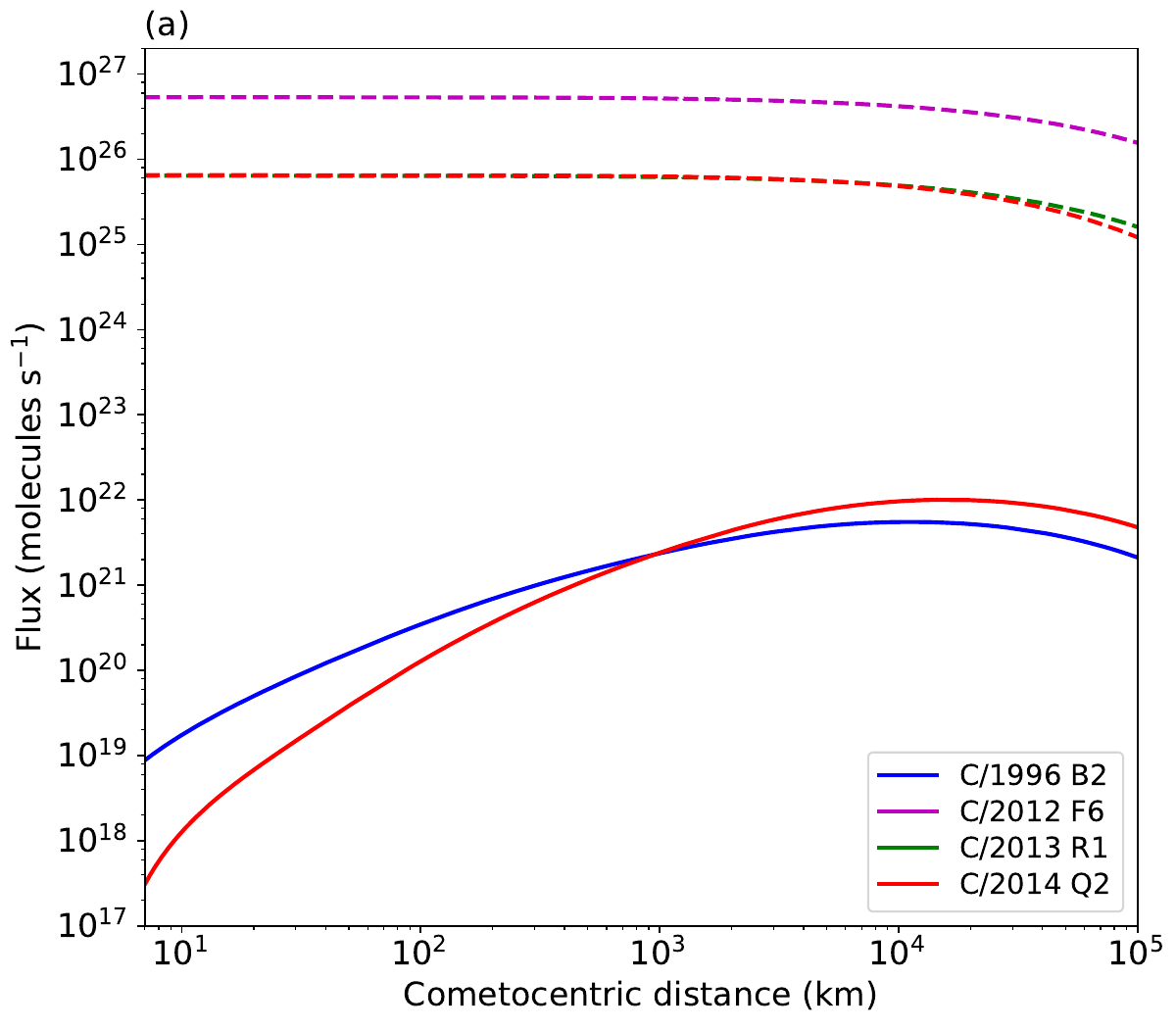}
\includegraphics[height=7cm, width=0.49\textwidth]{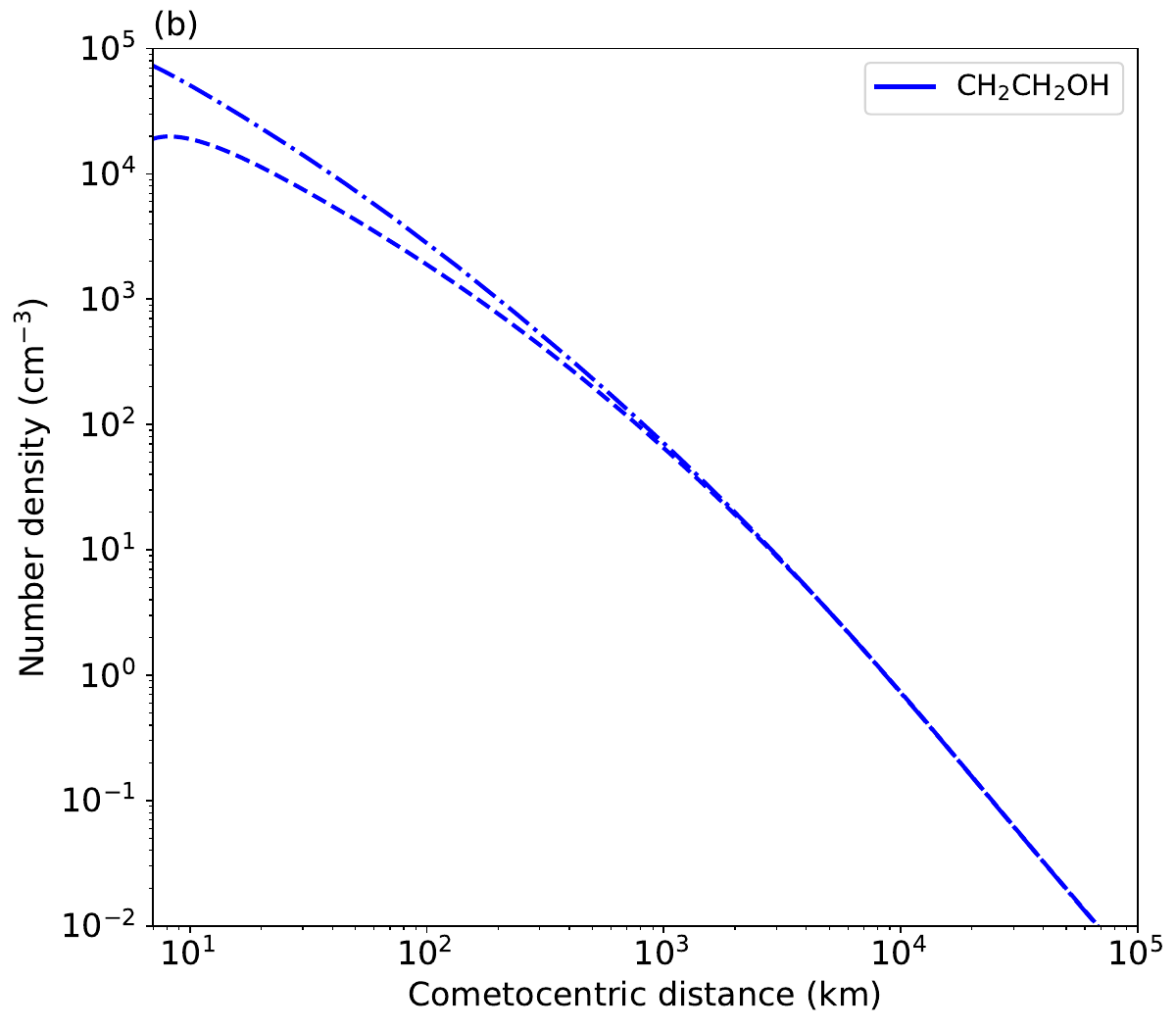}
\caption{(a) The coma flux profile of \ch{CH2OHCHO}. Solid lines indicate abundance due to coma chemistry while 
dashed lines indicate net abundance due to nucleus outgassing and coma chemistry.
(b) The cometocentric distance variation of the number density of \ch{CH2CH2OH} that leads to the formation of \ch{CH2OHCHO}
(line styles: same as Figure \ref{fig:HC3N}).}		
\label{fig:CH2OHCHO}
\end{figure}
	
Figure \ref{fig:CH2OHCHO}(a) shows the flux of \ch{CH2OHCHO} as a function of the cometocentric distance. The parent molecular flux of \ch{CH2OHCHO} 
lies in the range $6.4\times10^{25} - 5.4\times10^{26}$ molecules s$^{-1}$ for the different comets. The peak value of the production rate of 
\ch{CH2OHCHO} forming in the gas-phase coma of C/1996 B2 and C/2014 Q2 is $\sim 10^{22}$ molecules s$^{-1}$. In C/2012 F6 and C/2013 R1, the 
flux of \ch{CH2OHCHO} due to coma chemistry is quite low, of the order $\lesssim10^{12}$ molecules s$^{-1}$. \ch{CH2OHCHO} forms by the 
neutral-neutral reaction of atomic oxygen with \ch{CH2CH2OH} radicals (see Figure \ref{fig:CH2OHCHOa} in Appendix \ref{App_B}). The H-abstraction 
of ethanol can occur from different sites, leading to the formation of \ch{CH3CHOH} (discussed previously) and \ch{CH2CH2OH}. Since ethanol is 
outgassing from the nucleus of C/1996 B2 and 
C/2014 Q2, the formation rate and number density of \ch{CH2CH2OH} (upper limit for the case of C/1996
B2) is much higher in these two comets, which was also seen previously in the 
formation of \ch{CH3CHOH} from ethanol. In C/2012 F6 and C/2013 R1, the absence of ethanol as a parent leads to exceedingly low quantities 
of \ch{CH2OHCHO} that can form in the coma.
	
	\subsubsection{\ch{CH3CHO}} \label{subsubsection:CH3CHO}
	
	\begin{figure}
		\centering
		\includegraphics[height=7cm, width=0.49\textwidth]{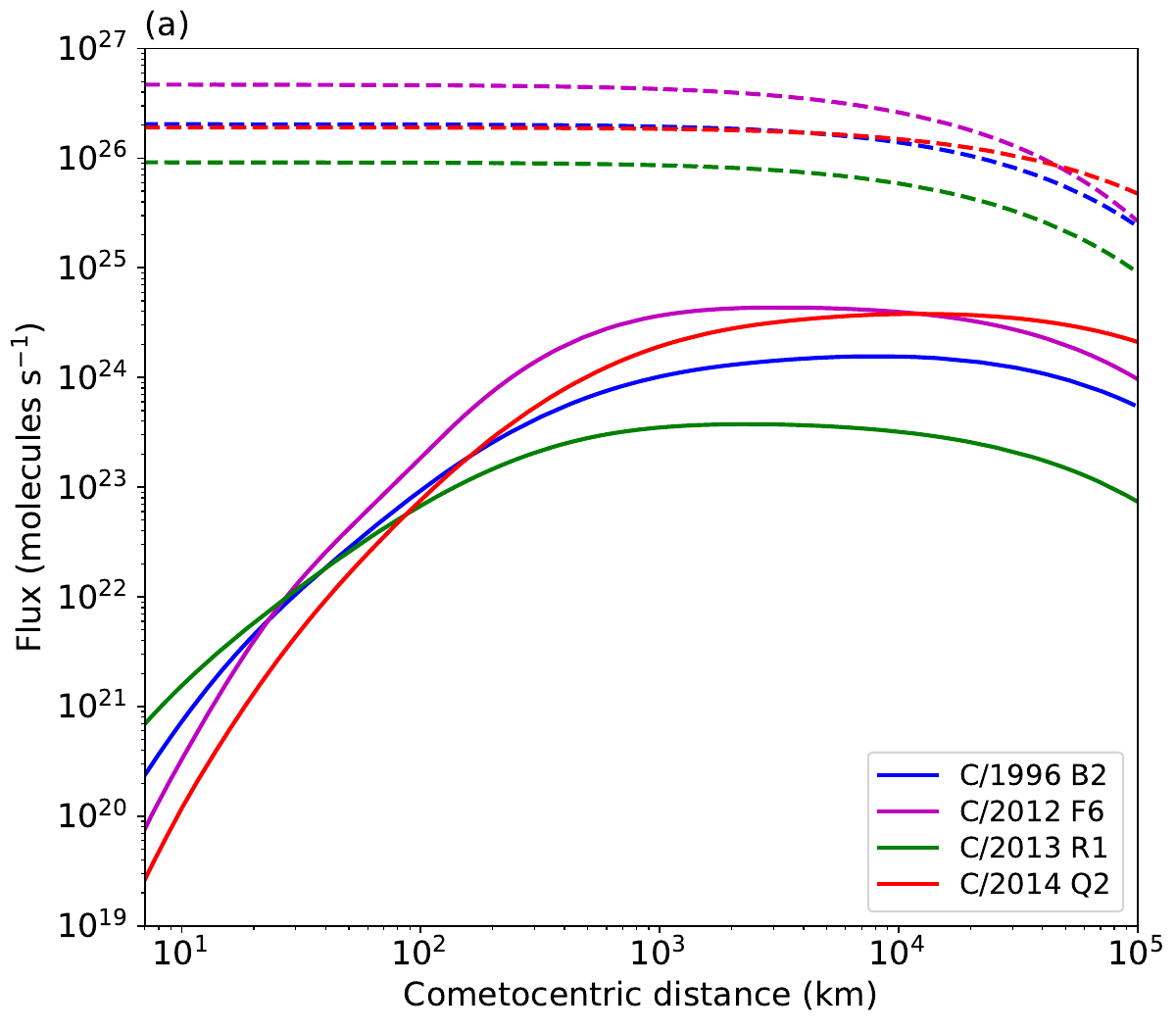}
		\includegraphics[height=7cm, width=0.49\textwidth]{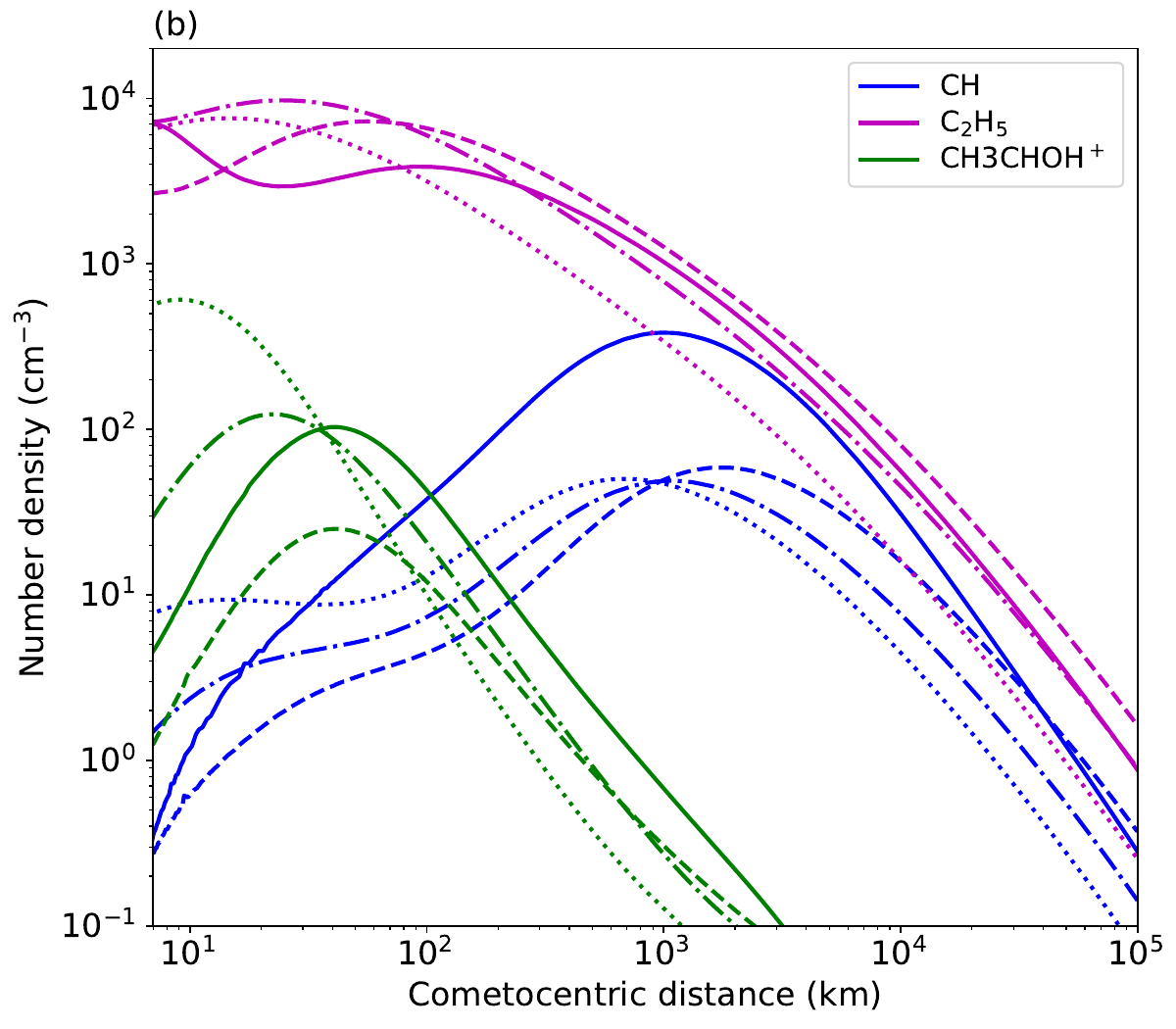}
		\caption{(a) The coma flux profile of \ch{CH3CHO}. Solid lines indicate abundance due to coma chemistry while 
			dashed lines indicate net abundance due to nucleus outgassing and coma chemistry.
			(b) The cometocentric distance variation of the number densities of species that lead to the formation of \ch{CH3CHO} 
			(line styles: same as Figure \ref{fig:HC3N}).}		
		\label{fig:CH3CHO}
	\end{figure}
	
	Figure \ref{fig:CH3CHO}(a) shows the flux of \ch{CH3CHO} as a function of the cometocentric distance. The flux of \ch{CH3CHO} due to 
	outgassing from the nucleus lies in the range of $9.1\times 10^{25} - 4.7\times10^{26}$ molecules s$^{-1}$ for the different comets. 
	The coma production rate of \ch{CH3CHO} acquires maximum value in the region of the coma from $10^3 - 10^4$ km, and the values lie 
	between $3.8\times 10^{23} - 4.4\times10^{24}$ molecules s$^{-1}$. 
	There are a number of reactions 
	that lead to the creation of \ch{CH3CHO} in the coma, and the relative contribution of these reactions towards $P_{\ch{CH3CHO}}$ 
	varies significantly in different regions of the coma and with changes in cometary compositions (see Figure \ref{fig:CH3CHOa} in Appendix \ref{App_B}). 
	Most of the \ch{CH3CHO} molecules are created in the coma by the following two neutral-neutral reactions:
	
	\begin{equation}
		\begin{split}
			&\ch{CH3OH} + \ch{CH} \rightarrow \ch{CH3CHO} + \ch{H} \\
			&\ch{C2H5} + \ch{O} \rightarrow \ch{CH3CHO} + \ch{H}. \\
		\end{split}
	\end{equation}
	The first of these reactions varies roughly as the inverse square of the temperature. Thus its contribution to $P_{\ch{CH3CHO}}$ 
	falls when the temperature is high, and the second reaction takes over.
	\ch{CH} and \ch{C2H5} are created by the photodissociation of \ch{CH4} and \ch{C2H6}, respectively. The production rate of \ch{CH4} 
	and \ch{C2H6} is the highest in C/2012 F6, and C/2014 Q2 also has similar production rates of these molecules. Thus, the flux of 
	\ch{CH3CHO} molecules is the highest in these comets. At low cometocentric distances, the proton transfer reaction of \ch{CH3CHOH+} 
	with \ch{NH3} can contribute upto $\sim 50\%$ towards $P_{\ch{CH3CHO}}$. This is the region where attenuation of the UV flux reduces
	the rate at which \ch{CH} forms, and the density of \ch{CH3CHOH+} ions is higher. There is some contribution to $P_{\ch{CH3CHO}}$ by 
	the dissociative recombination of \ch{C2H5OH+} and \ch{C2H5OH2+} ions in C/1996 B2 and C/2014 Q2 (labeled  `diss. recom' in Figure \ref{fig:CH3CHOa}). Since ethanol is present as a parent in these two comets, the ions created due 
	to photoionization and protonation of ethanol are abundant. 
	
	\subsubsection{\ch{CH3CN}} \label{subsubsection:CH3CN}
	
	\begin{figure}
		\centering
		\includegraphics[height=7cm, width=0.49\textwidth]{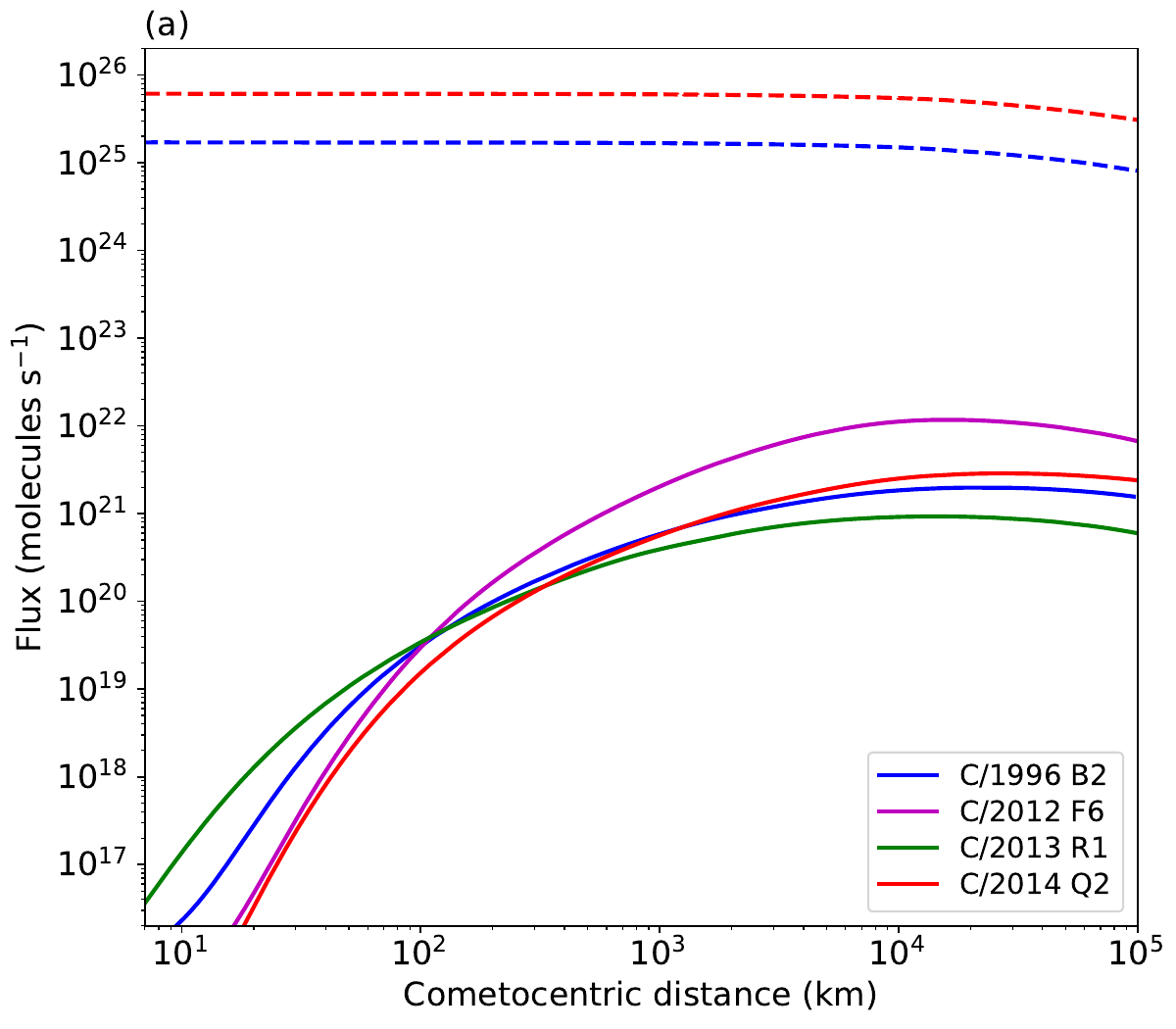}
		\includegraphics[height=7cm, width=0.49\textwidth]{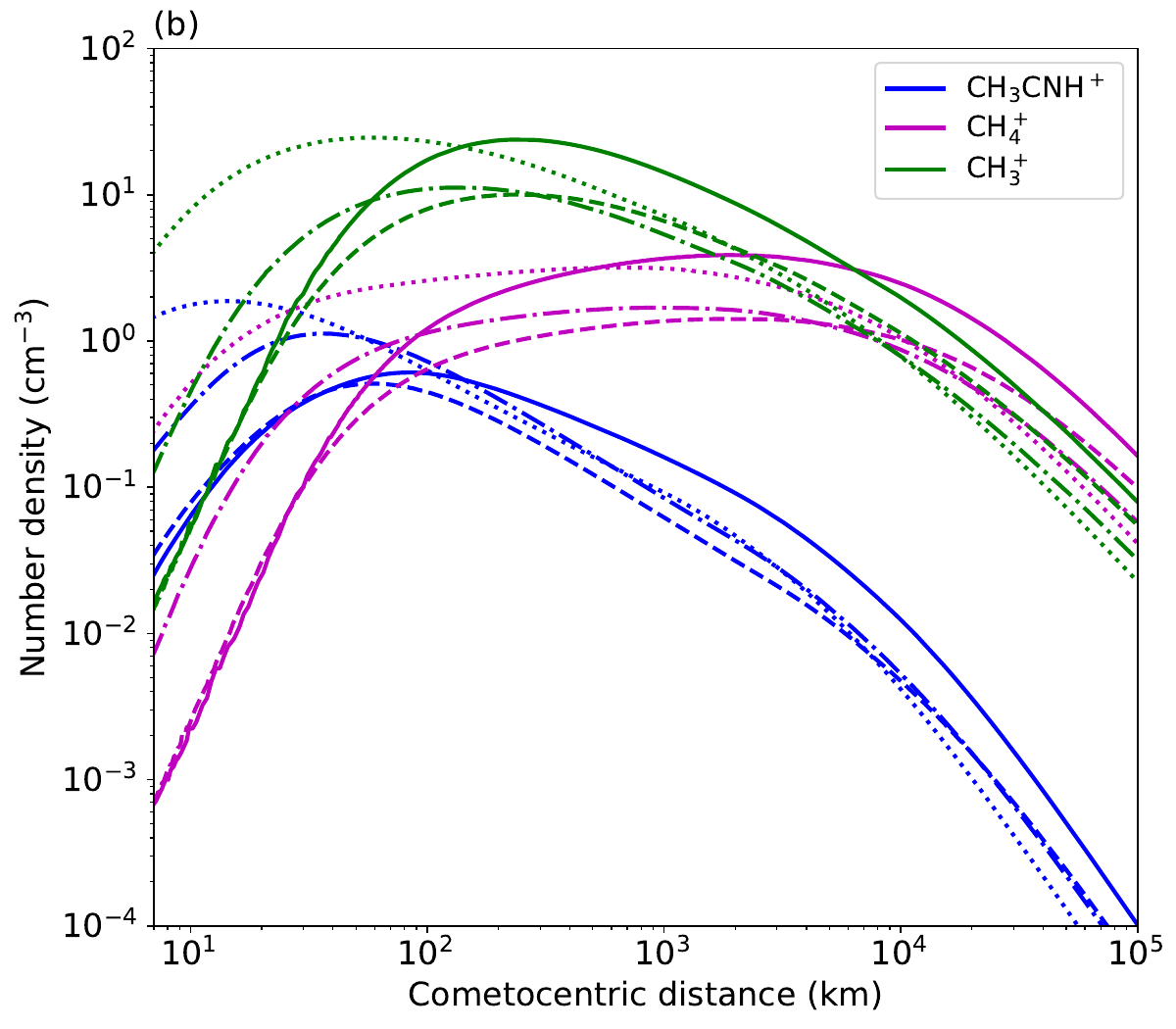}
		\caption{(a) The coma flux profile of \ch{CH3CN}. Solid lines indicate abundance due to coma chemistry while 
			dashed lines indicate net abundance due to nucleus outgassing and coma chemistry.
			(b) The cometocentric distance variation of the number densities of species that lead to the formation of \ch{CH3CN} 
			(line styles: same as Figure \ref{fig:HC3N}).}		
		\label{fig:CH3CN}
	\end{figure}
	
	Figure \ref{fig:CH3CN}(a) shows the flux of \ch{CH3CN} as a function of the cometocentric distance. \ch{CH3CN} is present as a parent 
	molecule in C/1996 B2 and C/2014 Q2, with production rates of $1.7\times10^{25}$ and $6.1\times10^{25}$ molecules s$^{-1}$ respectively. 
	The production rate of \ch{CH3CN} due to chemical reactions in the coma lies in the range from $9.2\times10^{20} - 1.2\times10^{22} $ 
	molecules s$^{-1}$. 
	Although there are a number of processes that form \ch{CH3CN} in the coma, the dissociative recombination of \ch{CH3CNH+} is the primary 
	reaction in most regions. \ch{CH3CNH+} is created by the ion-neutral 
	reaction of \ch{HCN} with \ch{CH3+} and \ch{CH4+} ions (for variation of formation rates, see Figure \ref{fig:CH3CNa} in Appendix \ref{App_B}). 
	The formation rate of \ch{CH3CNH+}, and consequently the flux of \ch{CH3CN} is the highest in C/2012 F6. This is because C/2012 F6 has the 
	highest production rate of \ch{HCN} and \ch{CH4} (which forms \ch{CH3+} and \ch{CH4+} ions by photoionization/photodissociative ionization). 
	The attenuation of UV flux reduces the density of \ch{CH3+} and \ch{CH4+} below 100 km, as seen previously for other molecules. Since this 
	effect is the weakest in C/2013 R1, the abundance of \ch{CH3CN} is the highest in this comet close to the nucleus, as before. Other bimolecular 
	reactions may form \ch{CH3CN} in the coma, though their contribution is limited to distances $<100$ km and in the outer coma region. 
	
\subsection{Other Organics}  \label{subsection:reslt_spec2}
\subsubsection{\ch{CH3OCH3} and \ch{CH3COOH}} \label{subsubsection:isomer}
	
\begin{figure}
\centering
\includegraphics[height=7cm, width=0.32\textwidth]{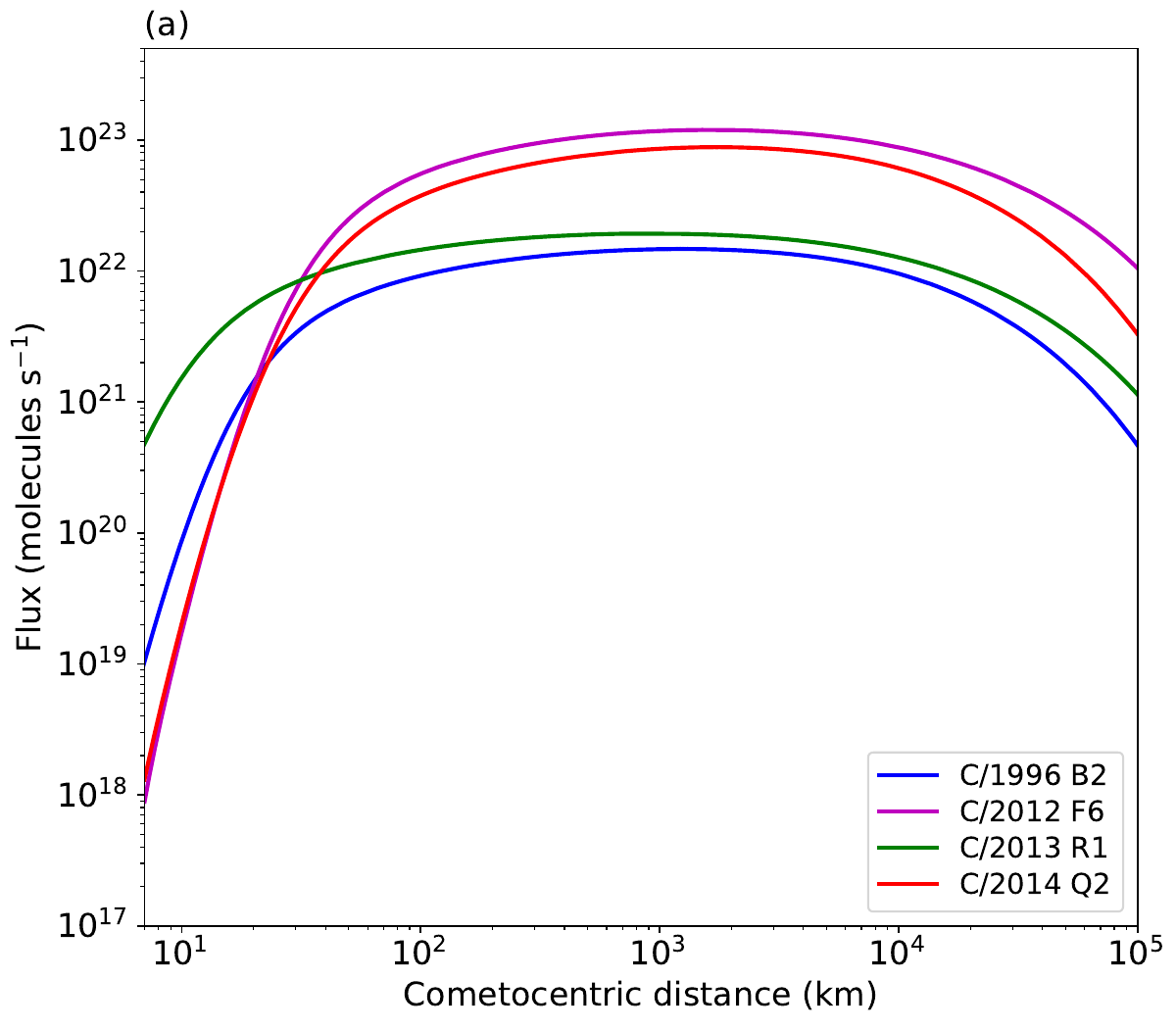}
\includegraphics[height=7cm, width=0.32\textwidth]{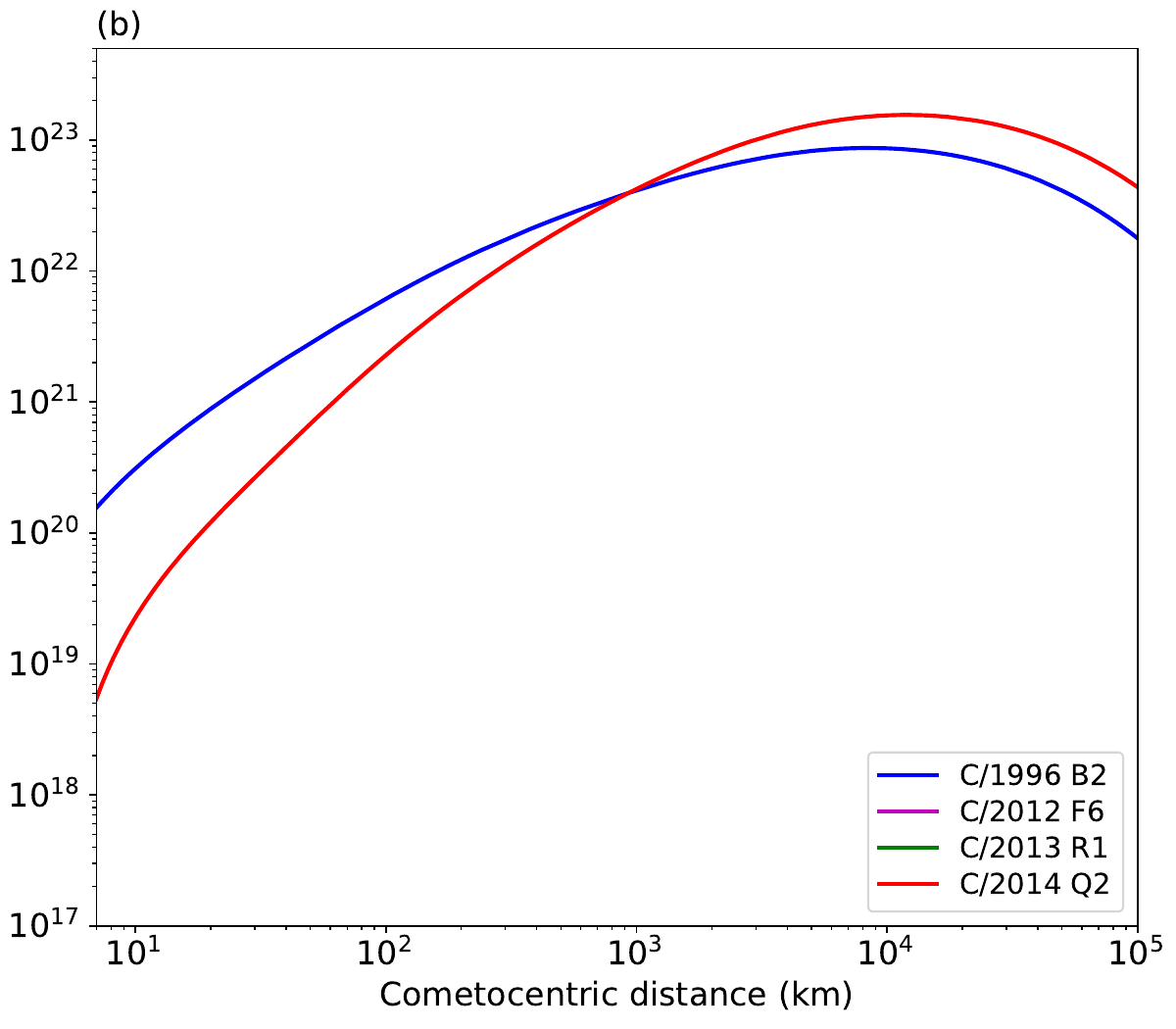}
\includegraphics[height=7cm, width=0.32\textwidth]{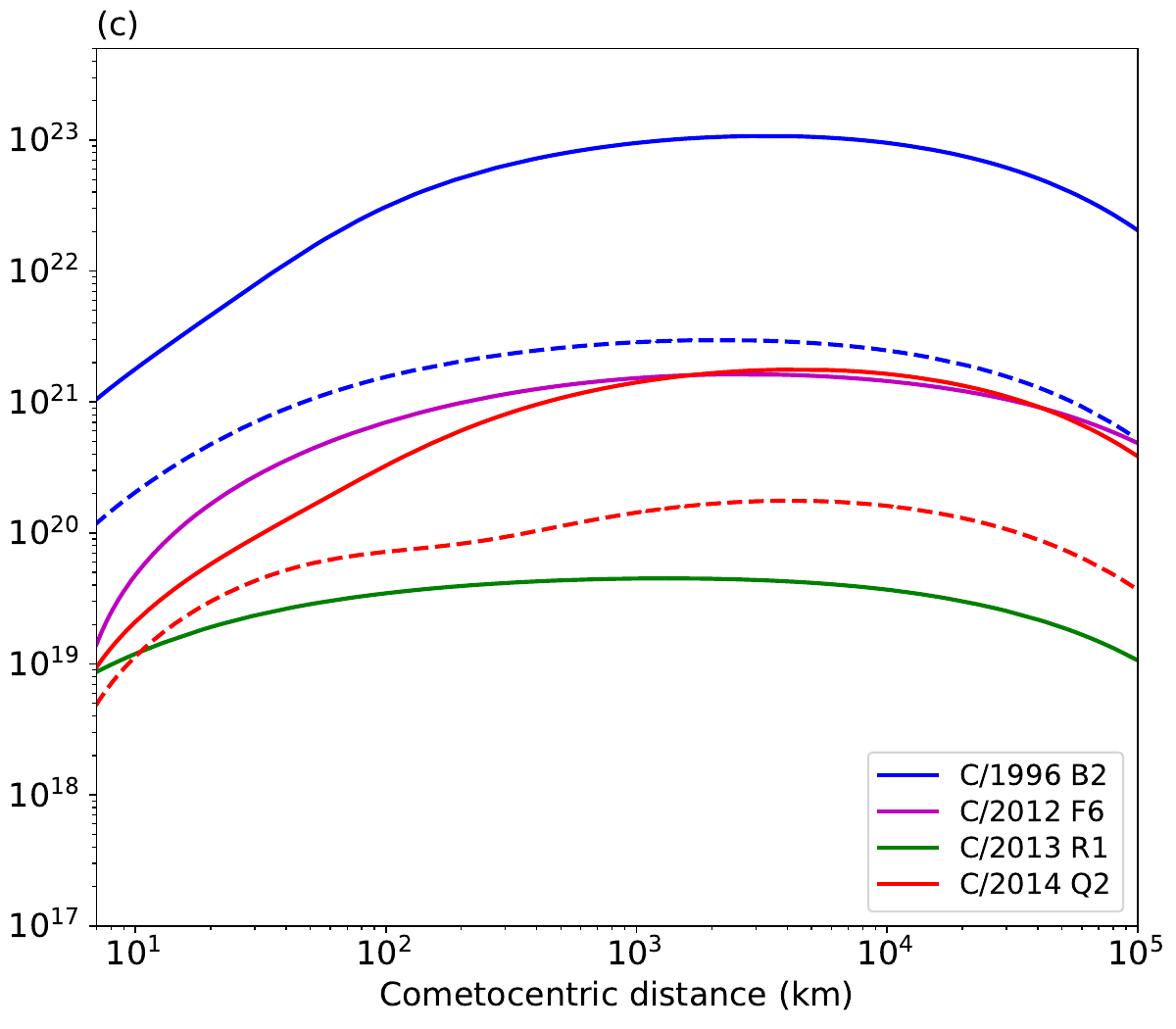}
		
\caption{Coma flux profiles of (a) \ch{CH3OCH3}, (b) \ch{CH3COOH}, and (c) \ch{HC5N}, resulting from coma chemistry. The dashed lines in panel (c) show the flux of \ch{HC5N} when \ch{HC3N} is not present as a parent in C/1996 B2 and C/2014 Q2.}		
\label{fig:CH3OCH3-CH3COOH}
\end{figure}
	
Organic molecules can be present in the coma in the form of isomers, for example, ethanol (\ch{C2H5OH}) and dimethyl 
ether (\ch{CH3OCH3}) are the isomeric forms of \ch{C2H6O}, while \ch{C2H4O2} has methyl formate (\ch{HCOOCH3}), glycolaldehyde 
(\ch{CH2OHCHO}) and acetic acid (\ch{CH3COOH}) as its isomers. 
The flux of \ch{CH3OCH3} (isomer of the parent molecule \ch{C2H5OH}) and \ch{CH3COOH} (isomer of the parent
molecule \ch{HCOOCH3}) forming in the coma due to gas phase formation reactions is shown in Figures \ref{fig:CH3OCH3-CH3COOH}(a) and 
\ref{fig:CH3OCH3-CH3COOH}(b), respectively. The coma produced flux of \ch{CH3OCH3} acquires a peak value beyond $\sim 1000$ km, 
ranging from $1.5\times 10^{22} - 1.2\times 10^{23}$ molecules s$^{-1}$ for different cometary compositions. The production rate 
of \ch{CH3COOH} in C/1996 B2 and C/2014 Q2 reaches peak values of about $\sim 10^{23}$ molecules s$^{-1}$
(upper limit for the case of C/1996 B2), while it is produced 
in low quantities in C/2012 F6 and C/2013 R1 ($<10^{14}$ molecules s$^{-1}$).
The methyl cation transfer reaction between \ch{CH3OH} and \ch{CH3OH2+} creates \ch{CH3OCH4+} or 
protonated dimethyl ether, which then undergoes dissociative recombination to form \ch{CH3OCH3} in the coma (formation rates shown in Figure 
\ref{fig:CH3OCH3-CH3COOHa}(a) in Appendix \ref{App_B}). C/2012 F6 and C/2014 Q2 have similar production rates of \ch{CH3OH}, which are also the highest 
among all four comets. Since methanol gets 
converted into \ch{CH3OH2+} and \ch{CH3OCH4+} ions, its presence drives the formation of \ch{CH3OCH3} in the coma. Thus, 
$P_{\ch{CH3OCH3}}$ is similar in C/2012 F6 and C/2014 Q2, and the resulting abundance of \ch{CH3OCH3} is the highest in 
these two comets at distances $>100$ km. Below this distance, the rate of formation of \ch{CH3OH2+} is the highest in C/2013 R1 
due to least attenuation of the UV flux, as discussed previously in Section \ref{subsection:reslt_spec}. Thus, \ch{CH3OCH3} has 
the highest rate of formation in C/2013 R1 at low cometocentric distances. 
The formation mechanism of \ch{CH3COOH} (see Figure \ref{fig:CH3OCH3-CH3COOHa}(b) in Appendix \ref{App_B}) in the coma 
is similar to that of its isomer \ch{CH2OHCHO}. Both these molecules are created when atomic oxygen causes the elimination of 
an H atom from H-abstracted ethanol radicals (\ch{CH3CHOH} forms acetic acid while \ch{CH2CH2OH} forms glycolaldehyde). It is 
to be noted, however, that the coma produced abundance of \ch{CH3COOH} is higher than that of \ch{CH2OHCHO}, since \ch{CH3CHOH} 
has a higher coma formation rate than \ch{CH2CH2OH}. The absence of ethanol as a parent in C/2012 F6 and C/2013 R1 leads to 
significantly lower formation rates of \ch{CH3COOH} in this two comets, as seen previously in the case of the coma formation 
of glycolaldehyde.
	
\subsubsection{\ch{HC5N}}
Cyanopolyynes are known to form in the gas phase, and Figure \ref{fig:CH3OCH3-CH3COOH}(c) shows the flux of \ch{HC5N} forming 
in the coma, (see Figure \ref{fig:CH3OCH3-CH3COOHa}(c) in Appendix \ref{App_B} for formation rate plot for the comets C/1996 B2 and C/2012 F6; 
the relative reaction rates in C/2014 Q2 follow a similar trend as C/1996 B2, while C/2013 R1 follows a similar trend as C/2012).
A dominant formation mechanism of \ch{HC5N} is the bimolecular reaction of \ch{HC3N} with \ch{C2H}. The presence of \ch{HC3N} 
as a parent molecule in C/1996 B2 and C/2014 Q2 results in higher values of $P_{\ch{HC5N}}$ in these two comets. However, 
if \ch{HC3N} outgassing is not considered, the resulting \ch{HC5N} abundance can reduce by $1-2$ orders of magnitude, as 
indicated by the dashed lines in Figure \ref{fig:CH3OCH3-CH3COOH}(c). In C/2012 F6 and C/2013 R1, \ch{HC5N} mainly forms by the reaction 
of \ch{C2H2} with the neutral radicals \ch{C3N} and \ch{C5N} (in Figure \ref{fig:CH3OCH3-CH3COOHa}(c), the lines labeled `\ch{C2H2}+neutral' 
show the sum of these two reactions). The reaction of \ch{CN} with \ch{C4H2} contributes upto $\sim 10\%$ 
towards the formation of \ch{HC5N} in certain regions of the coma. Other reactions forming \ch{HC5N} are the dissociative recombination 
of \ch{C5H4N+} and \ch{HC5NH+}. While \ch{C5H4N+} forms by ion-neutral reactions involving \ch{HC3N}, the protonation of \ch{HC5N} 
into \ch{HC5NH+} and its subsequent reconversion into \ch{HC5N} takes place (this interconversion cycle is seen previously in 
the case of \ch{NH2CHO} and \ch{HCOOCH3}). Thus, in C/1996 B2 and C/2014 Q2 the relative dissociative recombination rates are 
higher since these rates are influenced by \ch{HC3N} outgassing from the nucleus. 
	
\subsubsection{\ch{NH2CH2COOH}}
Glycine (\ch{NH2CH2COOH}) is among the most abundant amino acids that is found in CI and CM type carbonaceous chondrites. 
\cite{Crovisier2004} calculated an upper limit of 0.15\% relative to water for glycine in the comet C/1995 O1 (Hale-Bopp). \cite{Elsila2009} 
report on the detection of glycine along with its possible precursor molecules methylamine (\ch{CH3NH2}) and ethylamine (\ch{C2H5NH2}) 
in the dust samples of comet 81P/Wild brought back by the \textit{Stardust} mission, although there was suspected terrestrial contamination. 
\cite{Altwegg2016} report the confirmed presence of glycine, along with \ch{CH3NH2} and \ch{C2H5NH2}, in the coma of 67P/C-G, 
measured by the ROSINA/\textit{Rosetta}.
The relative abundance of glycine with respect to water was found to lie in the range $0-0.0025$, while those of methylamine and ethylamine
with respect to glycine are $1.0\pm0.5$ and $0.3\pm0.2$, respectively.
The presence of glycine in comets can be explained by its 
formation due to chemistry occurring in interstellar icy dust mantles or by the irradiation of ice by UV, and the subsequent preservation in 
cometary ices \citep{Meierhenrich2005,Bossa2010,Garrod2013}. A possible gas-phase formation route of glycine is acetic acid (\ch{CH3COOH}) 
and protonated hydroxylamine (\ch{NH3OH+}) reacting to form protonated glycine, which in turn undergoes recombination with electrons to form 
glycine, as proposed by \cite{Blagojevic2003}. However, we found this to be an insignificant mechanism of producing glycine in the coma, due to the 
presence of insufficient amounts of \ch{NH3OH+}. \cite{Garrod2013} finds that glycine can form within and upon dust-grain ice mantles by radical-radical 
addition mechanisms at temperatures $\sim 40-120$ K, which can be ejected into the coma by various mechanisms. Recently, it has been 
found that the 67P/C-G glycine observations can be explained by considering a distributed source of glycine embedded in water ice on dust particles 
ejected from the nucleus \citep{Hadraoui2019}.

	\subsection{Relative Importance of Chemical Reactions and Temperature Dependence}
The gas phase reactions that 
form neutral cometary organics are (1) neutral-neutral bimolecular reactions, (2) ion-neutral bimolecular reactions, 
and (3) dissociative recombination reactions. Protonated species form neutral molecules either by proton transfer or 
by dissociative recombination. While addition of \ch{H+} to neutral molecules form protonated species, they can also 
form by the  radiative association reactions of an ion and a neutral, or by methyl cation transfer reactions.
In general, the ion that is a reactant in the radiative association reaction is itself formed by the protonation of a 
neutral molecule, and thus carries an extra \ch{H+} cation. For example, in the case of the formation of \ch{HCOOH2+} 
by the association of \ch{H2O} and \ch{HCO+}, the \ch{HCO+} ions are formed by the protonation of \ch{CO}. Thus, an 
\ch{H+}-bearing species or protonated species combines radiatively with a neutral molecule, and the result is the 
formation of a heavier \ch{H+}-bearing ion. Although radiative association is a slow process, the neutral molecules 
that participate in radiative association may be abundant parent species, and the relative contribution of radiative 
association to $P_i$ can be higher than proton transfer reactions.

Dissociative recombination reactions are generally inversely dependent on the square root of the temperature, and the sharp rise 
in the electron temperature in the outer coma region lowers the recombination rate of ions.
However, it is seen for several species that bimolecular reactions have a higher relative contribution to $P_i$ in the inner 
regions of the coma, while the recombination reactions become relatively more dominant as we move outwards. This is because, 
as discussed previously, the percentage contribution of each reaction towards the formation of a molecule depends on the rate 
constants and temperature, and also on the availability of the reacting species. The flux of photoelectrons required for ions 
to undergo recombination is low in the inner regions of the coma, while increasing electron flux results in an increase in the 
relative reaction rate of the dissociative recombination reactions in the outer regions.

\section{Discussions} \label{section:discu}
The results presented in the previous section explore the gas-phase formation pathways of organics 
in an attempt to give an understanding of the coma versus nucleus origin.
We also see that a number of conditions are desirable to the formation of organic molecules. The coma produced abundances of molecules such 
as \ch{NH2CHO} and \ch{HCOOCH3} among others is the highest in C/2012 F6, since it has the highest production rate. Heliocentric distances 
$< 1$ au result in higher activity (leading to a higher coma density which increases the coma formation rate per unit
volume) and increased photolytic rates, which 
kickstart the coma chemistry. The photolytic rates scale roughly as $1/r^2$, and at heliocentric distances $> 1 $ au, the photolytic rates will
be lower, which will reduce the rate of formation of organics. Although the high coma density attenuates the UV flux on moving inwards towards 
the nucleus, this affects the formation of organics only at low cometocentric distances. In terms of composition, the presence 
of certain parent organics such as \ch{H2CO}, \ch{CH3OH} and hydrocarbons is desirable, since this will synthesize organic molecules of 
higher chemical complexity. Additionally, we have shown in our earlier work \citep{Ahmed2021} that the presence of a large amount of 
\ch{CO} also increases the abundance of large organics in the coma.
	
\subsection{Alternate Reaction Pathways}
Although we use the rates suggested by KIDA for obtaining the results of the previous section, alternate formation pathways 
are also recommended in literature. The main formation pathway for protonated methyl formate is through the reaction of \ch{H2CO} with 
\ch{CH3OH2+}. 
However, \cite{Horn2004} suggest that this reaction is not feasible due to the presence of a large barrier. 
Although, later studies by \cite{Cole2012} suggest that this reaction occurs with a branching ratio of 5\%. Also, an alternative ion-neutral 
reaction that has been suggested by \cite{Horn2004} is the reaction of \ch{HCOOH} with \ch{CH3OH2+}, which is included in our network.
Other reactions that may create protonated methyl formate are the radiative association of \ch{H2CO} with \ch{H2COH+} 
and of \ch{HCOOH} with \ch{CH3+}, and they have been tested in astrochemical models \citep{Horn2004, Vasyunin2013}. \cite{Horn2004} 
also modeled the hot core \ch{HCOOCH3} abundance by adopting a reduced rate for the reaction of \ch{H2CO} with \ch{CH3OH2+}, and 
this rate is 50 times smaller than the KIDA rate (they adopt this rate based on room temperature experimental results). By running 
models using this reduced rate, and by including the additional radiative association reactions in our network, we find that the 
main formation pathway is still the reaction of \ch{H2CO} with \ch{CH3OH2+}. However, the reduction in the reaction rate reduces 
the abundance of \ch{HCOOCH3} by 1-2 orders of magnitude. The combined contribution of the other reactions to $P_i$ is about 
10\% in this case. 
	
	Recent experiments conducted by \cite{Douglas2022} suggest the presence of a significant barrier for the formation of 
	formamide by the reaction of \ch{NH2} and \ch{H2CO}. By using the updated values of the Arrhenius coefficients suggested by 
	the authors, we find that the \ch{NH2CHO} abundance in our models is drastically reduced. An alternative pathway for \ch{NH2CHO} 
	formation from protonated formamide (which then undergoes dissociative recombination to create formamide), has been suggested 
	\citep{Barone2015}:
	
	\begin{equation}
		\ch{H2CO} + \ch{NH4+} \rightarrow \ch{NH2CHOH+} + \ch{H2}.
	\end{equation}
	However, the rate for this reaction is not reported. Using constant rate values between $10^{-9}$ cm$^3$ s$^{-1}$ and $10^{-12}$ cm$^3$ s$^{-1}$
	for this reaction we found that the resulting abundances are lower by an order of magnitude to four orders 
	of magnitude than those discussed in Section \ref{section:reslt}.

\begin{figure}[htb!]
	\centering
	\includegraphics[width=1.0\textwidth]{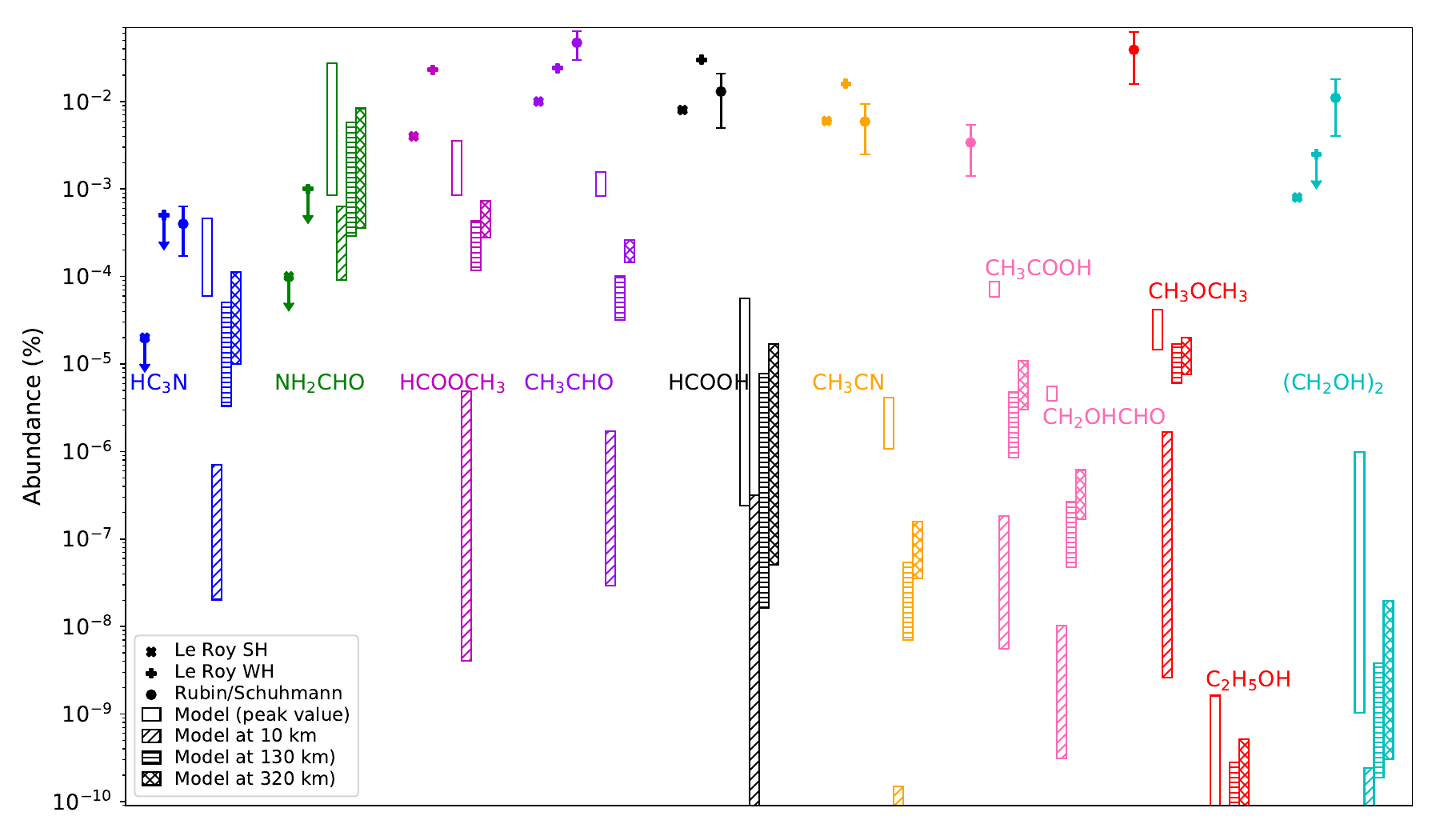}

\caption
{
The abundance of cometary organics observed in 67P/C-G compared with the abundances
obtained from our model runs (formation only due to gas phase chemistry). The scatter points
show the organic abundance (\%) with respect to water in 67P/C-G reported at $\sim$ 3 au (\citealp{LeRoy2015};
SH: summer hemisphere, WH: winter hemisphere) and $\sim$ 1.5 au \citep{Rubin2019b, Schuhmann2019}.
The water production rates of our modeled comets are 2-3 orders of magnitude higher than
that of 67P/C-G at 3 AU and a factor 2.6 to 17.3 higher at 1.5 AU. The unfilled vertical bars show our 
modeled range of peak abundances with respect to water; these values are obtained by dividing the peak production 
rates resulting from coma chemistry (discussed in Section \ref{section:reslt}) by the production rate of water. 
The textured bars show our modeled abundances due to coma chemistry with respect to water at 10 km, 130 km and 320 km 
(obtained by dividing the production rates at these distances by the production rate of water).
}		
\label{fig:tab3}
\end{figure}
\subsection{Comparison with in situ measurement of \textit{Rosetta}}
We only have a limited number of studies from in situ measurements of cometary coma, 
which provide precise information and can measure lower production rates. Therefore, 
comparing our model outcomes with those of the in situ measurements is relevant.   
{\it{Rosetta}} undertook a close-up investigation 
of 67P/C-G for nearly two years and measured a large number of simple through complex organic molecules. 
Our results do not apply directly to the in situ measurements; 
67P/C-G is a low activity comet and its outgassing rate is 1-3 orders of magnitude lower than the comets we have studied. However, we attempt to see how the peak abundances of organic molecules resulting solely from coma chemistry in comets showing relatively high activity (such as the ones investigated in this work) compare with the observed abundances of a low activity comet.
We compare
our modeled abundances with the coma abundance of 67P/C-G measured by {\it{Rosetta}} 
to understand if complex molecules that are produced only via coma chemistry can be detected 
or not via in situ observations.

The scattered points in Figure \ref{fig:tab3} show the relative abundances with respect to water of assorted 
organic molecules detected in 67P/C-G via in situ measurements. The abundances reported by \cite{LeRoy2015} 
are deduced by observations taken by ROSINA/\textit{Rosetta} at a high heliocentric distance $\sim 3$ au 
and at a cometocentric distance of 10 km. These abundances differ considerably in the summer and winter 
hemispheres, and it is difficult to infer which values 
best represent the bulk composition. Since the measurements are made at a large heliocentric distance, 
the effective sublimation of all the volatiles may not have started. The abundances obtained by \cite{Rubin2019b}
and \cite{Schuhmann2019b} are at a heliocentric distance $\sim 1.5$ au and at cometocentric distances between 
$130 - 320$ km, which is closer to perihelion and more representative of the cometary bulk abundance, as 
suggested by \cite{Calmonte2016}. 

In Figure \ref{fig:tab3}, we have also shown the abundance  (\%) with respect to water resulting from coma chemistry for the comets 
investigated in this work. As discussed previously (Section \ref{section:reslt}), the peak or maximum production of organic molecules due to gas phase chemistry is reached at distances ranging between $10^3 - 10^4$ km.
The untextured bars in Figure \ref{fig:tab3} show the range of values of the 
maximum relative abundance (\%) with respect to water resulting from coma chemistry (i.e. modeled peak production rates divided by the production rate of water). 
For the molecules \ch{CH2OHCHO} and \ch{CH3COOH}, only the model values obtained 
for the comets C/1996 B2 and C/2014 Q2 are shown. The textured bars show the relative abundances at three different cometocentric
distances: 10 km, 130 km and 320 km (cometocentric distances for which the abundances of organic molecules with respect to water 
are reported for 67P/C-G). 
Besides, the water production rates of our modeled comets are 2-3 orders
of magnitude higher than that of 67P/C-G at 3 AU \citep{LeRoy2015} and a factor 2.6 to 17.3 higher at 1.5 AU \citep{Hansen2016}.

The bulk abundance of \ch{HC3N} in 67P/C-G at 1.5 au is within the range of our modeled peak abundance percentage. For \ch{NH2CHO}, \cite{LeRoy2015} do not obtain a clear peak in the mass spectra and derive
upper limits for this molecule, and our peak abundance ratios are higher than these limits.
It is also interesting to note that the amount of \ch{NH2CHO} produced in our 
coma models of C/2012 F6, that exhibits higher activity, is nearly equal to or even exceeds the amount of 
\ch{NH2CHO} outgassing from the nucleus of the lower activity comet namely C/2013 R1. The relative abundance of
\ch{HCOOCH3} in summer hemisphere of 67P/C-G at large heliocentric distance is close to our peak abundance 
percentage range for this molecule. For all other molecules, our relative peak abundances are lower by an order of magnitude or more than those estimated for 67P/C-G.
For these molecules, we instead attempt to compare production rates to see how the peak production rates due to coma chemistry in a high activity comet compare with the outgassing rates of comet having in-situ observations.

From the long term monitoring of the water production rate of 67P/C-G during the mission duration, 
it is seen that the production rate increases from $\sim 2\times 10^{25}$ molecules s$^{-1}$ at 4 au 
and reaches a peak value of $3.5\times 10^{28}$ molecules s$^{-1}$ a few days after perihelion at 1.24 au 
\citep{Hansen2016}.  The water production rate of 67P/C-G changed significantly
therefore, the production rates of organic molecules in 67P/C-G 
calculated from the abundances given by \cite{LeRoy2015} is expected to be different from that of 
\cite{Rubin2019b}. We use the relative abundances of organic molecules given by \cite{LeRoy2015} (shown in Figure \ref{fig:tab3}) and estimate their  production rates by multiplying them with an approximate value of $10^{26}$ molecules s$^{-1}$ for the water production 
rate of 67P/C-G at 3 AU (from the empirical relation given by \cite{Hansen2016}). We obtain the lowest values of the production rates 
(considering both summer and winter hemisphere) as $8\times 10^{21}$ molecules s$^{-1}$ for \ch{HCOOH}, 
$8\times 10^{20}$ molecules s$^{-1}$ for \ch{(CH2OH)2}, $10^{22}$ molecules s$^{-1}$ for \ch{CH3CHO}, 
and $6\times 10^{21}$ molecules s$^{-1}$ for \ch{CH3CN}. Similarly, using an approximate value of $10^{28}$ 
molecules s$^{-1}$ for the water production rate of 67P/C-G at the time of measurements reported 
by \cite{Schuhmann2019b}, we obtain a production rate of $3.4\times 10^{23}$ molecules s$^{-1}$ for \ch{CH3COOH}. 
These values are within our range of peak production rates that we obtain due to coma chemistry. It is pertinent 
to mention that some of these species can be isomers. \cite{LeRoy2015} do not mention any isomers of 
\ch{HCOOCH3}, whereas \cite{Schuhmann2019b} identify \ch{CH3COOH} at mass 60 u/e and do not mention its 
isomers. \cite{Schuhmann2019b} also identify \ch{C2H5OH} as the compound at mass 46 u/e. It is possible 
to identify the structural isomers in the ROSINA-DFMS mass spectra by making a comparison 
with the fragmentation pattern as obtained from the NIST database \citep{LeRoy2015, Altwegg2017,Schuhmann2019b}. 
It can happen that there might be contamination from the other isomeric forms at 46 u/e and 60 u/e, 
but following \cite{Schuhmann2019b}, we consider the \ch{C2H4O2} abundance to be the abundance 
for \ch{CH3COOH} when making the comparison with our modeled values.
	
\cite{Cordiner2021} proposed that the abundance of \ch{HC3N} and \ch{NH2CHO} in cometary nuclei are significantly less than 
that implied by ground-based observations. Even without considering distributed sources that \cite{Cordiner2021} have used 
in their models, we see that the relative abundances of \ch{HC3N} and \ch{HCOOCH3} are within a factor of 2 of these abundances 
obtained for 67P/C-G, while the relative abundance of \ch{NH2CHO} is 1-2 orders of magnitude higher than the upper limit suggested 
by \cite{LeRoy2015}. For the other molecules, we obtain low relative abundances, though the peak production rates match the 
production rates of 67P/C-G at high heliocentric distance. 
Thus the abundance of some organic molecules created in the coma of comets is in the realm of detection via in situ/space-based observations, probably 
also for future ground-based observations. Therefore it is expected to detect molecules with increasing chemical complexity; however, it is to be kept in mind that 
large uncertainty exists in the rate constants thus more experiments and 
quantum mechanical calculations are needed to constrain the relevant reaction rates better.

\subsection{Way towards greater complexity}
Some of the molecular species that have been detected in comets due to outgassing from the nucleus or formation by gas-phase coma 
chemistry are of biological importance. \ch{HCN}, 
one of the main reservoirs of cometary volatile nitrogen, is a key precursor in the synthesis of amino acids and is likely 
to play an integral role in the creation of biomolecules \citep{Oro1991}. Formaldehyde (\ch{H2CO}) is a chemical precursor to sugars, 
and \ch{CH3OH} is the simplest alcohol that is the starting point from which more complex organics form in the ISM 
\citep{Garrod2008,Herbst2009,Oberg2009}. Glycolaldehyde is a two-carbon sugar precursor, and is an important biomarker molecule since 
it is postulated to be a building block for ribose, the backbone of RNA \citep{Jalbout2007}. Ethylene glycol is a chemically reduced 
variant of glycolaldehyde and is found in many of the same locations as glycolaldehyde \citep{Hollis2002,Li2017,Pagani2017}. Methyl formate 
is the structural isomer of glycolaldehyde and acetic acid, and is an important complex organic molecules that leads to the 
synthesis of bio-polymers. Formamide is a particularly promising organic molecule since it has an amide functional group which is essential 
in forming amino acid chains. It has been identified as a key precursor of prebiotic molecules, carboxylic acids and sugars 
\citep{Saladino2012, Saitta2014, Botta2018}. Glycine is the simplest amino acid and is a key building block for proteins and peptides 
\citep{Wincel2000}. Phosphorus complements the detection of glycine; phosphorus is an essential element in living organisms since it is 
required for the formation of adenosine 5$'$-triphosphate which is a driver of many biological processes. 
	
The elemental budget of organic molecules consists mostly of carbon, hydrogen, oxygen and nitrogen, with small quantities of heavy 
elements like phosphorus and sulphur. These elements are found in abundance in astrochemical regions in general, and cometary environments 
in particular. This raises the question of how organic chemistry works in these environments and what are the molecular abundances we can 
expect to find. Answering these questions will
prove vital in understanding the evolutionary history of the Solar System, and
the exogenous delivery of organics to terrestrial planets.
	
While most previous studies have proposed the formation of organics on cold grain surfaces, 
there are an increasing number of studies that demonstrate the importance of gas-phase 
chemical reactions towards the formation of organic molecules (for example, \citealp{Vasyunin2013, 
Balucani2015, Vasyunin2017, Skouteris2018}). New experimental results show that many reactions 
that were previously considered to be unimportant because of the presence of energy barriers, 
are actually quite promising. 
Accurate rate coefficients for a large number of chemical reactions are required to explain the
formation of organics in cometary environments. This provides an incentive to undertake 
theoretical and experimental studies to expand the gas-phase chemical networks.
Laboratory studies and rate measurements of additional organic forming reactions will aid in 
further understanding the chemistry of organic species, including molecules of increasing
complexity. Modeling studies of cometary organics are also useful in preparing a template for 
future in situ measurements.

\section{Conclusion} \label{section:concl}

We have studied the formation and evolution of assorted C-H-O and N-bearing organic molecules observed in the cometary 
coma of four comets using a multifluid chemical-hydrodynamical model having an updated chemical network and employing a large number of model runs.
Subsequently, we examined the extent COMs formation purely due to the gas-phase chemistry without them being parent volatiles.
	
The highlights of the results are as follows: 
\begin{enumerate}
\item
We found peak production rate for \ch{HC3N} to vary between 3.5 $\times$ 10$^{22}$ and 5.2 $\times$ 10$^{23}$ molecules s$^{-1}$, while for
\ch{NH2CHO} it varies between 3.9 $\times$ 10$^{23}$ and 7.8 $\times$ 10$^{25}$  molecules s$^{-1}$ due to gas-phase chemistry in the cometary
coma, which the in situ measurements can detect.
These two species were previously studied by \cite{Cordiner2021} for a fixed set of initial conditions; we found the maximum abundance
depends on the availability of reactant species such as \ch{CN} and \ch{C2H2} for \ch{HC3N} and \ch{H2CO} and \ch{NH2} for \ch{NH2CHO}.
Besides, they are also dependent on parameters such as photolytic rates because photochemistry creates radicals such as \ch{CN} and \ch{NH2}.
	
\item
Peak production rate of \ch{HCOOH}, \ch{CH3CHO},  \ch{HCOOCH3}, \ch{CH3COOH}, \ch{CH3OCH3}, \ch{HC5N}, and \ch{CH3CN} can have a value
$>$ 10$^{22}$ molecules s$^{-1}$ due to gas-phase coma chemistry, which assumes significance considering the fact the {\it Rosetta} detected production
rates nearly two orders of magnitude less than these values. Whereas \ch{(CH2OH)2} and \ch{CH2OHCHO} have production rate below
10$^{22}$ but above 10$^{21}$  molecules s$^{-1}$ due to gas-phase chemistry. However, \ch{C2H5OH} and simple amino acid glycine production are
inefficient ($<$ 10$^{20}$  molecules s$^{-1}$). Also, the abundance of these molecules can vary from one comet to the other depending on the
availability of reactant species which produce them.
		
\item Formation of radicals like \ch{CH2CH2OH} and \ch{CH3CHOH} is conditional, they are only produced when \ch{C2H5OH} 
is present as parent species, which undergoes an H-abstraction reaction to form them.
		
\end{enumerate}
Finally, abundances of organic molecules when they are present as parent volatiles is moderate to significantly large in comparison 
with their formation solely due to gas-phase chemistry. We find that the production rates of some of the coma-synthesised organic
molecules can reach peak values of $\sim$ 10$^{22}$ - 10$^{26}$ molecules s$^{-1}$; while, production rate as low as 10$^{21}$ 
molecules s$^{-1}$ has been detected in 67P/C-G by {\it Rosetta}. Thus the abundance of some organic molecules created in the coma 
of comets is in the realm of detection via in situ/space-based observations, probably also for future ground-based observations. 
It is expected that in the future, molecules with increasing chemical complexity will be detected - like the detections 
by {\it Rosetta}. Therefore, more experiments and quantum mechanical calculations will be needed to understand their formation and 
destruction.

	\section*{Acknowledgements}
 We like to thank anonymous referees for the constructive comments that improved this paper. 
 The computations were performed on the Param Vikram-1000 High Performance Computing Cluster of the Physical Research Laboratory (PRL), India.
The work done at PRL is supported by the Department of Space, Government of India. 
The support of the SERB (Government of India) Core Research Grant of Kinsuk Acharyya (File Number: CRG/2022/000975) is kindly acknowledged.  
The authors thank Vikas Soni for useful insights and discussions and Michael Weiler for the elctronic version of the chemical network.

	\section*{Data Availability}
	Data related to the reactions and physical parameters underlying this article are available in the article, in its appendix material (Appendix~\ref{App_A}), 
	as well as cited in online kinetic data repositories, especially the Kinetic Database for Astrochemistry (KIDA) via its URL https://astrophy.u-bordeaux.fr.
	In addition, data for the abundances of various COMs studied can be obtained from any of the authors via email.
	 
\bibliographystyle{cas-model2-names}



	\appendix 
	
\section{Chemical Reactions} \label{App_A}
The major gas-phase chemical reactions for organic molecules, that are used in our chemical network, are listed 
in Table \ref{table:reactions}.  A new reaction scheme for the formation of organics by the chemical 
activation of ethanol was added to our gas-phase 
network \citep{Skouteris2018}. The H-abstraction of ethanol can occur at three different sites, leading to the formation 
of the radicals \ch{CH2CH2OH}, \ch{CH3CHOH} and \ch{CH3CH2O}, though \ch{CH3CH2O} has not been experimentally detected 
\citep{Caravan2015}. The branching ratio for the formation of \ch{CH2CH2OH} and \ch{CH3CHOH} due to the H-abstraction 
reaction with \ch{OH} shows some variation with temperature, and we have chosen a value of 0.1:0.9. Subsequently, an 
\ch{O} atom can get added to these radicals, resulting in the formation of an intermediate, which can finally result 
in the cleavage of the \ch{C-C} bond, or the elimination of an \ch{H} atom or \ch{OH} radical. Glycolaldehyde forms 
as a result of the \ch{H}-atom elimination of the intermediate formed from \ch{CH2CH2OH}, while acetic acid forms from 
the similar process undergone by the intermediate resulting from \ch{CH3CHOH}. The \ch{C-C} bond cleavage of the 
\ch{CH3CHOH} and \ch{O} intermediate results in the creation of formic acid, while \ch{OH}-radical elimination forms
acetaldehyde. 
A similar H-abstraction reaction undergone by methanol 
results in the creation of \ch{CH2OH} radicals, which form ethylene glycol in the coma.
	
\begin{scriptsize}

\begin{longtable}{l c c c l l}
	\caption{ Major chemical reactions for organic molecules that are added in the network are listed here. Eq. 1 indicates 
that the reaction rates need to be calculated using Equation \ref{eq:arrhenius}. Eq. 2 indicates that Equations \ref{eq:ionpol1} 
or \ref{eq:ionpol2} need to be used, depending on the temperature.}	
		\label{table:reactions} \\
		
		\hline \multicolumn{1}{l}{\textbf{Reaction}} & \multicolumn{1}{c}{\textbf{$\alpha$}} & \multicolumn{1}{c}{\textbf{$\beta$}} & \multicolumn{1}{c}{\textbf{$\gamma$}} & \multicolumn{1}{l}{\textbf{Reference}} & \multicolumn{1}{l}{\textbf{Eq.}} \\ \hline 
		\endfirsthead
		
		\multicolumn{6}{c}%
		{{\bfseries \tablename\ \thetable{} -- continued from previous page}} \\
		
		\hline \multicolumn{1}{l}{\textbf{Reaction}} & \multicolumn{1}{c}{\textbf{$\alpha$}} & \multicolumn{1}{c}{\textbf{$\beta$}} & \multicolumn{1}{c}{\textbf{$\gamma$}} & \multicolumn{1}{l}{\textbf{Reference}} & \multicolumn{1}{l}{\textbf{Eq.}} \\ \hline 
		\endhead
		
		\hline 
		\endfoot
		
		\hline \hline
		\endlastfoot
		$\ch{OH} + \ch{C2H5OH} \rightarrow \ch{H2O} + \ch{CH2CH2OH}$ & 2.70(-12) & 0.00 & 0.00 &\cite{Skouteris2018} & 1\\

		$\ch{O} + \ch{C2H5OH} \rightarrow \ch{OH} + \ch{CH2CH2OH}$ & 1.58(-13) & 3.23(0) & 2.34(3) &\cite{Wu2007} & 1\\
		
		$\ch{H} + \ch{C2H5OH} \rightarrow \ch{H2} + \ch{CH2CH2OH}$ & 7.91(-13) & 2.81(0) & 3.77(3) &\cite{Sivaramakrishnan2010} & 1\\
		
		$\ch{CH3} + \ch{C2H5OH} \rightarrow \ch{CH4} + \ch{CH2CH2OH}$ & 1.06(-13) & 3.45(0) & 5.54(3) &\cite{Olm2016} & 1\\
		
		$\ch{OH} + \ch{C2H5OH} \rightarrow \ch{H2O} + \ch{CH3CHOH}$ & 2.40(-11) & 0.00 & 0.00 &\cite{Skouteris2018} & 1\\
		
		$\ch{O} + \ch{C2H5OH} \rightarrow \ch{OH} + \ch{CH3CHOH}$ & 3.11(-13) & 2.47(0) & 4.41(2) &\cite{Wu2007} & 1\\
		
		$\ch{H} + \ch{C2H5OH} \rightarrow \ch{H2} + \ch{CH3CHOH}$ & 6.24(-13) & 2.68(0) & 1.46(3) &\cite{Sivaramakrishnan2010} & 1\\
		
		$\ch{CH3} + \ch{C2H5OH} \rightarrow \ch{CH4} + \ch{CH3CHOH}$ & 8.87(-15) & 3.37(0) & 3.95(3) &\cite{Olm2016} & 1\\
		
		$\ch{H2O} + \ch{HCO+} \rightarrow \ch{HCOOH2+} + h\nu $ & 4.00(-13) &-1.30(0) & 0.00 &\cite{Herbst1985} & 1\\
		
		$\ch{CH4} + \ch{O2+} \rightarrow \ch{H} + \ch{HCOOH2+}$ & 3.80(-12) &-1.80(0) & 0.00 &\cite{Wakelam2015} & 1\\
		
		$\ch{HCOOH} + \ch{HCO+} \rightarrow \ch{CO} + \ch{HCOOH2+}$ & 1.00(0) & 1.01(-9) & 2.77(0) &\cite{Woon2009} & 2\\
		
		$\ch{HCOOH} + \ch{CH3OH2+} \rightarrow \ch{CH3OH} + \ch{HCOOH2+}$ & 3.70(-10) &-5.00(-1) & 0.00 &\cite{Freeman1978} & 1\\
		
		$\ch{HCOOH} + \ch{H2COH+} \rightarrow \ch{H2CO} + \ch{HCOOH2+}$ & 2.00(-9) &-5.00(-1) & 0.00 &\cite{Freeman1978} & 1\\
		
		$\ch{CH3CHO} + \ch{HCOOH2+} \rightarrow \ch{HCOOH} + \ch{CH3CHOH+}$ & 2.90(-9) &-5.00(-1) & 0.00 &\cite{Feng1994} & 1\\
		
		$\ch{CH3CN} + \ch{HCOOH2+} \rightarrow \ch{HCOOH} + \ch{CH3CNH+}$ & 4.07(-9) &-5.00(-1) & 0.00 &\cite{Feng1994} & 1\\
		
		$\ch{CH3OH} + \ch{HCOOH2+} \rightarrow \ch{HCOOH} + \ch{CH3OH2+}$ & 2.29(-9) &-5.00(-1) & 0.00 &\cite{Feng1994} & 1\\
		
		$\ch{NH3} + \ch{HCOOH2+} \rightarrow \ch{HCOOH} + \ch{NH4+}$ & 1.38(-9) &-5.00(-1) & 0.00 &\cite{Feng1994} & 1\\
		
		$\ch{H2O} + \ch{HCOOH2+} \rightarrow \ch{HCOOH} + \ch{H3O+}$ & 2.10(-11) &-5.00(-1) & 0.00 &\cite{VanDoren1986} & 1\\
		
		$\ch{HCOOH2+} + \ch{e-} \rightarrow \ch{H} + \ch{HCOOH}$ & 1.50(-7) &-5.00(-1) & 0.00 &\cite{Wakelam2015} & 1\\
		
		$\ch{O} + \ch{CH3CHOH} \rightarrow \ch{CH3} + \ch{HCOOH}$ & 3.90(-10) & 1.80(-1) & 4.90(-1) &\cite{Skouteris2018} & 1\\
		
		$\ch{H2CO} + \ch{CH3OH2+} \rightarrow \ch{H2} + \ch{HCOOCH4+}$ & 1.00(0) & 9.53(-10) & 5.15(0) &\cite{Woon2009} & 2\\
		
		$\ch{HCOOH} + \ch{CH3OH2+} \rightarrow \ch{H2O} + \ch{HCOOCH4+}$ & 1.70(-11) & 0.00 & 0.00 &\cite{Cole2012} & 1\\
		
		$\ch{HCOOCH3} + \ch{H3+} \rightarrow \ch{H2} + \ch{HCOOCH4+}$ & 4.05(-9) &-5.00(-1) & 0.00 &\cite{Wakelam2015} & 1\\
		
		$\ch{HCOOCH3} + \ch{HCO+} \rightarrow \ch{CO} + \ch{HCOOCH4+}$ & 1.55(-9) &-5.00(-1) & 0.00 &\cite{Wakelam2015} & 1\\
		
		$\ch{HCOOCH3} + \ch{H3O+} \rightarrow \ch{H2O} + \ch{HCOOCH4+}$ & 1.81(-9) &-5.00(-1) & 0.00 &\cite{Wakelam2015} & 1\\
		
		$\ch{HCOOCH4+} + \ch{e-} \rightarrow \ch{H} + \ch{HCOOCH3}$ & 1.50(-7) &-5.00(-1) & 0.00 &\cite{Wakelam2015} & 1\\
		
		$\ch{O} + \ch{CH3OCH2} \rightarrow \ch{H} + \ch{HCOOCH3}$ & 2.56(-10) & 1.50(-1) & 0.00 &\cite{Ruaud2015} & 1\\
		
		$\ch{CH4} + \ch{H2CO+} \rightarrow \ch{H} + \ch{CH3CHOH+}$ & 1.65(-11) & 0.00 & 0.00 &\cite{Wakelam2015} & 1\\
		
		$\ch{C2H5OH} + \ch{C+} \rightarrow \ch{CH} + \ch{CH3CHOH+}$ & 7.07(-10) &-5.00(-1) & 0.00 &\cite{Wakelam2015} & 1\\
		
		$\ch{C2H5OH} + \ch{H+} \rightarrow \ch{H2} + \ch{CH3CHOH+}$ & 3.30(-9) &-5.00(-1) & 0.00 &\cite{Wakelam2015} & 1\\
		
		$\ch{C2H5OH} + \ch{H3+} \rightarrow \ch{H2} + \ch{H2} + \ch{CH3CHOH+}$ & 5.00(-10) &-5.00(-1) & 0.00 &\cite{SungLee1992} & 1\\
		
		$\ch{CH3CHOH} + \ch{H+} \rightarrow \ch{H} + \ch{CH3CHOH+}$ & 3.00(-9) &-5.00(-1) & 0.00 &\cite{Skouteris2018} & 1\\
		
		$\ch{CH2CH2OH} + \ch{H+} \rightarrow \ch{H} + \ch{CH3CHOH+}$ & 3.00(-9) &-5.00(-1) & 0.00 &\cite{Skouteris2018} & 1\\	
		
		$\ch{H2CO} + \ch{CH3OH2+} \rightarrow \ch{H2O} + \ch{CH3CHOH+}$ & 2.10(-11) &-5.00(-1) & 0.00 &\cite{Karpas1989} & 1\\
		
		$\ch{CH3OCH3} + \ch{H+} \rightarrow \ch{H2} + \ch{CH3CHOH+}$ & 2.50(-9) &-5.00(-1) & 0.00 &\cite{Wakelam2015} & 1\\
		
		$\ch{CH3OCH3} + \ch{CH3+} \rightarrow \ch{CH4} + \ch{CH3CHOH+}$ & 3.50(-10) &-5.00(-1) & 0.00 &\cite{Wilson1994} & 1\\
		
		$\ch{CH3OCH3} + \ch{O2+} \rightarrow \ch{O2H} + \ch{CH3CHOH+}$ & 1.35(-9) &-5.00(-1) & 0.00 &\cite{McElroy2013} & 1\\
		
		$\ch{C3H6} + \ch{O2+} \rightarrow \ch{H} + \ch{CO} + \ch{CH3CHOH+}$ & 3.00(-11) & 0.00 & 0.00 &\cite{McElroy2013} & 1\\
		
		$\ch{CH3CHO} + \ch{H3+} \rightarrow \ch{H2} + \ch{CH3CHOH+}$ & 6.20(-9) &-5.00(-1) & 0.00 &\cite{Wakelam2015} & 1\\
		
		$\ch{CH3CHO} + \ch{HCO+} \rightarrow \ch{CO} + \ch{CH3CHOH+}$ & 2.50(-9) &-5.00(-1) & 0.00 &\cite{Wakelam2015} & 1\\
		
		$\ch{CH3CHO} + \ch{H3O+} \rightarrow \ch{H2O} + \ch{CH3CHOH+}$ & 2.86(-9) &-5.00(-1) & 0.00 &\cite{Wakelam2015} & 1\\
		
		$\ch{NH3} + \ch{CH3CHOH+} \rightarrow \ch{CH3CHO} + \ch{NH4+}$ & 1.80(-9) &-5.00(-1) & 0.00 &\cite{Wilson1994} & 1\\
		
		$\ch{CH3CHOH+} + \ch{e-} \rightarrow \ch{H} + \ch{CH3CHO}$ & 1.50(-7) &-5.00(-1) & 0.00 &\cite{Wakelam2015} & 1\\
		
		$\ch{C2H5OH2+} + \ch{e-} \rightarrow \ch{H} + \ch{H2} + \ch{CH3CHO}$ & 1.50(-7) &-5.00(-1) & 0.00 &\cite{Wakelam2015} & 1\\
		
		$\ch{C2H5OH+} + \ch{e-} \rightarrow \ch{H2} + \ch{CH3CHO}$ & 1.50(-7) &-5.00(-1) & 0.00 &\cite{Wakelam2015} & 1\\
		
		$\ch{CH} + \ch{CH3OH} \rightarrow \ch{CH3CHO} + \ch{H}$ & 2.49(-10) &-1.93(0) & 0.00 &\cite{McElroy2013} & 1\\
		
		$\ch{O} + \ch{CH3CHOH} \rightarrow \ch{OH} + \ch{CH3CHO}$ & 4.80(-11) & 1.90(-1) & 3.90(-1) &\cite{Skouteris2018} & 1\\
		
		$\ch{O} + \ch{C2H5} \rightarrow \ch{H} + \ch{CH3CHO}$ & 1.33(-10) & 0.00 & 0.00 &\cite{Wakelam2015} & 1\\
		
		$\ch{HCO} + \ch{CH3CO} \rightarrow \ch{CO} + \ch{CH3CHO}$ & 1.50(-11) & 0.00 & 0.00 &\cite{Hebrard2009} & 1\\
		
		$\ch{H2CO} + \ch{CH3CO} \rightarrow \ch{HCO} + \ch{CH3CHO}$ & 3.01(-13) & 0.00 & 6.51(3) &\cite{Hebrard2009} & 1\\
		
		$\ch{CH3OH} + \ch{CH3CO} \rightarrow \ch{CH3CHO} + \ch{CH2OH}$ & 2.17(-13) & 3.00(0) & 6.21(3) &\cite{Hebrard2009} & 1\\
		
		$\ch{H2} + \ch{CH3CO} \rightarrow \ch{H} + \ch{CH3CHO}$ & 2.21(-13) & 1.82(0) & 8.87(3) &\cite{Hebrard2009} & 1\\
		
		$\ch{CH4} + \ch{CH3CO} \rightarrow \ch{CH3} + \ch{CH3CHO}$ & 4.92(-14) & 2.88(0) & 1.08(4) &\cite{Hebrard2009} & 1\\
		
		$\ch{C2H6} + \ch{CH3CO} \rightarrow \ch{CH3CHO} + \ch{C2H5}$ & 1.95(-13) & 2.75(0) & 8.83(3) &\cite{Hebrard2009} & 1\\
		
		$\ch{CH3CO} + \ch{CH3CO} \rightarrow \ch{CH2CO} + \ch{CH3CHO}$ & 1.49(-11) & 0.00 & 0.00 &\cite{Hebrard2009} & 1\\
		
		$\ch{O2} + \ch{CH3OCH2} \rightarrow \ch{O2H} + \ch{CH3CHO}$ & 6.30(-14) & 0.00 & 5.50(2) &\cite{Sander2011} & 1\\
		
		$\ch{OH} + \ch{C3H6} \rightarrow \ch{CH3CHO} + \ch{CH3}$ & 2.08(-11) &-2.03(0) & 1.70(2) &\cite{McElroy2013} & 1\\
		
		$\ch{C2H4} + \ch{H3O+} \rightarrow \ch{C2H5OH2+} + h\nu $ & 2.40(-14) &-2.80(0) & 0.00 &\cite{Herbst1987} & 1\\
		
		$\ch{H2O} + \ch{C2H5+} \rightarrow \ch{C2H5OH2+} + h\nu $ & 4.10(-16) &-2.40(0) & 0.00 &\cite{Herbst1987} & 1\\
		
		$\ch{C2H5OH} + \ch{H3+} \rightarrow \ch{H2} + \ch{C2H5OH2+}$ & 1.30(-9) &-5.00(-1) & 0.00 &\cite{Wakelam2015} & 1\\
		
		$\ch{C2H5OH} + \ch{HCO+} \rightarrow \ch{CO} + \ch{C2H5OH2+}$ & 8.50(-10) &-5.00(-1) & 0.00 &\cite{Wakelam2015} & 1\\
		
		$\ch{C2H5OH} + \ch{H3O+} \rightarrow \ch{H2O} + \ch{C2H5OH2+}$ & 1.79(-9) &-5.00(-1) & 0.00 &\cite{Wakelam2015} & 1\\
		
		$\ch{NH3} + \ch{C2H5OH2+} \rightarrow \ch{C2H5OH} + \ch{NH4+}$ & 1.97(-9) &-5.00(-1) & 0.00 &\cite{Feng1995} & 1\\ 
		
		$\ch{C2H5OH2+} + \ch{e-} \rightarrow \ch{H} + \ch{C2H5OH}$ & 1.50(-7) &-5.00(-1) & 0.00 &\cite{Wakelam2015} & 1\\
		
		$\ch{(CH2OH)2} + \ch{H3+} \rightarrow \ch{H2} + \ch{(CH2OH)2H+}$ & 1.48(-9) & 0.00 & 0.00 &\cite{Garrod2007} & 1\\
		
		$\ch{(CH2OH)2} + \ch{H3O+} \rightarrow \ch{H2O} + \ch{(CH2OH)2H+}$ & 6.13(-10) & 0.00 & 0.00 &\cite{Garrod2007} & 1\\
		
		$\ch{(CH2OH)2} + \ch{HCO+} \rightarrow \ch{CO} + \ch{(CH2OH)2H+}$ & 5.07(-10) & 0.00 & 0.00 &\cite{Garrod2007} & 1\\
		
		$\ch{(CH2OH)2H+} + \ch{e-} \rightarrow \ch{(CH2OH)2} + \ch{H}$ & 1.50(-8) &-5.00(-1) & 0.00 &\cite{Garrod2007} & 1\\
		
		$\ch{CH2OH} + \ch{CH2OH} \rightarrow \ch{(CH2OH)2} $ & 1.60(-11) & 0.00 & 0.00 &\cite{Tsang1987} & 1\\
		
		$\ch{CH2OHCHO} + \ch{H3+} \rightarrow \ch{H2} + \ch{CH2OHCHOH+}$ & 5.22(-9) &-5.00(-1) & 0.00 &\cite{Garrod2007} & 1\\
		
		$\ch{CH2OHCHO} + \ch{H3O+} \rightarrow \ch{H2O} + \ch{CH2OHCHOH+}$ & 2.19(-9) &-5.00(-1) & 0.00 &\cite{Garrod2007} & 1\\
		
		$\ch{CH2OHCHO} + \ch{HCO+} \rightarrow \ch{CO} + \ch{CH2OHCHOH+}$ & 1.82(-9) &-5.00(-1) & 0.00 &\cite{Garrod2007} & 1\\
		
		$\ch{CH2OHCHOH+} + \ch{e-} \rightarrow \ch{CH2OHCHO} + \ch{H}$ & 1.50(-8) &-5.00(-1) & 0.00 &\cite{Garrod2007} & 1\\
		
		$\ch{O} + \ch{CH2CH2OH} \rightarrow \ch{H} + \ch{CH2OHCHO}$ & 1.10(-10) & 1.60(-1) & 5.50(-1) &\cite{Skouteris2018} & 1\\
		
		$\ch{HNC} + \ch{CH3+} \rightarrow \ch{CH3CNH+} + h\nu $ & 9.00(-9) &-5.00(-1) & 0.00 &\cite{Loison2014} & 1\\
		
		$\ch{HCN} + \ch{CH3+} \rightarrow \ch{CH3CNH+} + h\nu $ & 2.00(-10) &-3.00(0) & 0.00 &\cite{Herbst1985} & 1\\
		
		$\ch{HCN} + \ch{CH4+} \rightarrow \ch{H} + \ch{CH3CNH+}$ & 3.30(-9) & 0.00 & 0.00 &\cite{Plessis2010} & 1\\
		
		$\ch{CH3CN} + \ch{H3+} \rightarrow \ch{H2} + \ch{CH3CNH+}$ & 1.00(0) & 2.91(-9) & 6.58(0) &\cite{Woon2009} & 2\\
		
		$\ch{CH3CN} + \ch{HCO+} \rightarrow \ch{CO} + \ch{CH3CNH+}$ & 1.00(0) & 1.18(-9) & 6.58(0) &\cite{Woon2009} & 2\\
		
		$\ch{CH3CN} + \ch{H3O+} \rightarrow \ch{H2O} + \ch{CH3CNH+}$ & 1.00(0) & 1.35(-9) & 6.58(0) &\cite{Woon2009} & 2\\
		
		$\ch{CH3CN} + \ch{HCO2+} \rightarrow \ch{CO2} + \ch{CH3CNH+}$ & 1.00(0) & 1.05(-9) & 6.58(0) &\cite{Woon2009} & 2\\
		
		$\ch{CH3CN} + \ch{C2H2+} \rightarrow \ch{C2H} + \ch{CH3CNH+}$ & 8.36(-10) & 0.00 & 0.00 &\cite{Anicich2003} & 1\\
		
		$\ch{CH3CN} + \ch{C2H5+} \rightarrow \ch{C2H4} + \ch{CH3CNH+}$ & 3.80(-9) & 0.00 & 0.00 &\cite{Anicich2003} & 1\\
		
		$\ch{CH3CN} + \ch{HCNH+} \rightarrow \ch{HCN} + \ch{CH3CNH+}$ & 3.80(-9) & 0.00 & 0.00 &\cite{Anicich2003} & 1\\
		
		$\ch{CH3CN} + \ch{C3H+} \rightarrow \ch{C3} + \ch{CH3CNH+}$ & 4.50(-10) & 0.00 & 0.00 &\cite{Anicich2003} & 1\\
		
		$\ch{CH3CN} + \ch{C4H7+} \rightarrow \ch{C4H6} + \ch{CH3CNH+}$ & 5.00(-10) & 0.00 & 0.00 &\cite{McElroy2013} & 1\\
		
		$\ch{CH3CN} + \ch{HC3NH+} \rightarrow \ch{HC3N} + \ch{CH3CNH+}$ & 3.60(-9) &-5.00(-1) & 0.00 &\cite{Raksit1984} & 1\\
		
		$\ch{CH3CN} + \ch{N2H+} \rightarrow \ch{N2} + \ch{CH3CNH+}$ & 4.10(-9) &-5.00(-1) & 0.00 &\cite{Mackay1976} & 1\\
		
		$\ch{CNC+} + \ch{C2H6} \rightarrow \ch{CH3CN} + \ch{C2H3+}$ & 1.20(-10) & 0.00 & 0.00 &\cite{Anicich2003} & 1\\
		
		$\ch{CH3CNH+} + \ch{e-} \rightarrow \ch{H} + \ch{CH3CN}$ & 1.30(-7) &-5.00(-1) & 0.00 &\cite{Loison2014} & 1\\
		
		$\ch{CN} + \ch{CH4} \rightarrow \ch{H} + \ch{CH3CN}$ & 7.21(-15) & 2.64(0) & 7.80(1) &\cite{Hebrard2009} & 1\\
		
		$\ch{C2H4} + \ch{N(^2D)} \rightarrow \ch{H} + \ch{CH3CN}$ & 2.52(-10) & 0.00 & 0.00 &\cite{Hebrard2009} & 1\\
		
		$\ch{CN} + \ch{C3H6} \rightarrow \ch{C2H3} + \ch{CH3CN}$ & 1.73(-10) & 0.00 &-1.02(2) &\cite{Hebrard2009} & 1\\
		
		$\ch{NH2CHO} + \ch{H3+} \rightarrow \ch{H2} + \ch{NH2CHOH+}$ & 1.00(0) & 2.84(-9) & 6.62(0) &\cite{Woon2009} & 2\\
		
		$\ch{NH2CHO} + \ch{HCO+} \rightarrow \ch{CO} + \ch{NH2CHOH+}$ & 1.00(0) & 1.14(-9) & 6.62(0) &\cite{Woon2009} & 2\\
		
		$\ch{NH2CHO} + \ch{H3O+} \rightarrow \ch{H2O} + \ch{NH2CHOH+}$ & 1.00(0) & 1.30(-9) & 6.62(0) &\cite{Woon2009} & 2\\
		
		$\ch{NH2CHO} + \ch{N2H+} \rightarrow \ch{N2} + \ch{NH2CHOH+}$ & 1.00(0) & 1.14(-9) & 6.62(0) &\cite{Woon2009} & 2\\
		
		$\ch{NH2CHOH+} + \ch{e-} \rightarrow \ch{H} + \ch{NH2CHO}$ & 1.50(-7) &-5.00(-1) & 0.00 &\cite{Wakelam2015} & 1\\
		
		$\ch{NH2} + \ch{H2CO} \rightarrow \ch{H} + \ch{NH2CHO}$ & 1.00(-10) & 0.00 & 0.00 &\cite{Wakelam2015} & 1\\
		
		$\ch{CH4} + \ch{C2N+} \rightarrow \ch{H2} + \ch{HC3NH+}$ & 2.10(-10) & 0.00 & 0.00 &\cite{Wakelam2015} & 1\\
		
		$\ch{C2H2} + \ch{HCN+} \rightarrow \ch{H} + \ch{HC3NH+}$ & 1.35(-10) & 0.00 & 0.00 &\cite{Anicich2003} & 1\\
		
		$\ch{C2H4} + \ch{CN+} \rightarrow \ch{H2} + \ch{HC3NH+}$ & 6.50(-11) & 0.00 & 0.00 &\cite{Anicich2003} & 1\\
		
		$\ch{HCN} + \ch{C2H2+} \rightarrow \ch{H} + \ch{HC3NH+}$ & 1.33(-10) & 0.00 & 0.00 &\cite{Wakelam2015} & 1\\
		
		$\ch{HNC} + \ch{C2H2+} \rightarrow \ch{H} + \ch{HC3NH+}$ & 1.30(-10) & 0.00 & 0.00 &\cite{Wakelam2015} & 1\\
		
		$\ch{CH3CN} + \ch{C3H+} \rightarrow \ch{C2H2} + \ch{HC3NH+}$ & 1.05(-9) & 0.00 & 0.00 &\cite{Anicich2003} & 1\\
		
		$\ch{H2} + \ch{HC3N+} \rightarrow \ch{H} + \ch{HC3NH+}$ & 5.00(-12) & 0.00 & 0.00 &\cite{Wakelam2015} & 1\\
		
		$\ch{CH4} + \ch{HC3N+} \rightarrow \ch{CH3} + \ch{HC3NH+}$ & 1.77(-10) & 0.00 & 0.00 &\cite{Anicich2003} & 1\\
		
		$\ch{C2H4} + \ch{HC3N+} \rightarrow \ch{C2H3} + \ch{HC3NH+}$ & 1.34(-10) & 0.00 & 0.00 &\cite{Anicich2003} & 1\\
		
		$\ch{H2O} + \ch{HC3N+} \rightarrow \ch{OH} + \ch{HC3NH+}$ & 6.70(-10) & 0.00 & 0.00 &\cite{Anicich2003} & 1\\
		
		$\ch{H2O} + \ch{C4N+} \rightarrow \ch{CO} + \ch{HC3NH+}$ & 5.00(-1) & 7.45(-10) & 5.41(0) &\cite{Woon2009} & 2\\
		
		$\ch{CH2CHCN} + \ch{H+} \rightarrow \ch{H2} + \ch{HC3NH+}$ & 7.50(-9) &-5.00(-1) & 0.00 &\cite{Wakelam2015} & 1\\
		
		$\ch{HC3N} + \ch{C2H3+} \rightarrow \ch{C2H2} + \ch{HC3NH+}$ & 1.00(0) & 1.35(-9) & 5.44(0) &\cite{Woon2009} & 2\\
		
		$\ch{HC3N} + \ch{H3+} \rightarrow \ch{H2} + \ch{HC3NH+}$ & 1.00(0) & 3.37(-9) & 5.44(0) &\cite{Woon2009} & 2\\
		
		$\ch{HC3N} + \ch{CH5+} \rightarrow \ch{CH4} + \ch{HC3NH+}$ & 1.00(0) & 1.58(-9) & 5.44(0) &\cite{Woon2009} & 2\\
		
		$\ch{HC3N} + \ch{C2H4+} \rightarrow \ch{C2H3} + \ch{HC3NH+}$ & 1.00(0) & 1.33(-9) & 5.44(0) &\cite{Woon2009} & 2\\
		
		$\ch{HC3N} + \ch{HCO+} \rightarrow \ch{CO} + \ch{HC3NH+}$ & 1.00(0) & 1.32(-9) & 5.44(0) &\cite{Woon2009} & 2\\
		
		$\ch{HC3N} + \ch{N2H+} \rightarrow \ch{N2} + \ch{HC3NH+}$ & 1.00(0) & 1.32(-9) & 5.44(0) &\cite{Woon2009} & 2\\
		
		$\ch{HC3N} + \ch{H3O+} \rightarrow \ch{H2O} + \ch{HC3NH+}$ & 1.00(0) & 1.52(-9) & 5.44(0) &\cite{Woon2009} & 2\\
		
		$\ch{HC3N} + \ch{CH2+} \rightarrow \ch{CH} + \ch{HC3NH+}$ & 4.10(-9) & 0.00 & 0.00 &\cite{Anicich2003} & 1\\
		
		$\ch{HC3N} + \ch{CH4+} \rightarrow \ch{CH3} + \ch{HC3NH+}$ & 2.50(-9) & 0.00 & 0.00 &\cite{Anicich2003} & 1\\
		
		$\ch{HC3N} + \ch{C2H+} \rightarrow \ch{C2} + \ch{HC3NH+}$ & 1.41(-9) & 0.00 & 0.00 &\cite{Anicich2003} & 1\\
		
		$\ch{HC3N} + \ch{C2H5+} \rightarrow \ch{C2H4} + \ch{HC3NH+}$ & 3.55(-9) & 0.00 & 0.00 &\cite{Anicich2003} & 1\\
		
		$\ch{HC3N} + \ch{C3H5+} \rightarrow \ch{C3H4} + \ch{HC3NH+}$ & 1.00(-10) & 0.00 & 0.00 &\cite{Plessis2010} & 1\\
		
		$\ch{HC3N} + \ch{HCN+} \rightarrow \ch{CN} + \ch{HC3NH+}$ & 2.21(-9) & 0.00 & 0.00 &\cite{Anicich2003} & 1\\
		
		$\ch{HC3N} + \ch{HCNH+} \rightarrow \ch{HCN} + \ch{HC3NH+}$ & 3.40(-9) & 0.00 & 0.00 &\cite{Anicich2003} & 1\\
		
		$\ch{C2H2} + \ch{HC3N+} \rightarrow \ch{HC3N} + \ch{C2H2+}$ & 1.28(-10) & 0.00 & 0.00 &\cite{Anicich2003} & 1\\
		
		$\ch{C2H4} + \ch{HC3N+} \rightarrow \ch{HC3N} + \ch{C2H4+}$ & 5.36(-10) & 0.00 & 0.00 &\cite{Anicich2003} & 1\\
		
		$\ch{C4H2} + \ch{HC3N+} \rightarrow \ch{HC3N} + \ch{C4H2+}$ & 8.90(-10) & 0.00 & 0.00 &\cite{Anicich2003} & 1\\
		
		$\ch{NH3} + \ch{HC3N+} \rightarrow \ch{HC3N} + \ch{NH3+}$ & 1.70(-9) & 0.00 & 0.00 &\cite{Anicich2003} & 1\\
		
		$\ch{H2O} + \ch{C4N+} \rightarrow \ch{HC3N} + \ch{HCO+}$ & 5.00(-1) & 7.45(-10) & 5.41(0) &\cite{Woon2009} & 2\\
		
		$\ch{CH3CN} + \ch{C3H+} \rightarrow \ch{HC3N} + \ch{C2H3+}$ & 6.00(-10) & 0.00 & 0.00 &\cite{Anicich2003} & 1\\
		
		$\ch{NH3} + \ch{HC3NH+} \rightarrow \ch{HC3N} + \ch{NH4+}$ & 2.00(-9) & 0.00 & 0.00 &\cite{Anicich2003} & 1\\
		
		$\ch{HC3NH+} + \ch{e-} \rightarrow \ch{H} + \ch{HC3N}$ & 6.00(-7) &-5.80(-1) & 0.00 &\cite{Loison2017} & 1\\
		
		$\ch{HC5NH+} + \ch{e-} \rightarrow \ch{C2H} + \ch{HC3N}$ & $1.20\times10^{-7}$ & $-0.70$ & $0.00$ &\cite{Wakelam2015} & 1\\
		
		$\ch{C2H} + \ch{HCN} \rightarrow \ch{H} + \ch{HC3N}$ & 5.30(-12) & 0.00 & 7.70(2) &\cite{Hebrard2009} & 1\\
		
		$\ch{C2H} + \ch{HNC} \rightarrow \ch{H} + \ch{HC3N}$ & 2.00(-10) & 0.00 & 0.00 &\cite{Loison2014} & 1\\
		
		$\ch{CN} + \ch{C2H2} \rightarrow \ch{H} + \ch{HC3N}$ & 2.72(-10) &-5.20(-1) & 1.90(1) &\cite{Wakelam2015} & 1\\
		
		$\ch{H2} + \ch{C3N} \rightarrow \ch{H} + \ch{HC3N}$ & 1.20(-11) & 0.00 & 9.98(2) &\cite{Hebrard2009} & 1\\
		
		$\ch{CH2} + \ch{C3N} \rightarrow \ch{CH} + \ch{HC3N}$ & 3.00(-11) & 0.00 & 0.00 &\cite{Hebrard2009} & 1\\
		
		$\ch{C3N} + \ch{CH4} \rightarrow \ch{CH3} + \ch{HC3N}$ & 5.73(-12) & 0.00 & 6.75(2) &\cite{Hebrard2009} & 1\\
		
		$\ch{C2H2} + \ch{C3N} \rightarrow \ch{C2H} + \ch{HC3N}$ & 2.19(-10) & 0.00 & 0.00 &\cite{Harada2010} & 1\\
		
		$\ch{C3N} + \ch{C2H3} \rightarrow \ch{C2H2} + \ch{HC3N}$ & 1.60(-12) & 0.00 & 0.00 &\cite{Hebrard2009} & 1\\
		
		$\ch{C3N} + \ch{C2H5} \rightarrow \ch{HC3N} + \ch{C2H4}$ & 3.00(-12) & 0.00 & 0.00 &\cite{Hebrard2009} & 1\\
		
		$\ch{C3N} + \ch{C2H6} \rightarrow \ch{HC3N} + \ch{C2H5}$ & 2.08(-11) & 2.20(-1) &-5.80(1) &\cite{Hebrard2009} & 1\\
		
		$\ch{CH4} + \ch{C4N+} \rightarrow \ch{H2} + \ch{HC5NH+}$ & 1.00(-10) & 0.00 & 0.00 &\cite{Wakelam2015} & 1\\
		
		$\ch{HC5N} + \ch{HCO+} \rightarrow \ch{CO} + \ch{HC5NH+}$ & 8.70(-9) &-5.00(-1) & 0.00 &\cite{Wakelam2015} & 1\\
		
		$\ch{C2H4} + \ch{HC5N+} \rightarrow \ch{C2H3} + \ch{HC5NH+}$ & 9.00(-10) & 0.00 & 0.00 &\cite{Anicich2003} & 1\\
		
		$\ch{HC5NH+} + \ch{e-} \rightarrow \ch{H} + \ch{HC5N}$ & 9.20(-7) &-7.00(-1) & 0.00 &\cite{Wakelam2015} & 1\\
		
		$\ch{C5H4N+} + \ch{e-} \rightarrow \ch{H} + \ch{H2} + \ch{HC5N}$ & 1.50(-7) &-5.00(-1) & 0.00 &\cite{Wakelam2015} & 1\\
		
		$\ch{CN} + \ch{C4H2} \rightarrow \ch{H} + \ch{HC5N}$ & 2.72(-10) &-5.20(-1) & 1.90(1) &\cite{Wakelam2015} & 1\\
		
		$\ch{C2H} + \ch{HC3N} \rightarrow \ch{H} + \ch{HC5N}$ & 1.06(-10) &-2.50(-1) & 0.00 &\cite{Wakelam2015} & 1\\
		
		$\ch{C2H2} + \ch{C3N} \rightarrow \ch{H} + \ch{HC5N}$ & 1.30(-10) & 0.00 & 0.00 &\cite{Hebrard2009} & 1\\
		
		$\ch{C2H2} + \ch{C5N} \rightarrow \ch{C2H} + \ch{HC5N}$ & 2.19(-10) & 0.00 & 0.00 &\cite{Harada2010} & 1\\
		
		$\ch{CH3OH} + \ch{CH3+} \rightarrow \ch{CH3OCH4+} + h\nu $ & 7.80(-12) &-1.10(0) & 0.00 &\cite{Herbst1987} & 1\\
		
		$\ch{CH3OH} + \ch{CH3OH2+} \rightarrow \ch{H2O} + \ch{CH3OCH4+}$ & 1.00(-10) &-1.00(0) & 0.00 &\cite{Wakelam2015} & 1\\
		
		$\ch{CH3OCH3} + \ch{H3+} \rightarrow \ch{H2} + \ch{CH3OCH4+}$ & 3.00(-9) &-5.00(-1) & 0.00 &\cite{Wakelam2015} & 1\\
		
		$\ch{CH3OCH3} + \ch{HCO+} \rightarrow \ch{CO} + \ch{CH3OCH4+}$ & 1.20(-9) &-5.00(-1) & 0.00 &\cite{Wakelam2015} & 1\\
		
		$\ch{CH3OCH3} + \ch{H3O+} \rightarrow \ch{H2O} + \ch{CH3OCH4+}$ & 1.37(-9) &-5.00(-1) & 0.00 &\cite{Wakelam2015} & 1\\
		
		$\ch{CH3OCH4+} + \ch{e-} \rightarrow \ch{H} + \ch{CH3OCH3}$ & 1.19(-7) &-7.00(-1) & 0.00 &\cite{Hamberg2010} & 1\\
		
		$\ch{CH3COOH} + \ch{H3+} \rightarrow \ch{H2} + \ch{CH3COOH2+}$ & 3.33(-9) &-5.00(-1) & 0.00 &\cite{Garrod2007} & 1\\
		
		$\ch{CH3COOH} + \ch{H3O+} \rightarrow \ch{H2O} + \ch{CH3COOH2+}$ & 7.41(-9) &-5.00(-1) & 0.00 &\cite{Garrod2007} & 1\\
		
		$\ch{CH3COOH} + \ch{HCO+} \rightarrow \ch{CO} + \ch{CH3COOH2+}$ & 1.16(-9) &-5.00(-1) & 0.00 &\cite{Garrod2007} & 1\\
		
		$\ch{CH3COOH2+} + \ch{e-} \rightarrow \ch{CH3COOH} + \ch{H}$ & 1.50(-8) &-5.00(-1) & 0.00 &\cite{Garrod2007} & 1\\
		
		$\ch{O} + \ch{CH3CHOH} \rightarrow \ch{H} + \ch{CH3COOH}$ & 2.20(-10) & 1.60(-1) & 5.90(-1) &\cite{Skouteris2018} & 1\\
		
		$\ch{CH3COOH} + \ch{NH3OH+} \rightarrow \ch{H2O} + \ch{NH2CH2COOH2+}$ & 1.00(-9) & 0.00 & 0.00 &\cite{Blagojevic2003} & 1\\
		
		$\ch{NH2CH2COOH2+} + \ch{e-} \rightarrow \ch{NH2CH2COOH} + \ch{H}$ & 1.50(-8) &-5.00(-1) & 0.00 &\cite{Blagojevic2003} & 1\\
		
	\end{longtable}
\end{scriptsize}

\section{Coma Formation Rate Plots} \label{App_B}
We have quantified the gas phase formation of any species in the coma by means of the coma formation rate plots, as shown in the following figures.
The relative 
reaction rate can be multiplied by $P_{i}$ to obtain the rate per unit volume (cm$^{-3}$ s$^{-1}$) at which any 
particular reaction leads to the creation of the species. While  a species can form in the coma by a number of 
processes, only those reactions that have a relative rate higher than 1\% are shown.
	
\newpage
\begin{figure}
\centering
\includegraphics[height=7cm, width=0.48\textwidth]{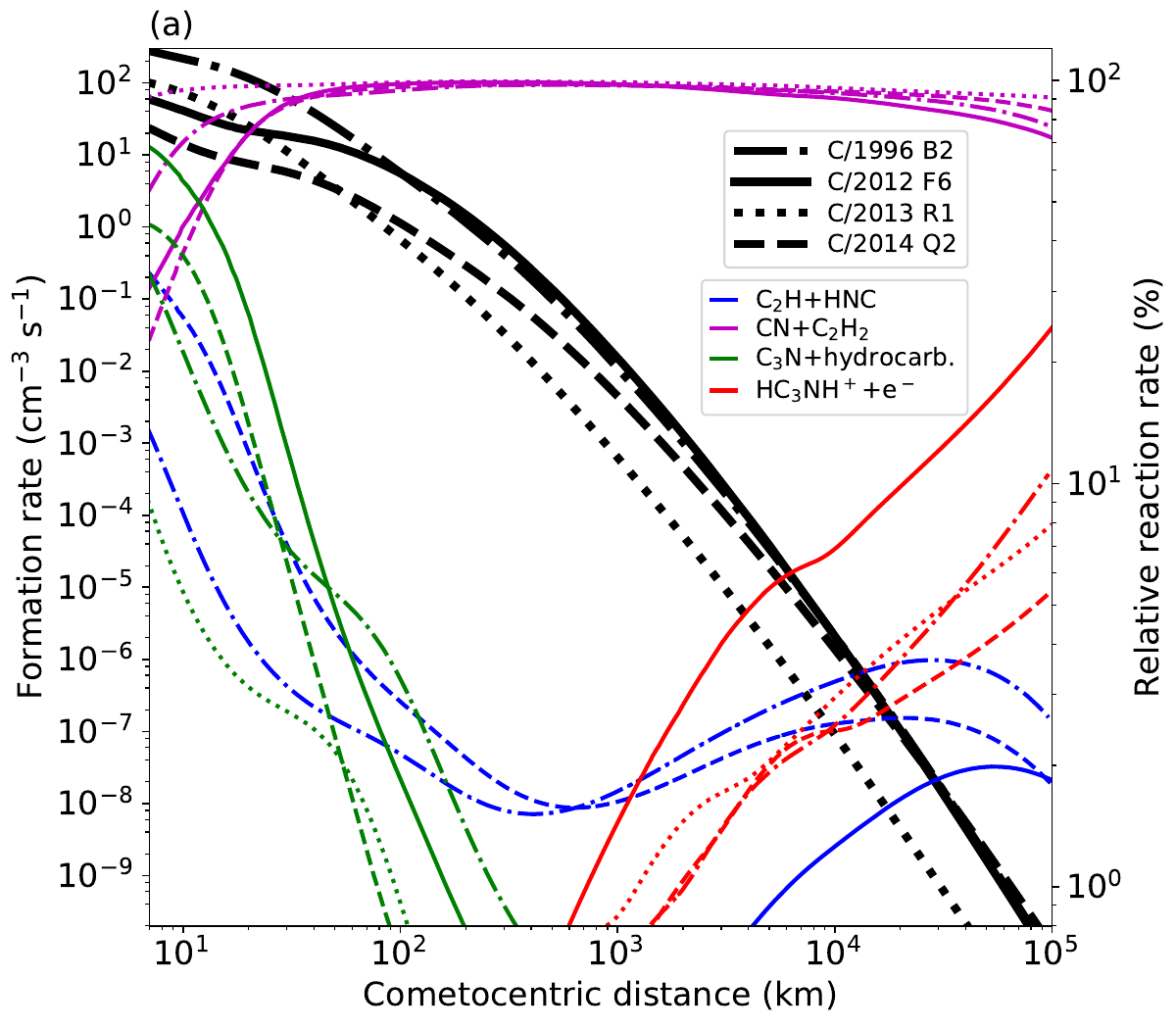}
\includegraphics[height=7cm, width=0.48\textwidth]{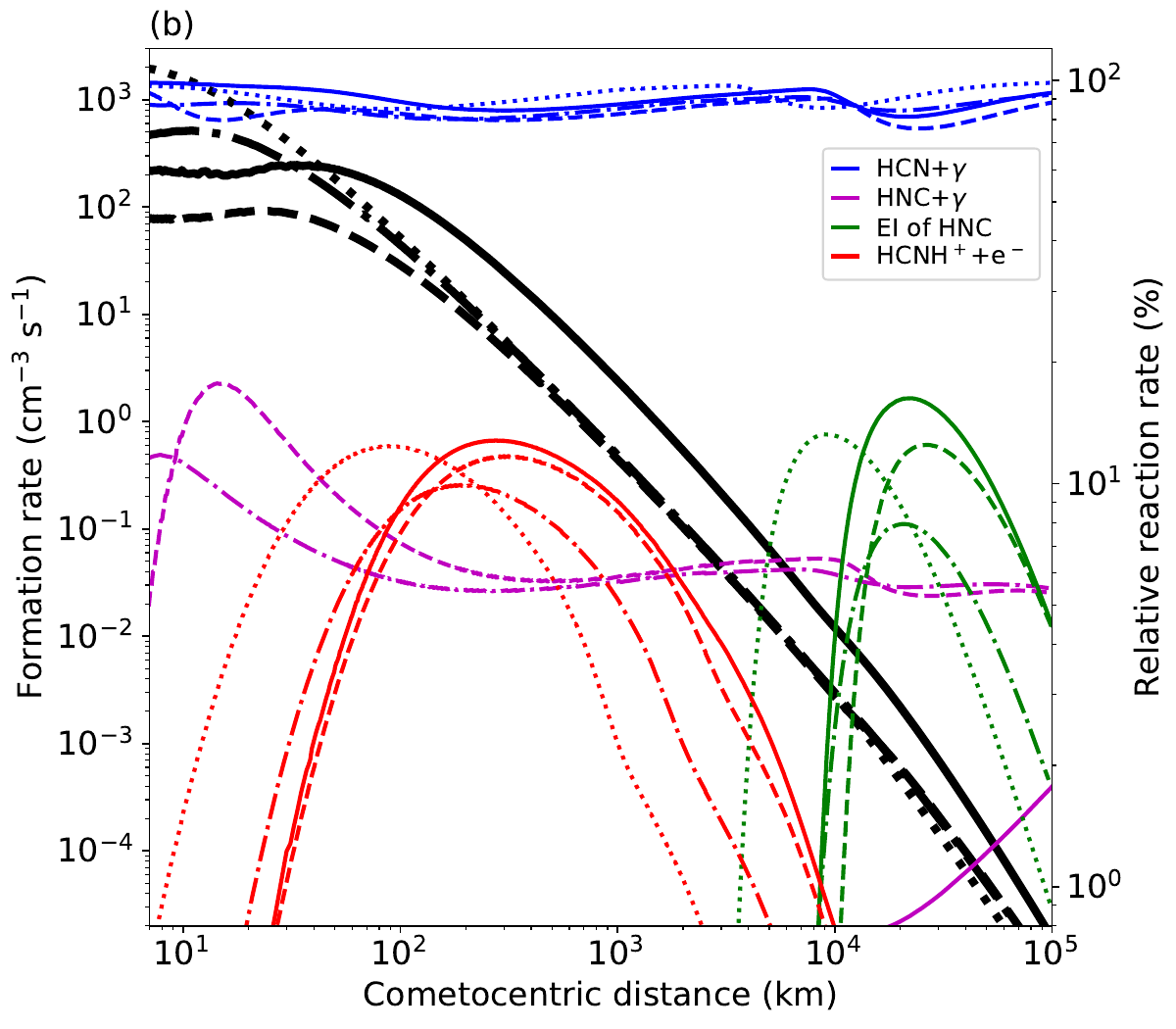}
		
\caption{  The coma formation rate plots for (a) \ch{HC3N} and (b) \ch{CN}. The black lines (left y-axis) show 
		the cometocentric distance variation of the formation rate ($P_{i}$) of the species. The colored lines
		 (right y-axis) show the cometocentric distance variation of the relative rates (\%) of the chemical
		  reactions that form the species. The line styles indicate different cometary compositions 
		  (dotted-dashed: C/1996 B2, solid: C/2012 F6, dotted: C/2013 R1 and dashed: C/2014 Q2). 
	}		
		\label{fig:HC3Na}
	\end{figure}

	\begin{figure}
		\centering
		{\includegraphics[height=7cm, width=0.47\textwidth]{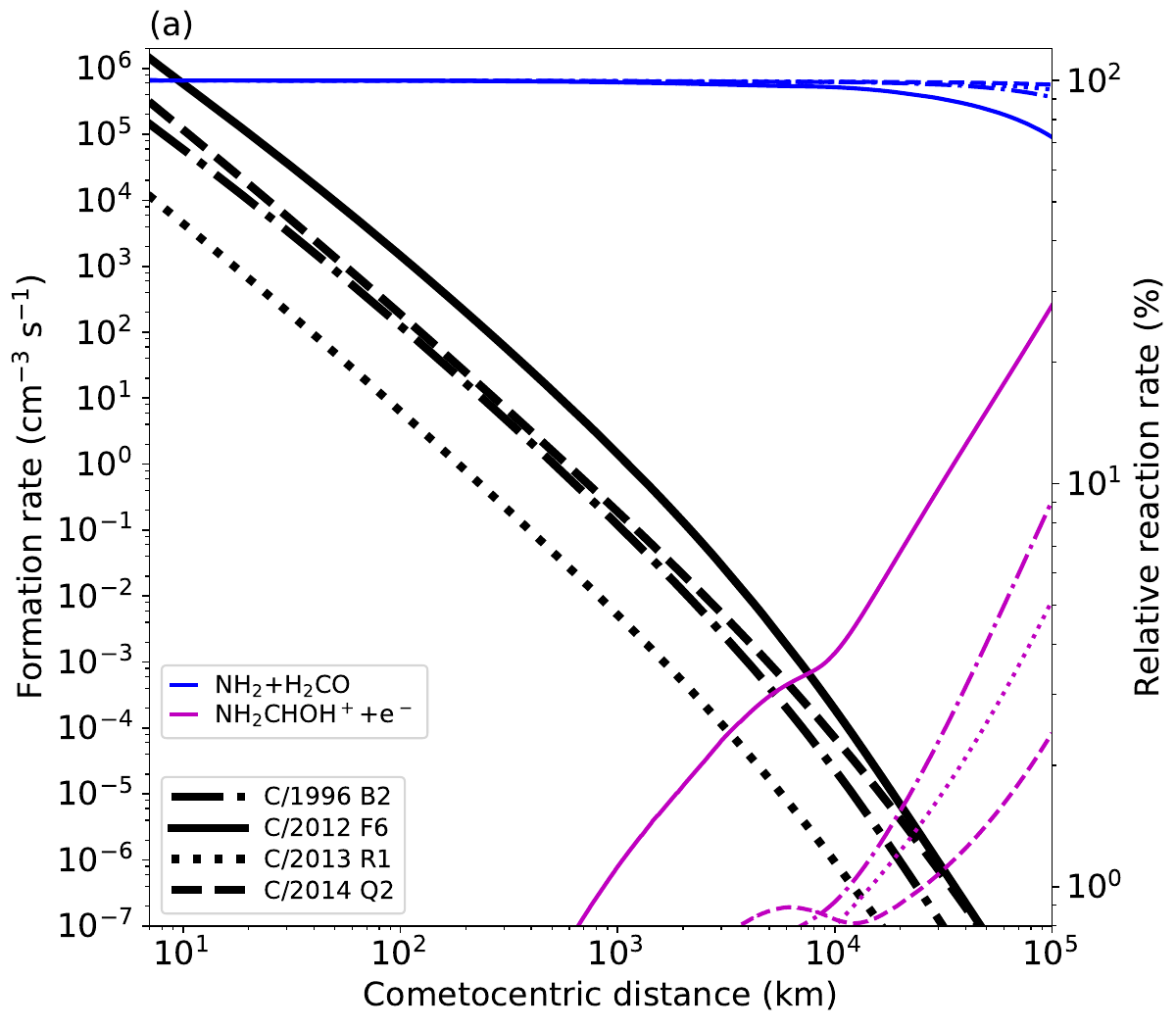}}
		{\includegraphics[height=7cm, width=0.47\textwidth]{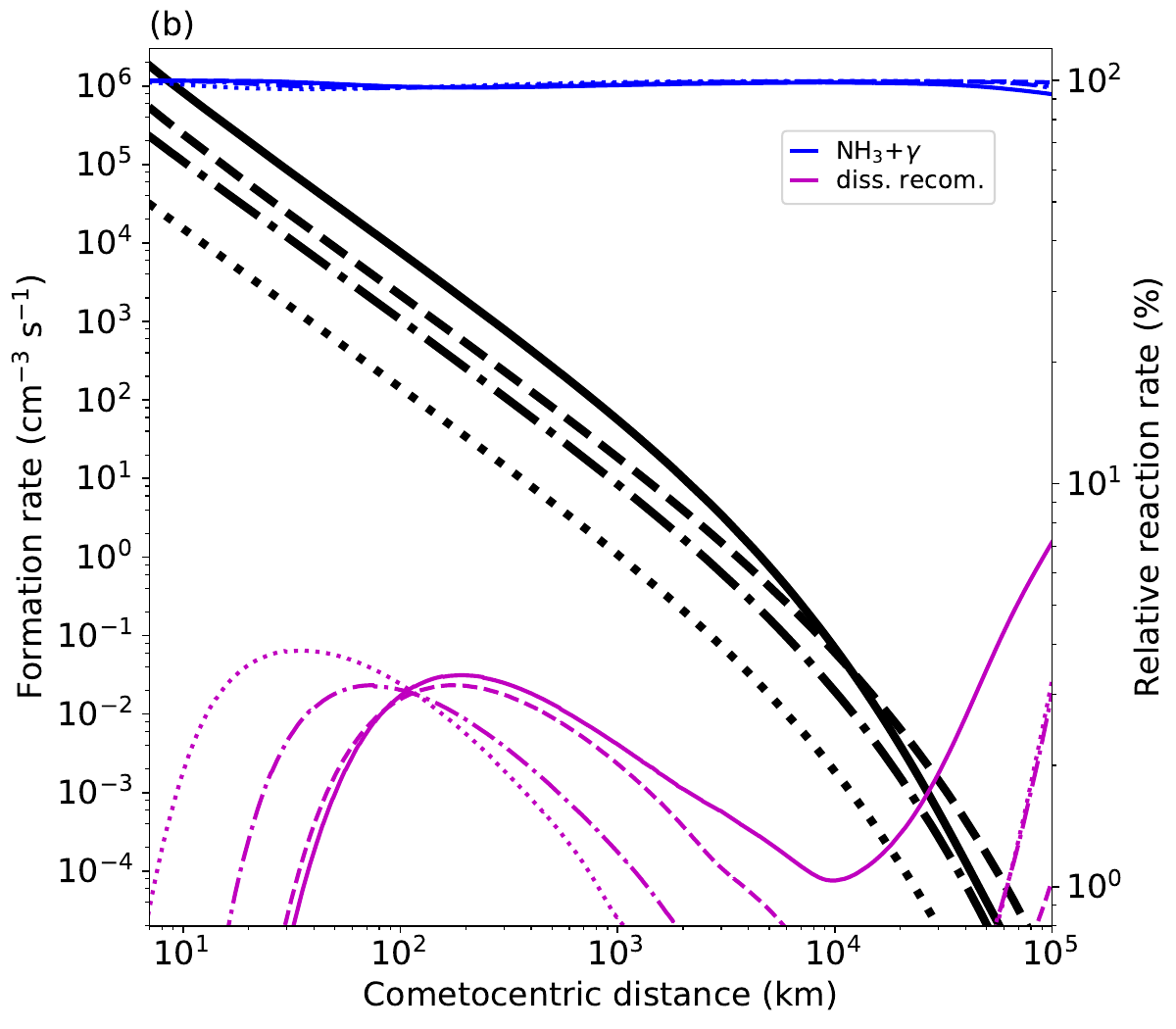}}
		{\includegraphics[height=7cm, width=0.47\textwidth]{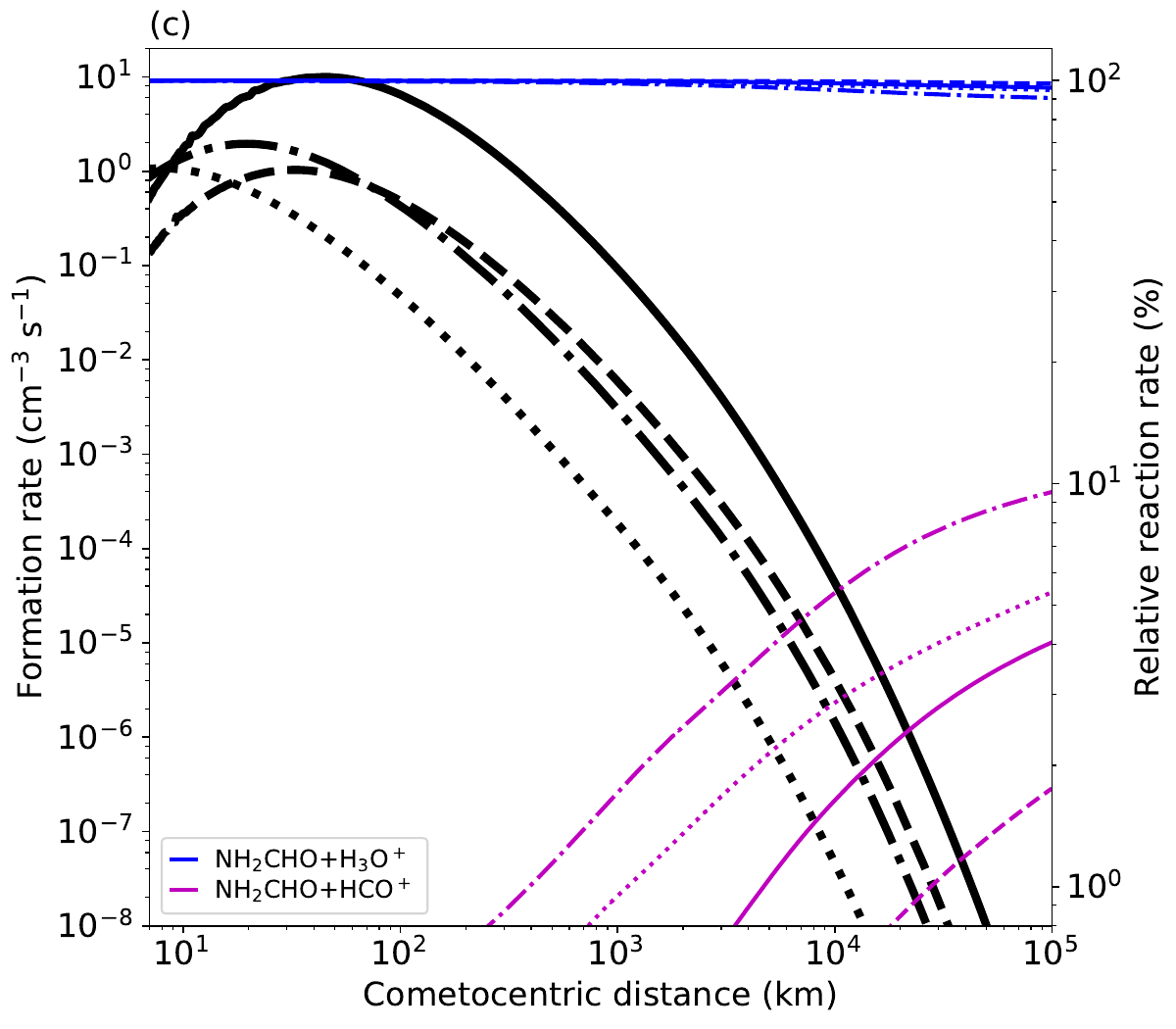}}
		\caption{ Same as Figure \ref{fig:HC3Na} but for (a) \ch{NH2CHO}, (b) \ch{NH2} and (c) \ch{NH2CHOH+}.}		
		\label{fig:NH2CHOa}
	\end{figure}

	\begin{figure}
		\centering
		\includegraphics[height=7cm, width=0.48\textwidth]{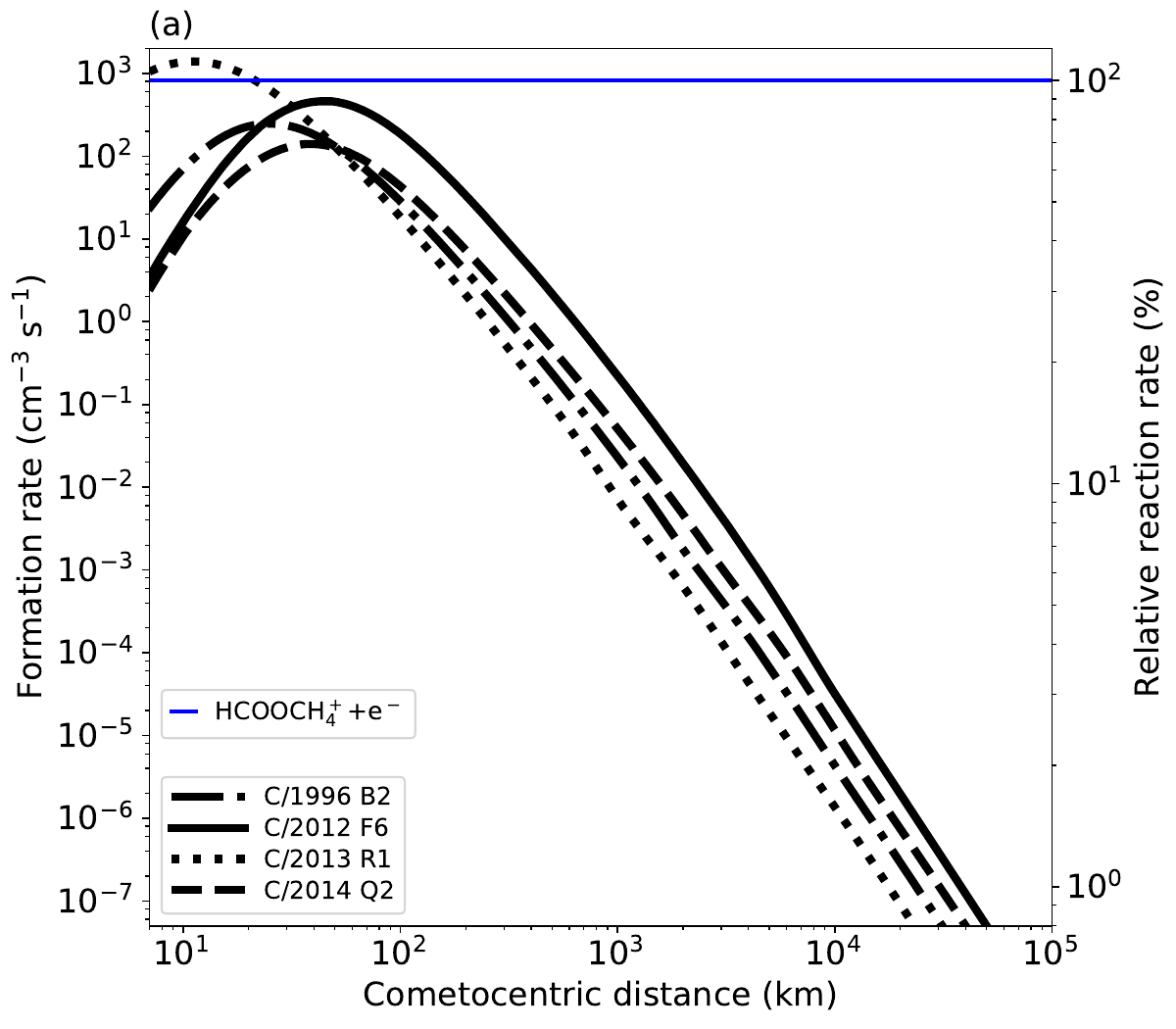}	
		\includegraphics[height=7cm, width=0.48\textwidth]{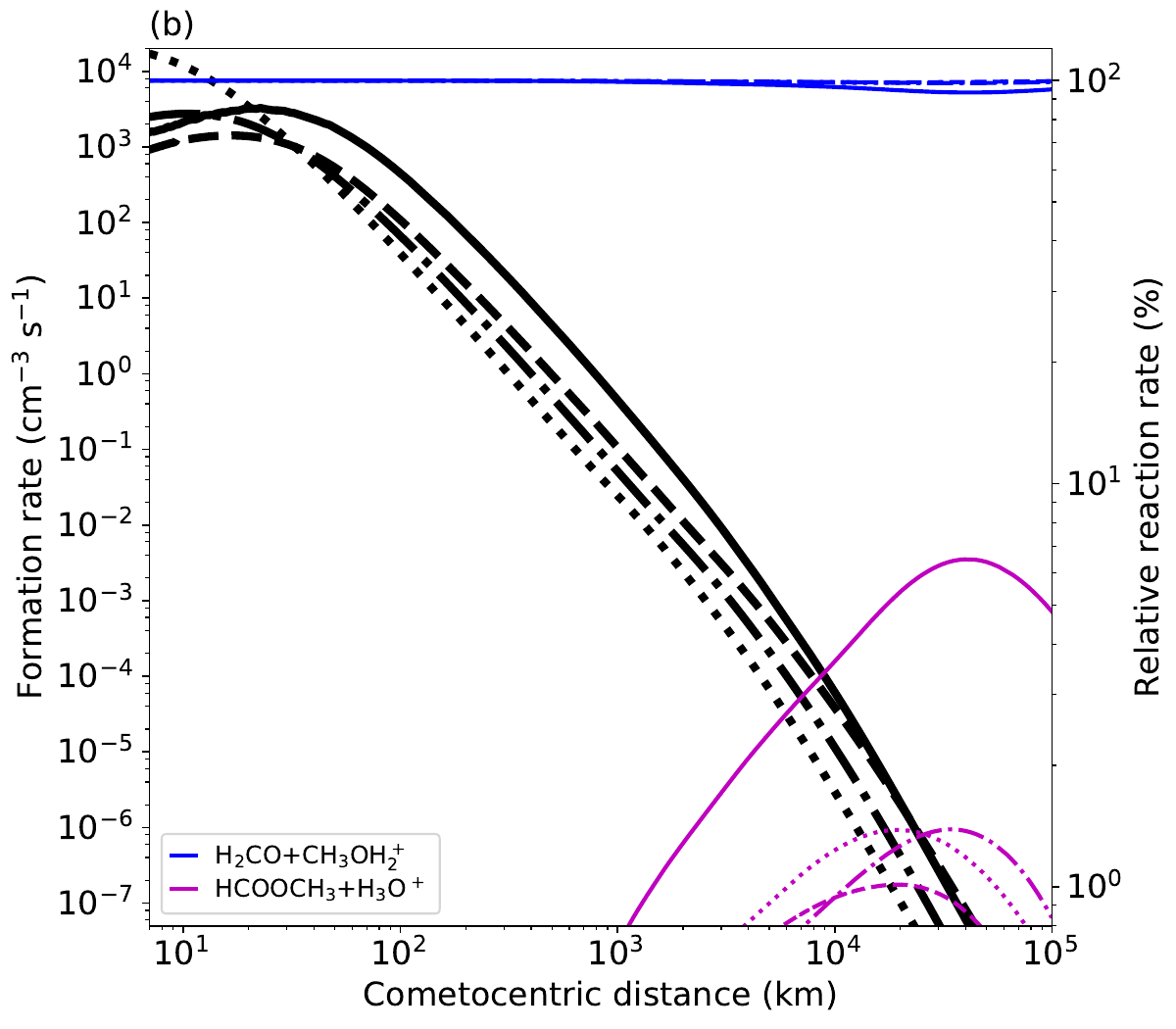}
		
		\caption{
			Same as Figure \ref{fig:HC3Na} but for (a) \ch{HCOOCH3} and (b) \ch{HCOOCH4+}.}		
		\label{fig:HCOOCH3a}
	\end{figure}

	\begin{figure}
		\centering
		\includegraphics[height=7cm, width=0.48\textwidth]{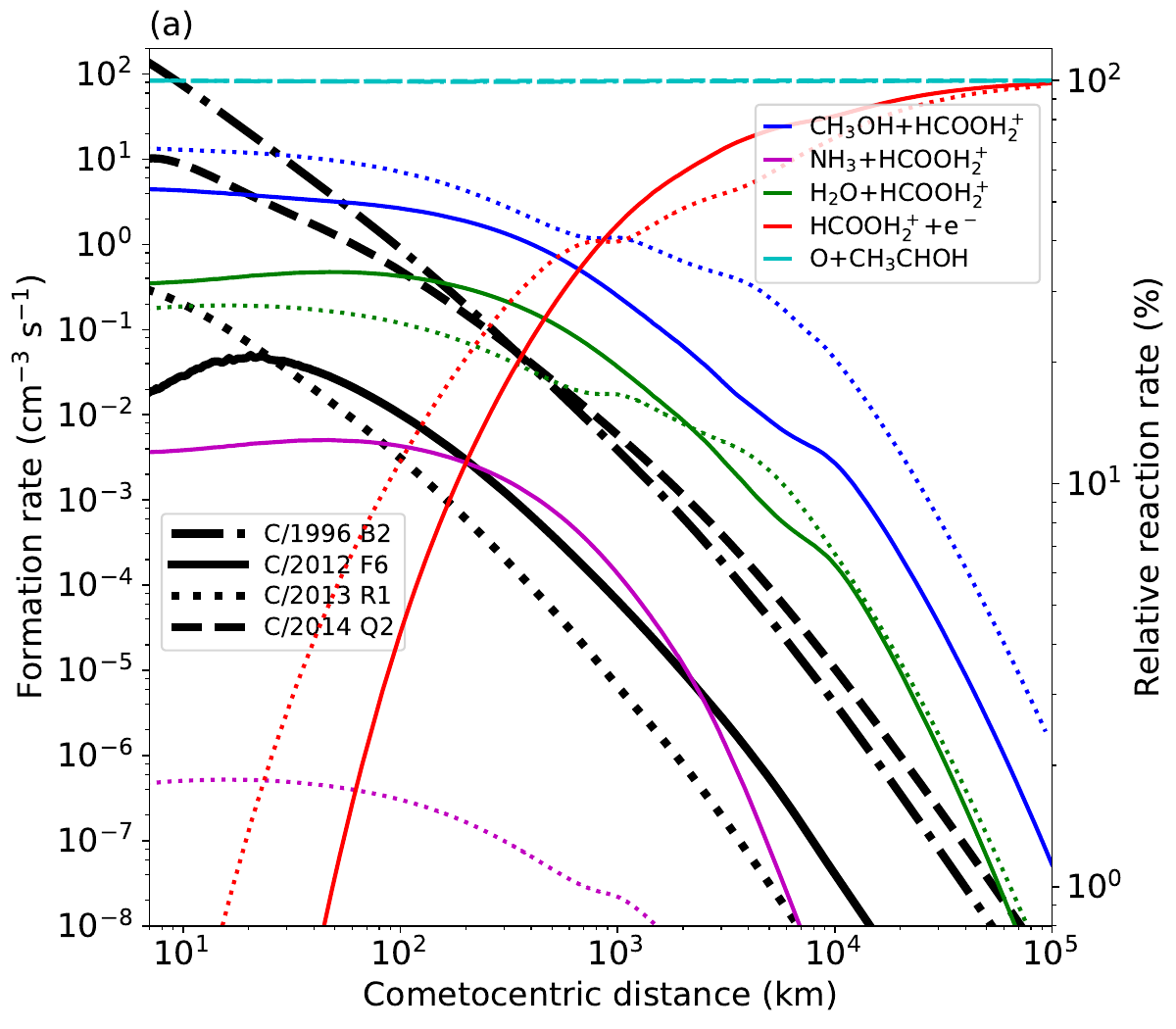}
		\includegraphics[height=7cm, width=0.48\textwidth]{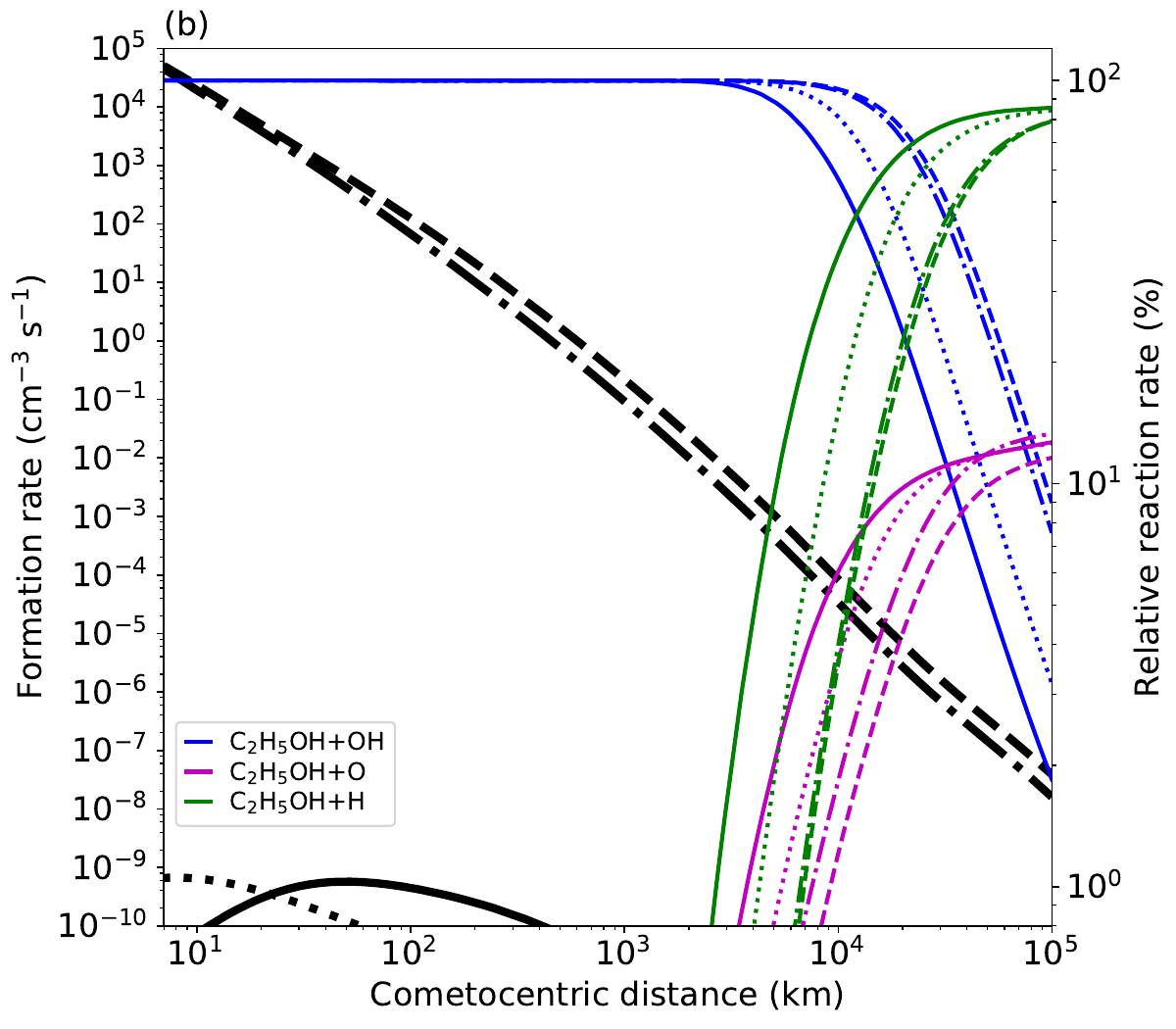}
		\caption{Same as Figure \ref{fig:HC3Na} but for (a) \ch{HCOOH} and (b) \ch{CH3CHOH}.}		
		\label{fig:HCOOHa}
	\end{figure}
	
	\begin{figure}
		\centering
		\includegraphics[height=7cm, width=0.48\textwidth]{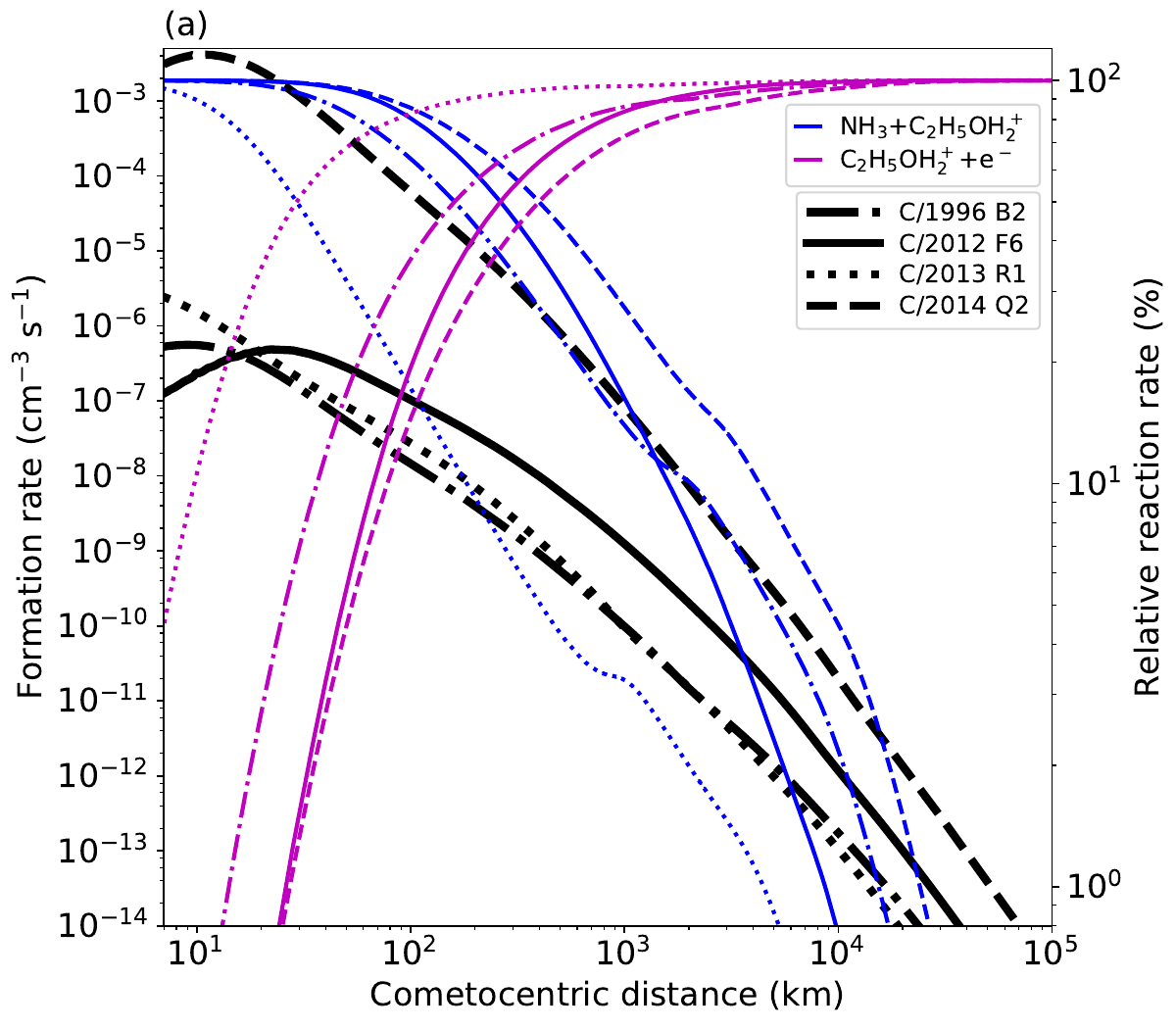}
		\includegraphics[height=7cm, width=0.48\textwidth]{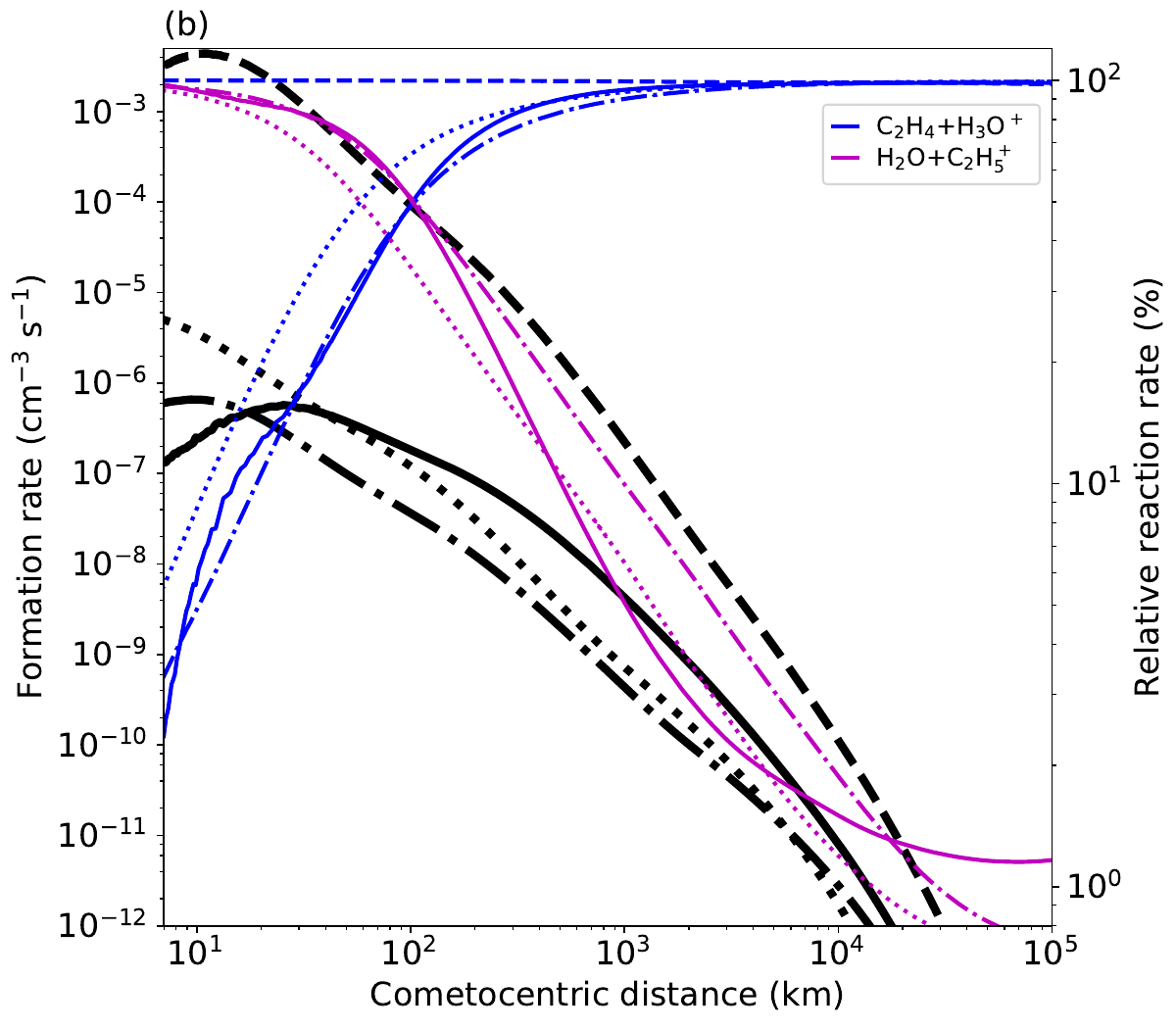}
		\caption{Same as Figure \ref{fig:HC3Na} but for (a) \ch{C2H5OH} and (b) \ch{C2H5OH2+}.}		
		\label{fig:C2H5OHa}
	\end{figure}

	\begin{figure}
		\centering
		\includegraphics[height=7cm, width=0.48\textwidth]{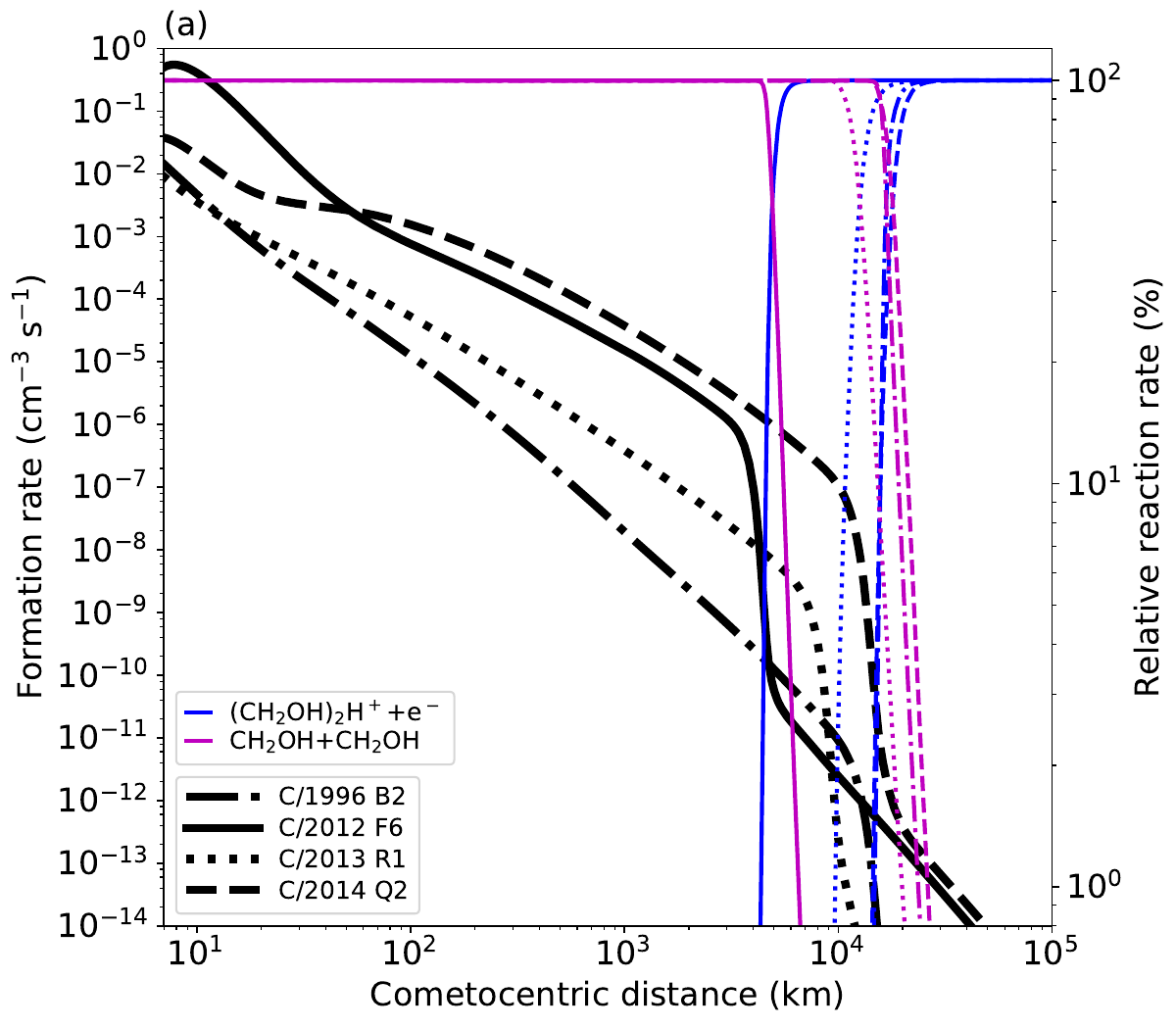}
		\includegraphics[height=7cm, width=0.48\textwidth]{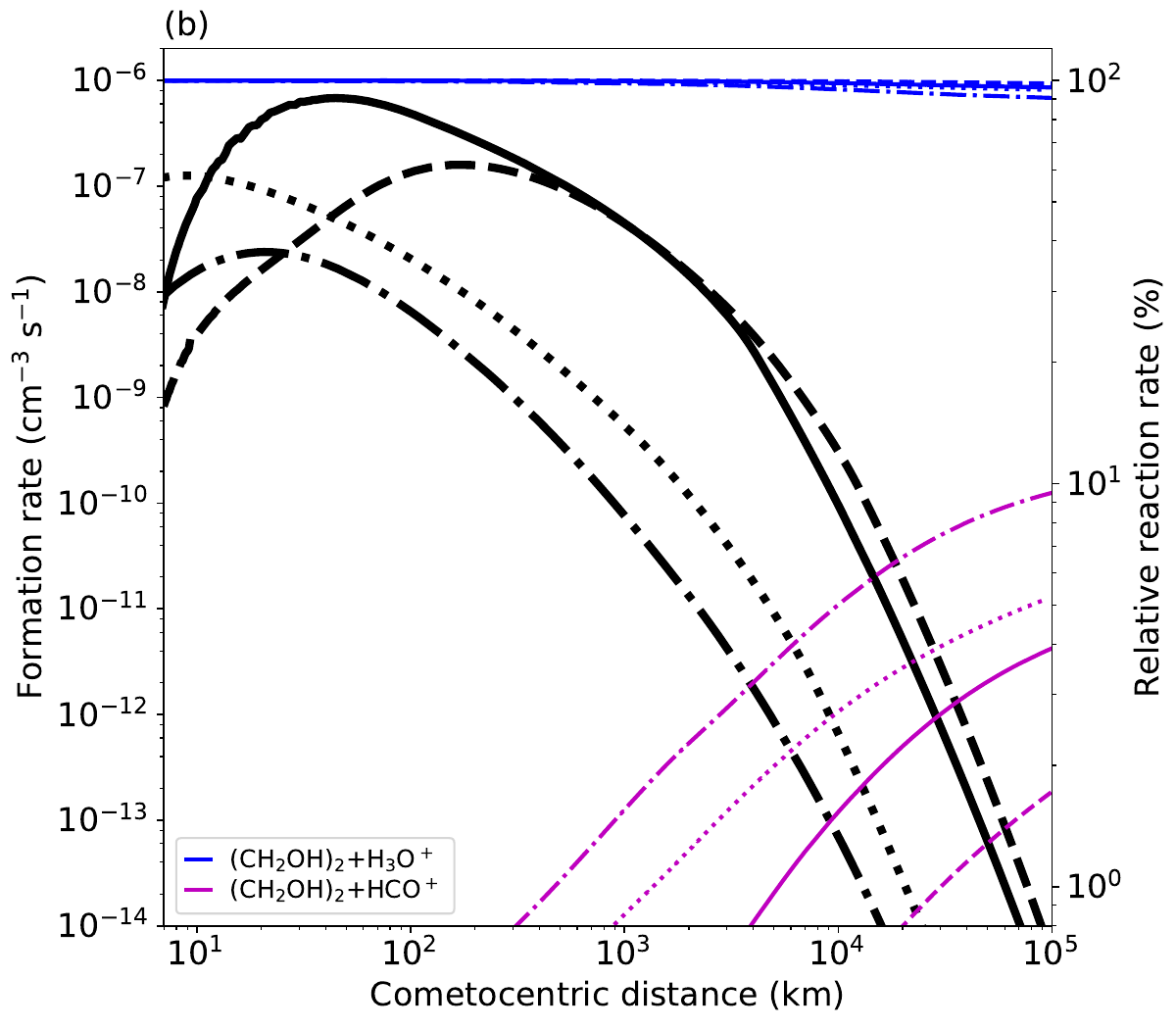}
		\caption{Same as Figure \ref{fig:HC3Na} but for (a) \ch{(CH2OH)2} and (b) \ch{(CH2OH)2H+}.}		
		\label{fig:CH2OH_2a}
	\end{figure}

	\begin{figure}
		\centering
		\includegraphics[height=7cm, width=0.48\textwidth]{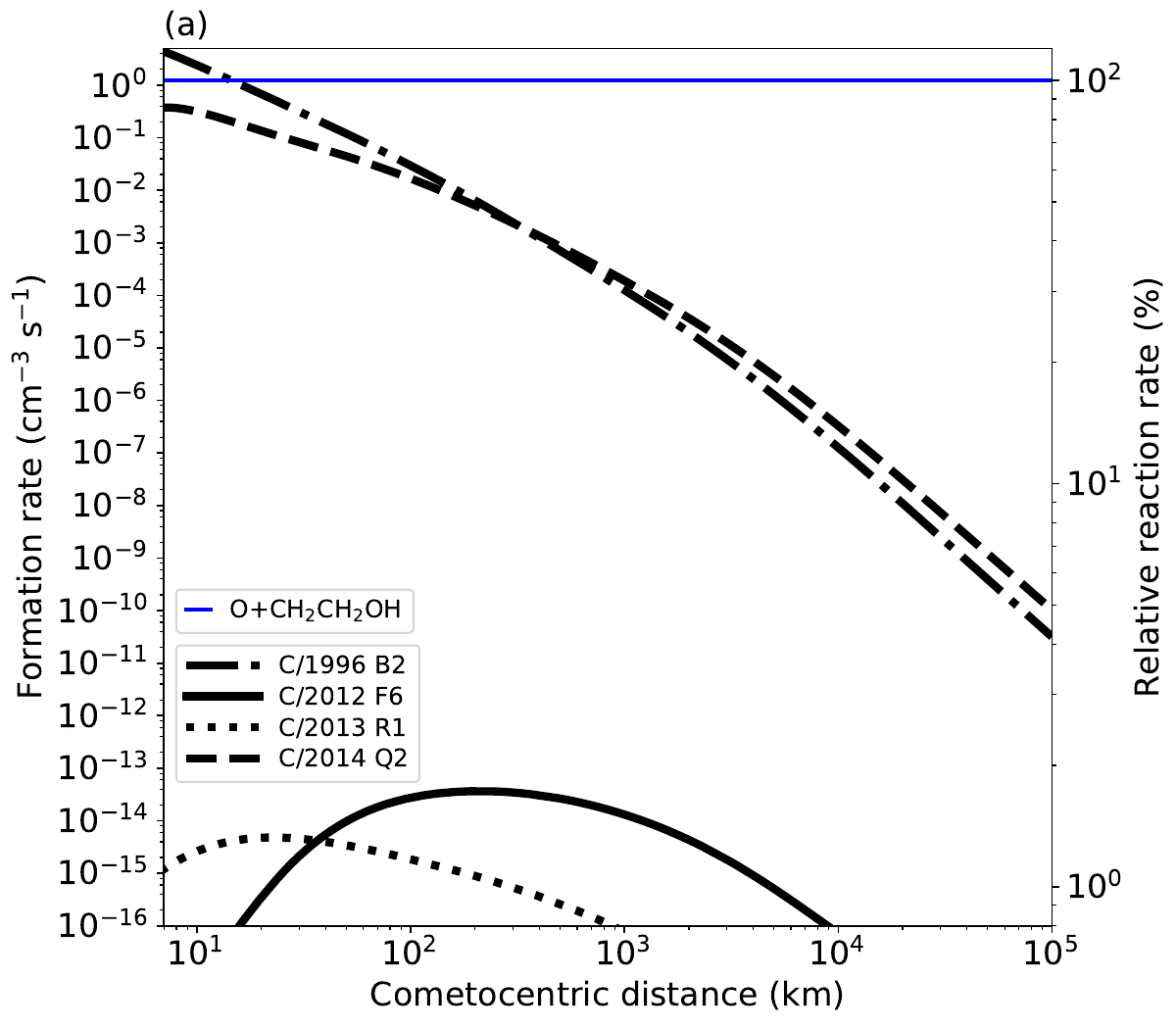}
		\includegraphics[height=7cm, width=0.48\textwidth]{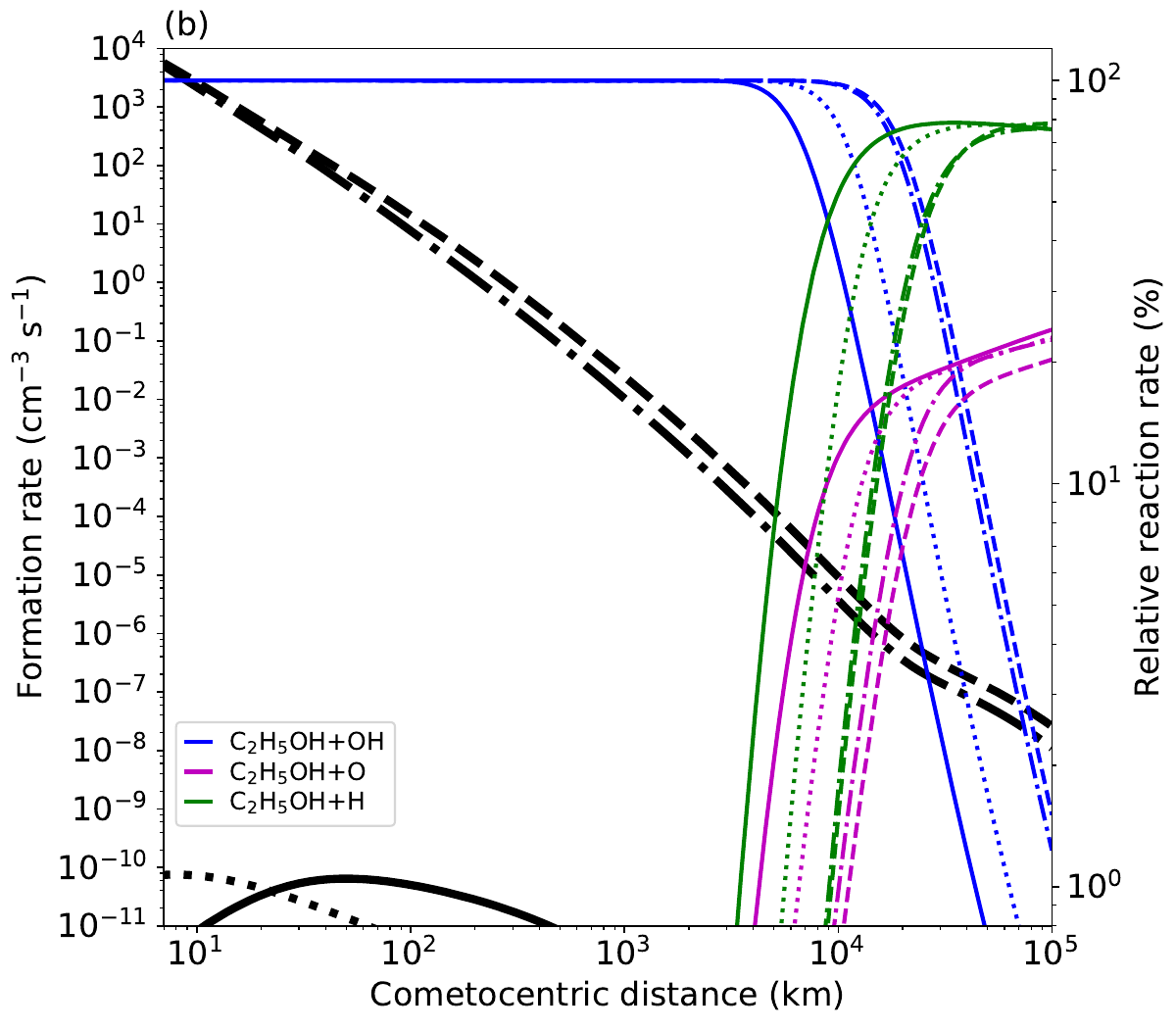}
		\caption{Same as Figure \ref{fig:HC3Na} but for (a) \ch{CH2OHCHO} 
			and (b) \ch{CH2CH2OH}.}		
		\label{fig:CH2OHCHOa}
	\end{figure}
	
	\begin{figure}
		\centering
		\includegraphics[height=7cm, width=0.48\textwidth]{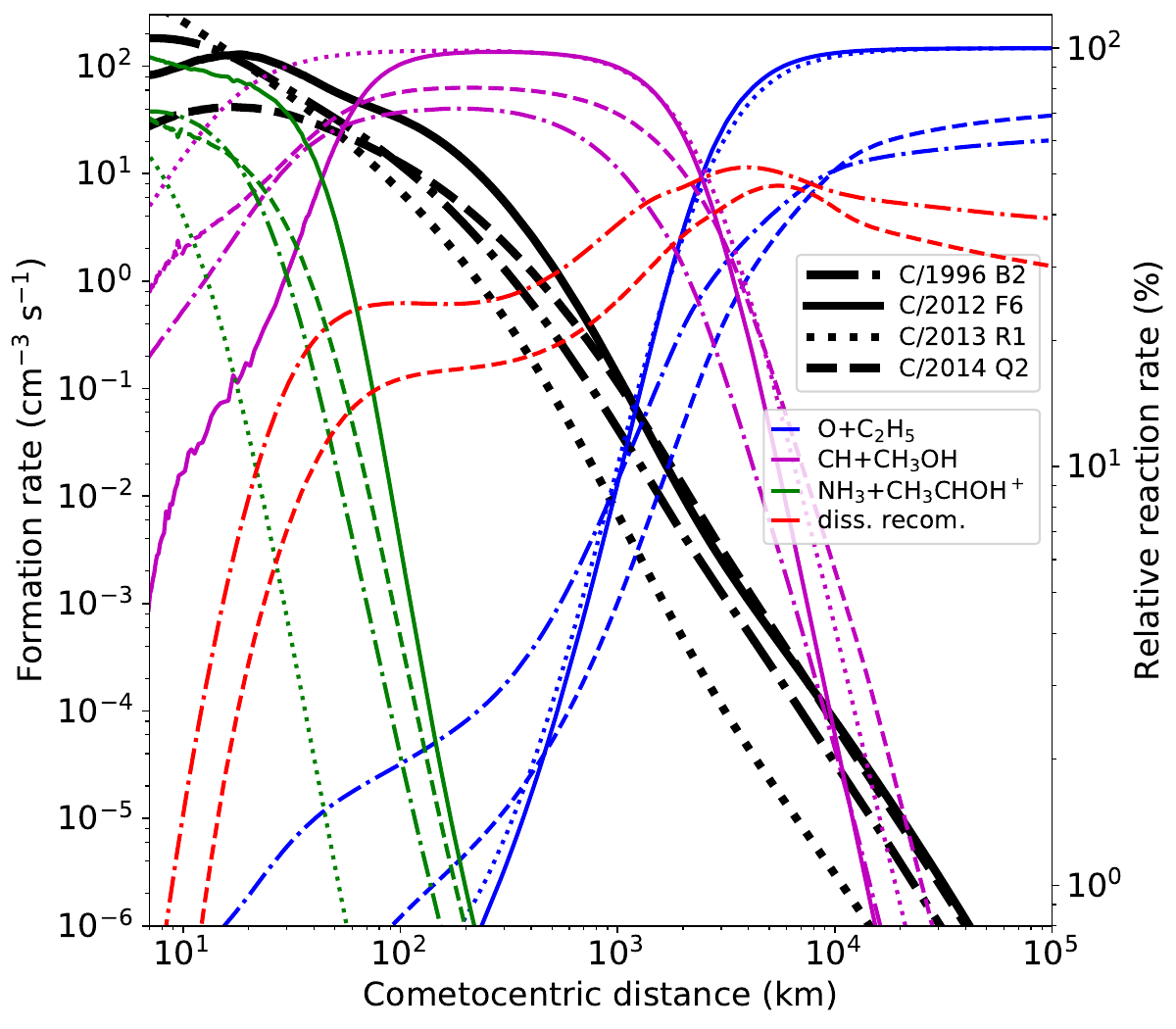}
		\caption{Same as Figure \ref{fig:HC3Na} but for \ch{CH3CHO}.}		
		\label{fig:CH3CHOa}
	\end{figure}

	\begin{figure}
		\centering
		\includegraphics[height=7cm, width=0.48\textwidth]{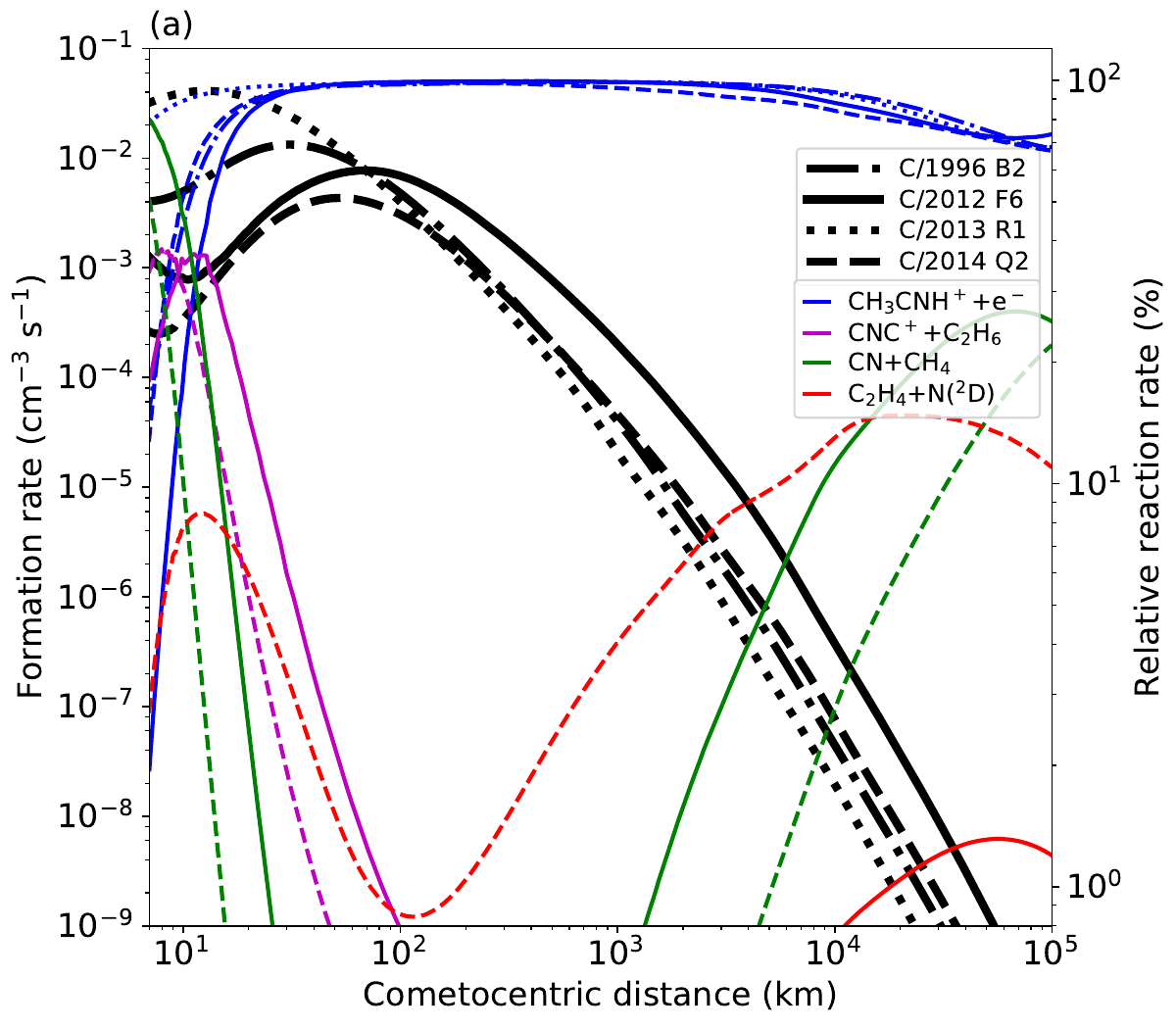}
		\includegraphics[height=7cm, width=0.48\textwidth]{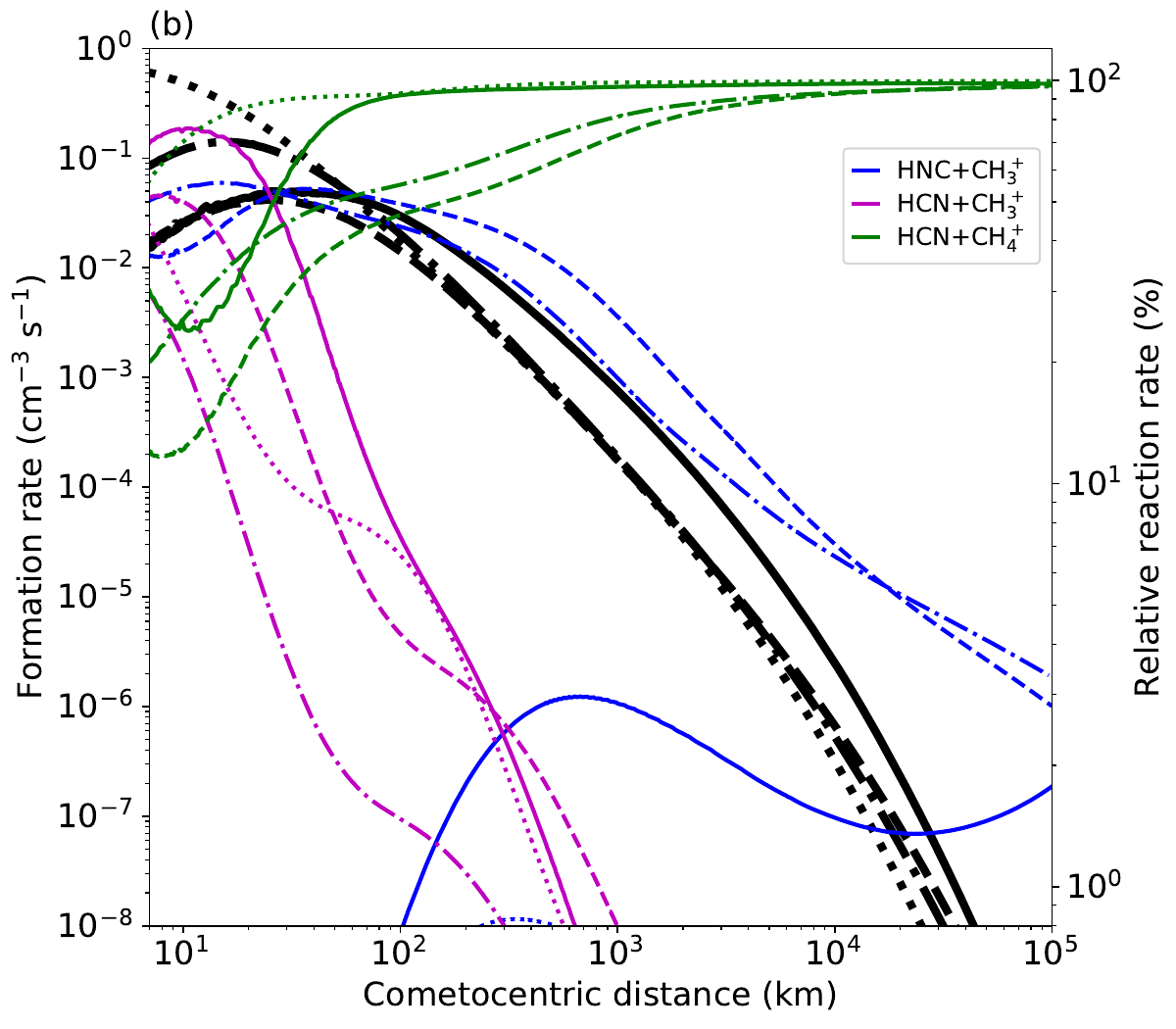}
		\caption{Same as Figure \ref{fig:HC3Na} but for 
			(a) \ch{CH3CN} and (b) \ch{CH3CNH+}.}		
		\label{fig:CH3CNa}
	\end{figure}
	
	\begin{figure}
		\centering
		\includegraphics[height=7cm, width=0.47\textwidth]{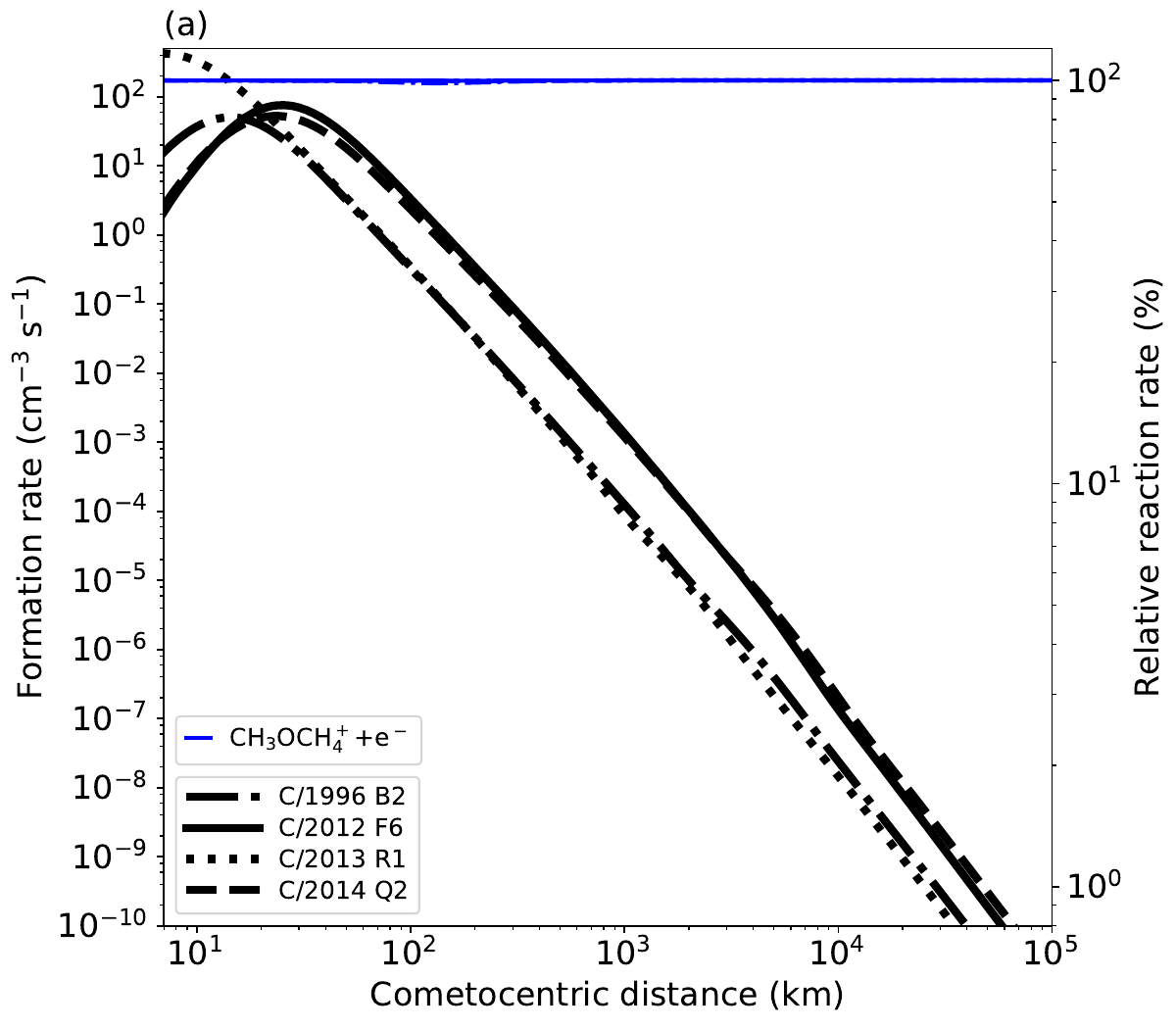}
		\includegraphics[height=7cm, width=0.47\textwidth]{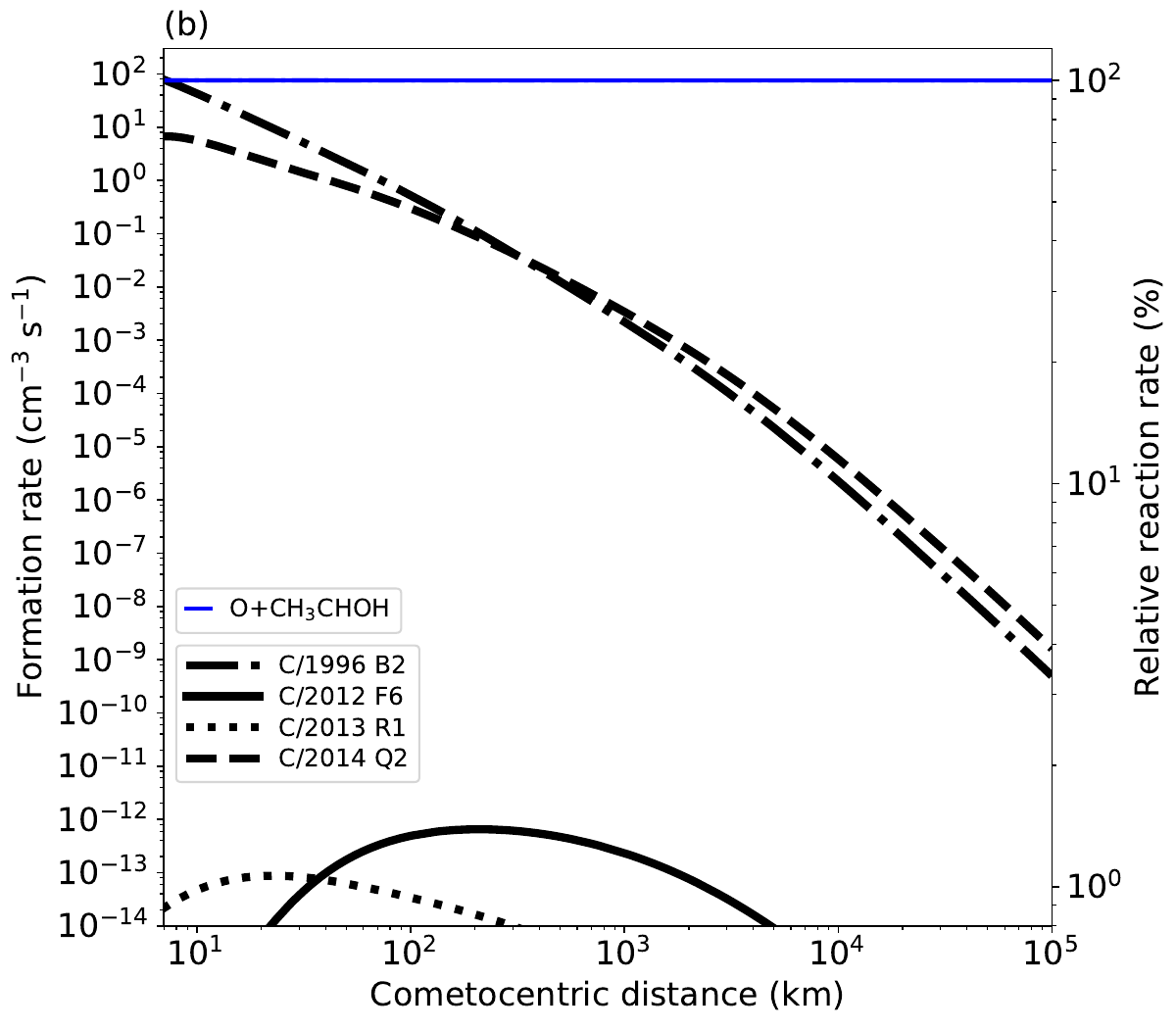}
		\includegraphics[height=7cm, width=0.47\textwidth]{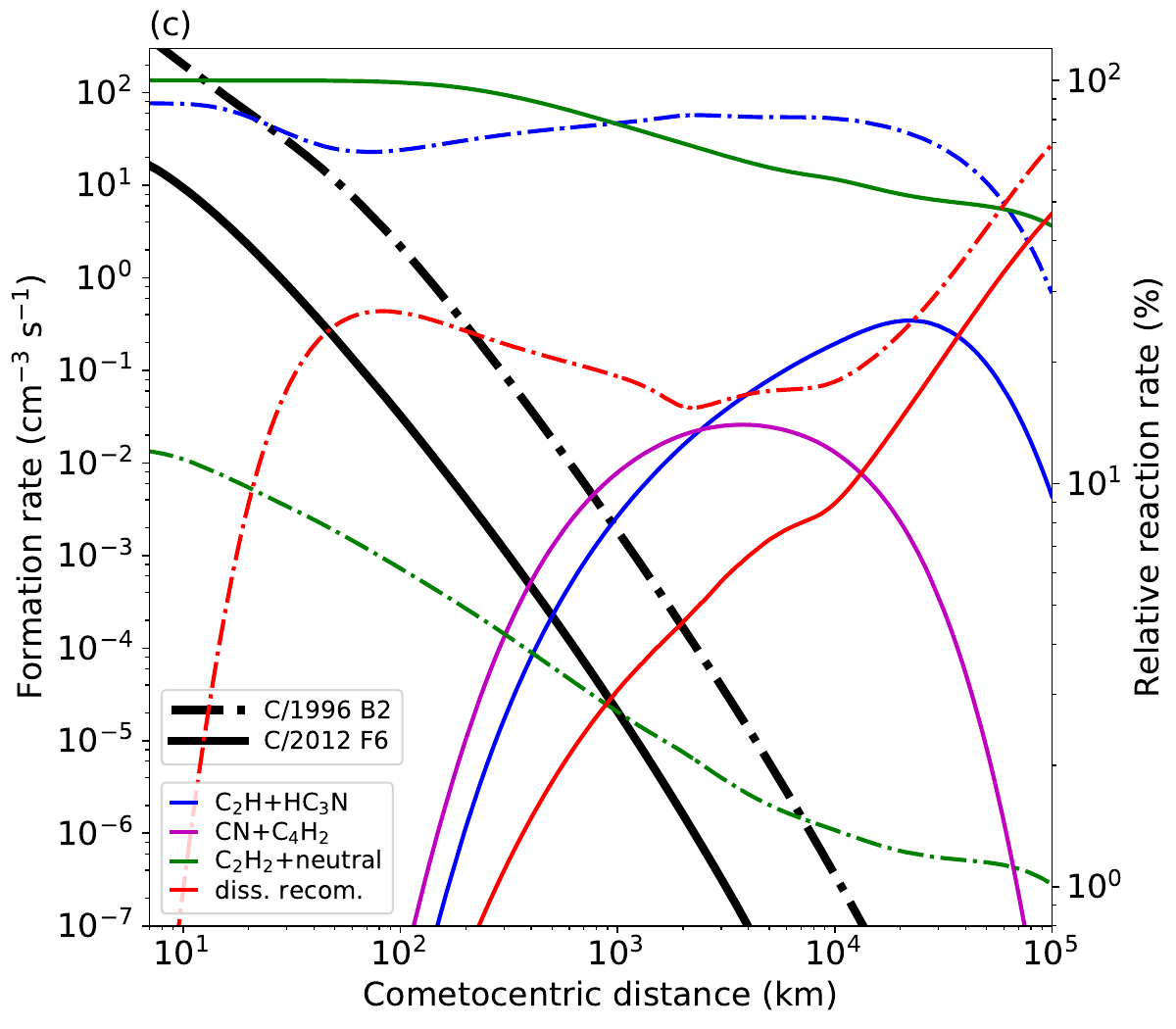}
		\caption{Same as Figure \ref{fig:HC3Na} but for (a) \ch{CH3OCH3}, (b) \ch{CH3COOH}, and (c) \ch{HC5N}.}		
		\label{fig:CH3OCH3-CH3COOHa}
	\end{figure}

\clearpage

\section{Effect of Varying Solar Zenith Angle}  \label{App_C}
\noindent The second term on the right hand side of Equation \ref{eq:energy_balance} denotes how much of the incident 
solar flux is used up in the sublimation of volatiles from the nucleus. This term is almost independent 
of temperature, while the excess heat resulting from a decrease in the solar zenith angle $\phi$ leads to 
an increase in the surface temperature. The variation of the surface temperature with $\phi$ is shown in 
Figure \ref{fig:zenith} (also, first panel in the top row of Figure \ref{fig:zenith_all}). While the 
increase in temperature is steeper when $\phi$ reduces from $60^\circ$ to $30^\circ$, the temperature 
becomes nearly constant for $\phi < 30^\circ$. An increase in the initial gas temperature causes the 
particles to expand at a higher initial velocity, which leads to a reduction in the initial density 
of water (and thereby other parent volatiles) as denoted in Figure \ref{fig:zenith_all} (top row, 
second panel). The extent of reduction is not the same in all four comets and depends on factors 
including production rate and heliocentric distance. Figure \ref{fig:zenith_all} shows by what factor 
the number densities of the cometary organics reduce when $\phi = 30^\circ$ as opposed to when 
$\phi = 60^\circ$. The ratio $n_{\phi=30} / n_{\phi=60}$ lies mostly between $0.6-0.8$, though 
it may leave this range in some cases such as \ch{CH2OHCHO} and \ch{CH3COOH}. In the case of \ch{CH3CN}, 
the density actually increases close to the nucleus. Here, we also assume that the production rate is 
uniform over the entire nucleus surface. For non-uniform production rates, a net effective area can be 
calculated from which sublimation is occurring \citep{Marshall2019}. This in turn will modify the 
number density by some factor, as described for the case of the change in solar zenith angle. 

\begin{figure}[htb!]
	\centering
	\includegraphics[width=0.8\textwidth]{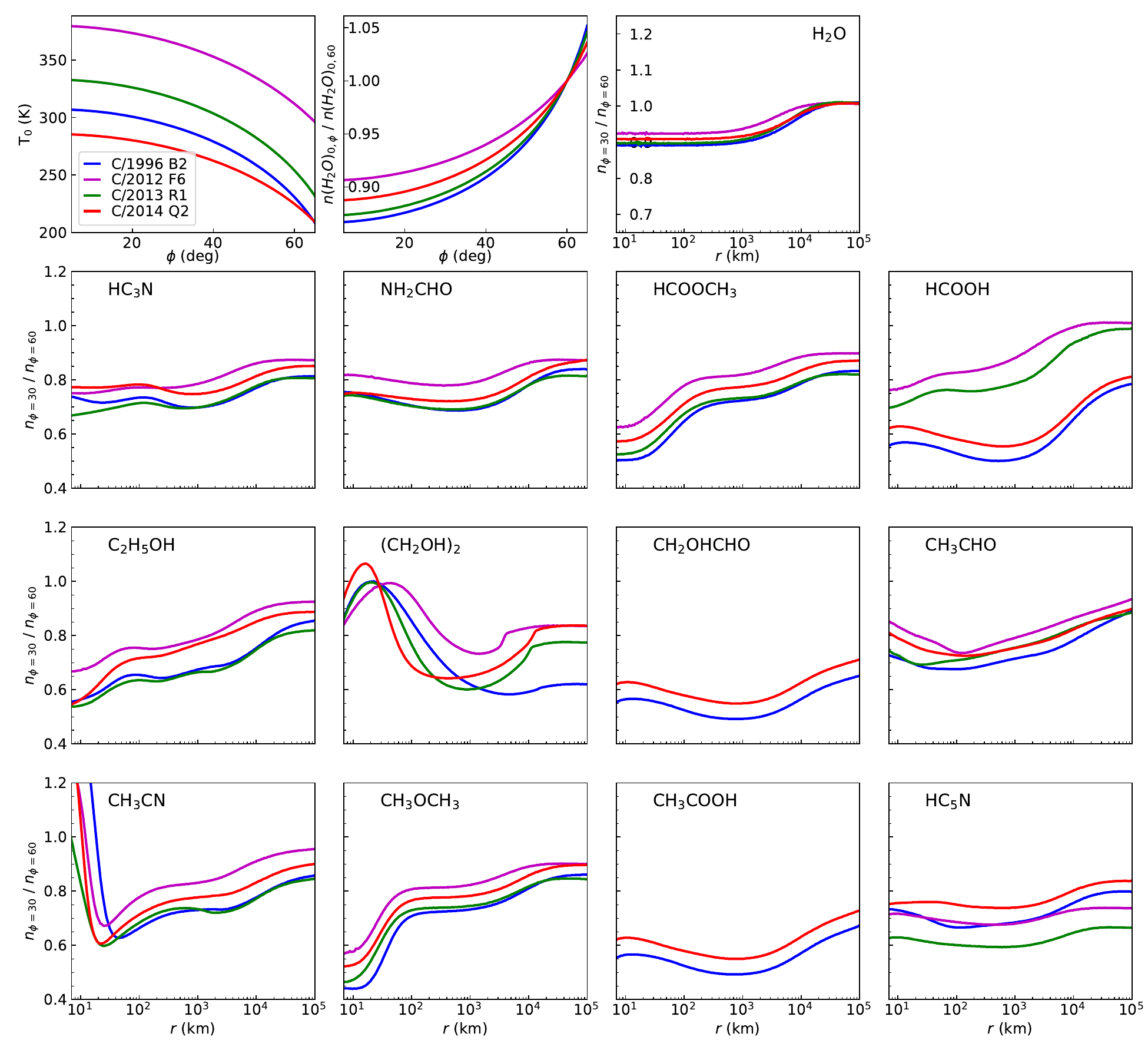}	
	\caption{The first panel in the top row shows the variation of the initial temperature $T_0$ with solar zenith angle $\phi$. The second panel shows the variation of the ratio $n(\ch{H2O})_{0,\phi} / n(\ch{H2O})_{0,60}$, where $n(\ch{H2O})_{0,\phi}$ is the initial number density of \ch{H2O} at an angle $\phi$ while $n(\ch{H2O})_{0,60}$ is the density when $\phi = 60^\circ$. The other panels show the cometocentric distance variation (for the species indicated in each panel) of the ratio $n_{\phi=30} / n_{\phi=60}$, where $n_{\phi=30}$ and $n_{\phi=60}$ are the species number density when the solar zenith angles are $30^\circ$ and $60^\circ$, respectively.}
	\label{fig:zenith_all}
\end{figure}

\clearpage

\section{Comparison of Model Results with Other Studies}  \label{App_D}
\noindent We used the chemical network of \cite{Weiler2006Thesis} in our model in order to benchmark our results for
C/1996 B2 (Hyakutake). The input conditions for
these model runs are the same as used by \cite{Weiler2006Thesis}; these are $Q_{\ch{H2O}} = 1.7 \times 10^{29}$
molecules s$^{-1}$, $r_h = 1$ au, and radius of nucleus = 2.2 km. The abundance of the other volatiles with respect to
water are: \ch{CO} = 20\%, \ch{CO2} = 6\%, \ch{CH3OH} = 2\%, \ch{H2CO} = 1\%, \ch{CH4} = 0.7\%, \ch{C2H2} = 0.1\%,
\ch{C2H6} = 0.4\%, \ch{NH3} = 1\%, \ch{HCN} = 0.1\%, and \ch{N2} = 0.04\%. As mentioned in Section \ref{subsection:reslt_temp},
our ion temperatures are higher than that of \cite{Weiler2006Thesis} by factors $\sim 2 -4$ because we use non-zero 
values of the reaction enthalpy ($\Delta E$) of ion-neutral reactions in our network. Thus, we ran two models: in the first
one we used the exact network of \cite{Weiler2006Thesis}, while in the second one, we used the network of \cite{Weiler2006Thesis}
but with the values of $\Delta E$ that we use in our network. The resulting temperature profiles are shown in Figure
\ref{fig:T_weiler}. We see that in the first case, our temperature profiles match the profiles obtained by \cite{Weiler2006Thesis} (left panel of Figure \ref{fig:T_weiler}). In the second case, 
the ion temperature is higher than \cite{Weiler2006Thesis} (right panel of Figure \ref{fig:T_weiler}).
However, the difference by a factor of $\sim 2 -4$ does not affect the modeled abundances of the ions in the coma.
The rates of dissociative recombination reactions are sensitive to temperature changes; these rates depend on $T_e$ 
and our electron temperatures match well in both cases. 

Although we assume that all of $\Delta E$ goes towards increasing the kinetic energy of the products, this is not necessarily true. In ion-neutral reactions, the kinetic energy would decrease due to the attractive potential between the ion and neutral species. For all reactions, including photodissociation, part of the energy is also expended in increasing the internal energy of the products (electronic and ro-vibrational modes). It would require detailed quantum mechanical calculations to find how $\Delta E$ is distributed as kinetic and internal energies of the products, and this distribution is different for each reaction. The two ion temperatures show in Figure \ref{fig:T_weiler} (right panel) can be thought of as the lower and upper limits of the ion temperature calculated using zero and non-zero values of $\Delta E$, respectively, and the computed temperature would lie between these limits if we were to consider that the kinetic energy gained by the products is lower than $\Delta E$. In case of the neutral temperature, cooling dominates over photolytic heating in the inner regions. In the outer regions, if we consider that part of the excess energy goes towards increasing the internal energy of the molecules, then we would get a lower temperature. Since most of the collisional chemistry occurs in the inner regions, a reduction in the neutral temperature in the outer regions would not affect our computed species densities.

\begin{figure}[htb!]
	\centering
	\includegraphics[width=\textwidth]{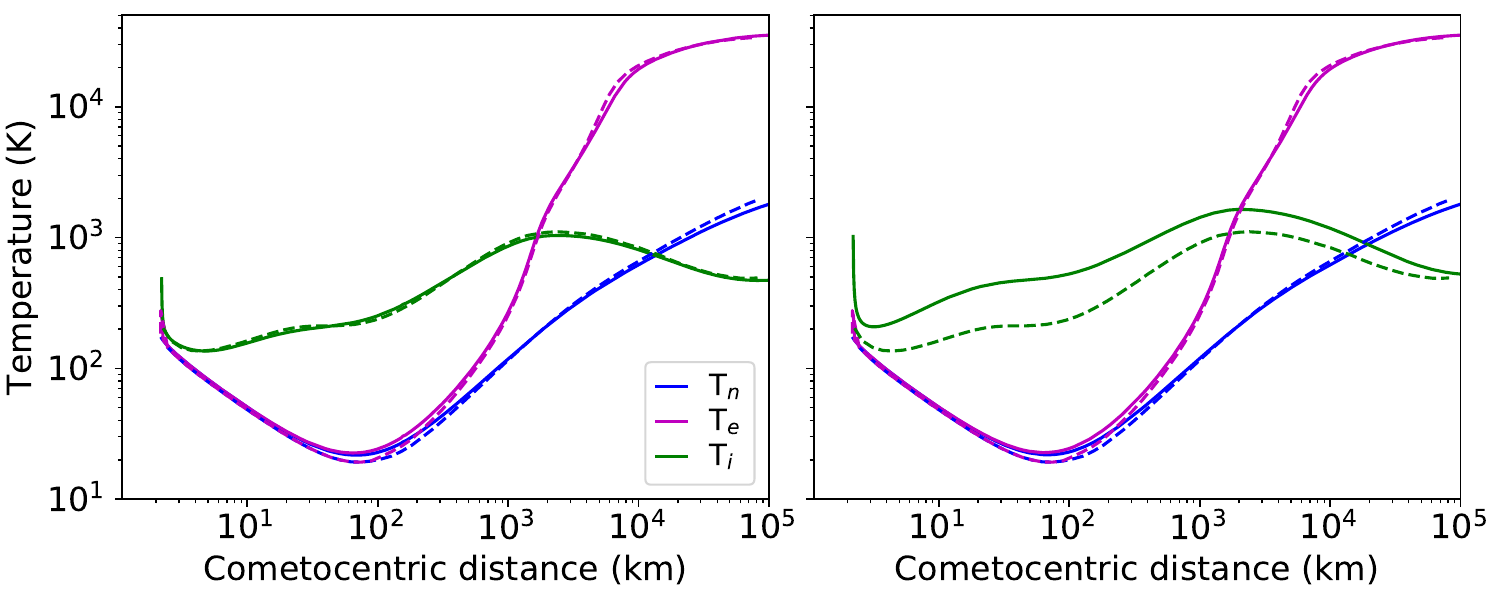}	
	\caption{Left panel: The solid lines show the temperature profiles for C/1996 B2 (Hyakutake) when we use the network of \cite{Weiler2006Thesis} in our model. Right panel: 
	The solid lines show the temperature profiles with the network of \cite{Weiler2006Thesis} but with non-zero $\Delta E$. In both panels, the dashed lines show the temperature profiles obtained by \cite{Weiler2006Thesis}.}
	\label{fig:T_weiler}
\end{figure}

The comets studied in this work have activities in the range of $\sim 1-7 \times 10^{29}$ molecules s$^{-1}$.
Since 1P/Halley also showed activities in this range, we tested the soundness of our model by running it for
1P/Halley with the input conditions given by \cite{Haider2005}. These input conditions are a total gas production rate of $1.3\times 10^{30}$ molecules s$^{-1}$ at 1 au and a nucleus size of 6 km.
The abundance of the other volatiles with respect to
water are: \ch{NH3} = 1.5\%, \ch{CH4} = 0.5\%, \ch{CO2} = 3\%, \ch{CS2} = 0.1\%, \ch{N2} = 0.1\%, \ch{C2H2} = 2\%,
\ch{H2S} = 0.41\%, \ch{CH3OH} = 1.7\%, \ch{SO2} = 0.1\% and \ch{HCN} = 0.1\%. Our modeled temperature and velocity profiles are shown by the solid lines in the left panel of Figure \ref{fig:Halley}. 
The red scatter points in the left panel show the bulk ion speed for the water group ions from the HIS sensor of the ion mass spectrometer (IMS) of \textit{Giotto} \citep{Altwegg1993}.
The green scatter points show the ion temperatures derived from the HIS count rates of masses 18 and 19 corresponding to \ch{H2O+} and \ch{H3O+}, respectively \citep{Balsiger1986}.
The crossing of the contact surface by the spacecraft (at about 4800 km from the comet nucleus; \citealp{Schwenn1987}) is characterized by a sudden drop in the ion temperature. The contact surface separates the cold ions moving with an outflow speed $\geqq 1$ km s$^{-1}$ from the hotter ions and the magnetic field drops to almost zero inside this boundary. These effects cannot be captured in our model and a full MHD treatment is required to correctly model the plasma boundaries \citep{Benna2007,Rubin2014}. Our model cannot distinguish between the cold and hot ion populations, which is why we see a discrepancy with the observed ion temperatures. Including a large chemical network in an MHD model is computationally expensive and beyond the scope of the current work.


The solid lines in the right panel of Figure \ref{fig:Halley} show the total ion density along with the
densities of the water group ions (\ch{H3O+}, \ch{H2O+}, \ch{OH+}) and the protonated high-proton affinity neutrals
(\ch{NH4+}, \ch{CH3OH2+}), as obtained from our model. The yellow scatter points show the total ion density profile (summed over 12-56 amu/e mass-to-charge ratio) obtained from HIS data \citep{Altwegg1993}. 
The other scatter points show the ion density profiles modeled by 
\cite{Haider2005} in order to fit the ion peaks obtained by \textit{Giotto}'s IMS. Some of the profiles agree reasonably well. 
It is not possible to obtain an exact match due to differences in reaction networks. Differences in densities may also arise depending on the parameters used in the energy balance equation, (cf. Appendix \ref{App_C}).

\begin{figure}
	\centering
	\includegraphics[height=7cm, width=0.48\textwidth]{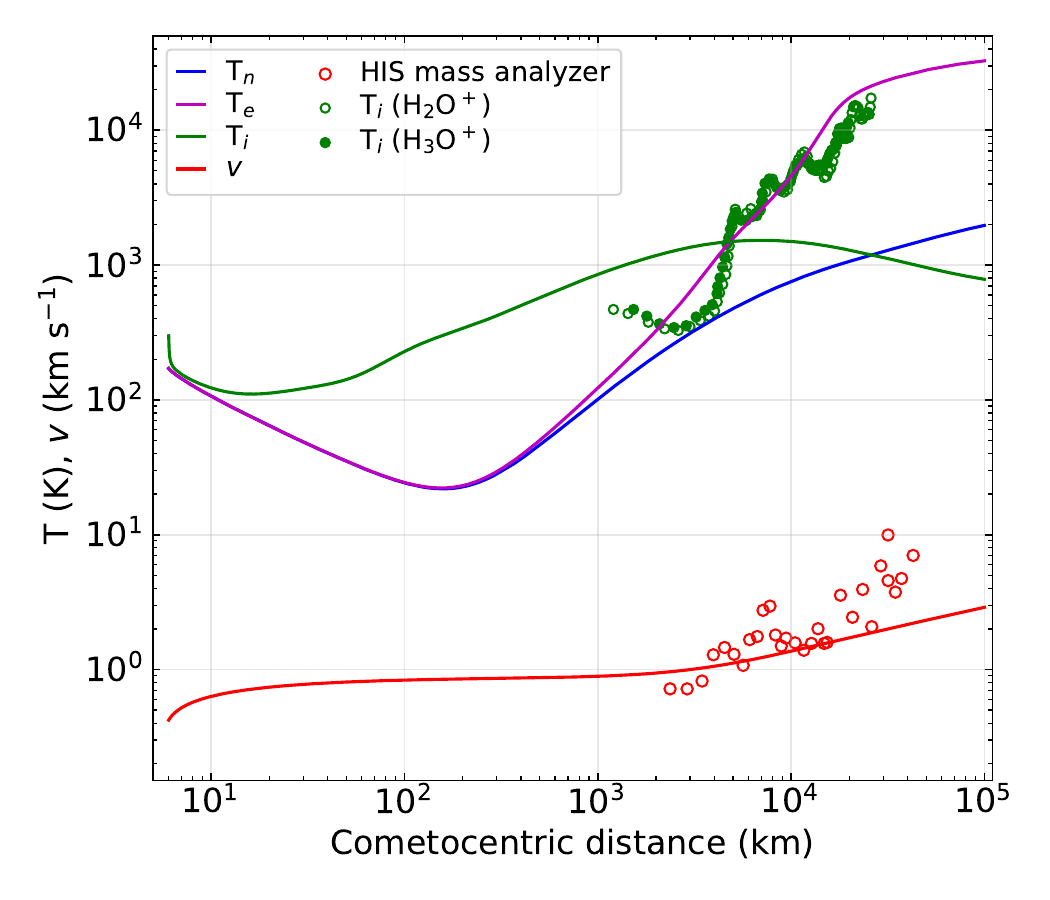}
	\includegraphics[height=7cm, width=0.48\textwidth]{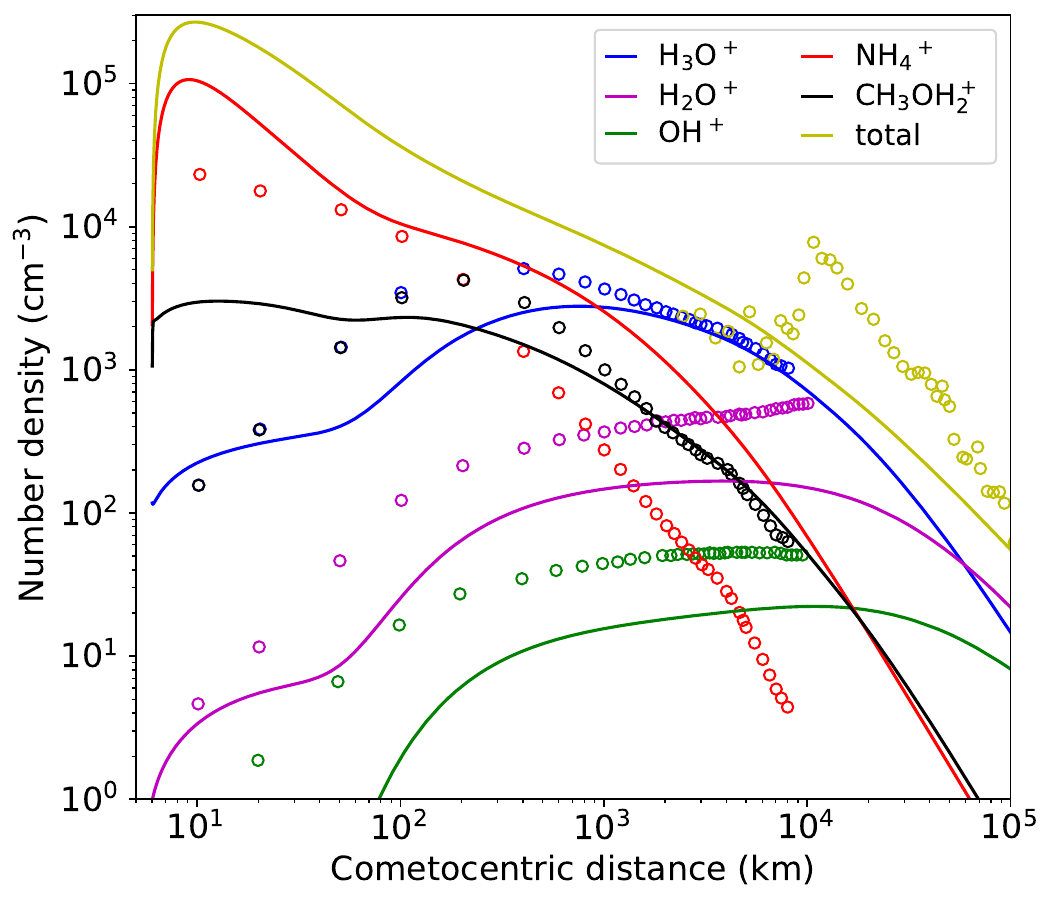}
	\caption{The temperature and velocity profiles (left panel) and total and assorted ion densities (right panel) for 1P/Halley. The solid lines show our modeled values. 
	In the left panel, the green scatter points show the ion temperature derived by \cite{Balsiger1986} from an analysis of the HIS count rates of mass 18 (\ch{H2O+}; unfilled circles) and mass 19 (\ch{H3O+}; filled circles).
	The red scatter points show the bulk ion speed for the water group ions from the HIS mass analyzer \citep{Altwegg1993}.
	In the right panel, the yellow scatter points show the total ion density summed over the mass/charge range of 12-56 amu/e measured by HIS \citep{Altwegg1993} while the
	other scatter points show the assorted ion densities from the model of \cite{Haider2005}. }		
	\label{fig:Halley}
\end{figure}

\end{document}